\documentclass[12pt]{article}

\usepackage{amssymb}
\usepackage{epsfig}
\usepackage{cite}
\usepackage{inputenc}
\inputencoding{latin1}

\def\ariadne{{\sc Ariadne}}
\def\lepto{{\sc Lepto}}
\def\jetset{{\sc Jetset}}
\def\herwig{{\sc HERWIG}}
\def\Pslash{{P\!\!\!\!/}}
\def\tPslash{$\Pslash$}
\def\as{\alpha_s}
\def\tas{$\as$}
\def\tordas{${\cal O}(\as)$}

\def\zcut{{z_{\mbox{{\scriptsize cut}}}}}
\def\tzcut{$\zcut$}
\def\f2{F_2}
\def\tf2{$\f2$}
\def\q2{Q^2}
\def\tq2{$\q2$}
\def\et{E_\perp}
\def\tet{$\et$}
\def\kt{k_\perp}
\def\tkt{$\kt$}
\def\k2t{k_\perp^2}
\def\tk2t{$\k2t$}
\def\alb{\bar{\alpha}}

\def\eq#1{eq.~(\ref{#1})}

\def\eqs#1{eqs.~(\ref{#1})}

\def\refc#1{ref.~\cite{#1}}
\def\refs#1{refs.~\cite{#1}}
\def\fig#1{fig.~\ref{#1}}
\def\figs#1{figs.~\ref{#1}}
\def\laeq{\,\lower3pt\hbox{$\buildrel < \over\sim$}\,}
\def\bigtimes{\scalebox{1.5}{$\times$}}

\newcommand{\ZETF}[3]{{\it Zh.~Eksp.~Teor.~Fiz.} {\bf #1} ({#3}) {#2}}
\newcommand{\PL}[3]{{\it Phys.~Lett.} {\bf #1} ({#3}) {#2}}
\newcommand{\PLB}[3]{{\it Phys.~Lett.} {\bf B#1} ({#3}) {#2}}
\newcommand{\YF}[3]{{\it Yad.~Fiz.} {\bf #1} ({#3}) {#2}}
\newcommand{\SJNP}[3]{{\it Sov.~J.~Nucl.~Phys.} {\bf #1} ({#3}) {#2}}
\newcommand{\JETP}[3]{{\it Sov.~Phys.~JETP} {\bf #1} ({#3}) {#2}}
\newcommand{\NPB}[3]{{\it Nucl.~Phys.} {\bf B#1} ({#3}) {#2}}
\newcommand{\PRD}[3]{{\it Phys.~Rev.} {\bf D#1} ({#3}) {#2}}

\newcommand{\ZPC}[3]{{\it Z.~Phys.} {\bf  C#1} ({#3}) {#2}}

\newcommand{\CPC}[3]{{\it Comput.~Phys.~Comm.} {\bf #1} ({#3}) {#2}}

\newcounter{Aenumct}
\newenvironment{Aenumerate}{\begin{list}{\Alph{Aenumct}}%
{\usecounter{Aenumct}\Alph{Aenumct}}}{\end{list}}

\begin{document}

\begin{titlepage}
  \renewcommand{\thefootnote}{\fnsymbol{footnote}}
  \begin{flushright}
    LU-TP 97-21\\
    NORDITA-97/54 P\\
    hep-ph/9709424\\
    September 1997\\
    Revised\footnote{The original publication was based
      on results from an implementation containing an error. In this
      revised version this error has been corrected, some of the
      beyond leading-log assumptions have been revised and so have some
      of the results.} December 1997

  \end{flushright}
  \begin{center}

    \vskip 10mm
    {\LARGE\bf The Linked Dipole Chain Monte Carlo}
    \vskip 15mm

    {\large Hamid Kharraziha}\\
    Dept.~of Theoretical Physics\\
    Sölvegatan 14a\\
    S-223 62  Lund, Sweden\\
    hamid@thep.lu.se

    \vskip 10mm

    {\large Leif Lönnblad}\\
    NORDITA\\
    Blegdamsvej 17\\
    DK-2100 København Ø, Denmark\\
    leif@nordita.dk

  \end{center}
  \vskip 30mm
  \begin{abstract}
    We present an implementation of the Linked Dipole Chain model for
    deeply inelastic $ep$ scattering into the framework of the
    \ariadne\ event generator. Using this implementation we obtain
    results both for the inclusive structure function as well as for
    exclusive properties of the hadronic final state.

  \end{abstract}

\end{titlepage}

\addtocounter{footnote}{-1}

\section{Introduction}


With the HERA collider, a new kinematical regime has been opened up
for studying deeply inelastic $ep$ scattering (DIS) on partons
carrying a very small momentum fraction $x$ of the proton. Much
theoretical and experimental effort has been made to increase our
understanding of the dynamics of such small-$x$ partons, and much
progress has been made, although many problems still need to be
solved. On the theoretical side, a major issue is how to handle the
resummation of large logarithms of $x$ and $Q^2$ in a consistent way,
while a big obstacle for the experimental analysis has been the lack
of theoretically well founded event generators which are able to
describe the measurements made in this kinematical region.


The conventional way of describing the QCD evolution of the partons
within a hadron, is to look at ladder diagrams where an incoming
parton with large momentum and low virtuality undergo successive
splittings, thus reducing its momentum and increasing its virtuality.
The contributions from splittings where the parton retains only a
small fraction $z\ll 1$ of its momentum or where its virtuality is
increased by a large factor $\nu$, are enhanced by large logarithms of
$1/z$ or $\nu$ respectively, therefore such ladders need to be
resummed to all orders.


The so-called DGLAP\cite{DGLAP} evolution equations handles the
resummation of logarithms of the virtuality by summing all ladder
diagrams where the virtualities, or the transverse momenta \tkt, are
strongly ordered along the chain, while the BFKL\cite{BFKL} equations
perform a $\ln(1/z)$ resummation, summing ladders with strongly
ordered momentum fractions, but unordered in \tkt. Looking only at the
inclusive structure function \tf2, both of these approaches are able
to explain the steep rise with $1/x$ measured at
HERA\cite{HERAWS1}. This is true also for the very small-$x$ region
where the DGLAP equations are thought to be unreliable. However, the
prediction from these approximations rely heavily on the assumption of
the input parton distributions from which the evolution is started,
and the absence of a small-$x$ enhancement in the evolution can be
compensated with a steeply rising input distribution.

From the \tf2\ measurement alone it is therefore difficult to estimate
the relative importance of the different resummation approaches, and
several suggestions have been made to instead look at different
details of the hadronic final state to get a better understanding of
the dynamics of QCD evolution \cite{HERAWS2}. Neither the BFKL or DGLAP
equations are, however, suitable for describing non-inclusive event
properties since that may destroy cancellations between real and
virtual diagrams which are essential for the different approaches. In
contrast, the so-called CCFM\cite{CCFM} evolution equations are
designed to explicitly describe exclusive final-state properties by
very carefully handling interferences between initial- and final-state
splittings in the ladder, based on an angular ordered description. It
can be shown to reproduce both the BFKL and DGLAP equations in their
respective regions of validity.


To properly analyze measured properties of the hadronic final states
found at HERA, it is important to have event generators which
reproduce these properties to a satisfactory level. Unfortunately this
has not been the case. The conventional generators, such as
\herwig\cite{HERWIG} and \lepto\cite{LEPTO}, are based on leading-log
initial state parton showers derived from the DGLAP equations, and
predict much too small partonic activity in the direction of the
incoming proton (hereafter also referred to as the forward direction)
for small-$x$ events \cite{ForwardJets,ForwardParticles}.  This is
expected, as the cascades are strongly ordered in transverse momenta,
and are therefore limited by the smallness of $Q^2$ in these events.
In contrast, the dipole cascade implemented in the
\ariadne\cite{ARIADNE} program, where the generated partons are
unordered in transverse momenta, is able to describe the final-state
properties in the proton direction quite well. From this one may
suspect that the resummation in the BFKL equations indeed are
important, but since there is no clear relationship between the
semi-classical soft radiation model\cite{BGLP} in Ariadne and the BFKL
equation besides the \tkt\ non-ordering, no firm statement can be
made.


It is therefore important to construct an event generator implementing
eg.\ the CCFM evolution, which can make reliable prediction about
exclusive properties of the hadronic final state. Attempts has been
made in this direction\cite{SMALLX}, but several difficulties have been
encountered. A major obstacle with the CCFM equation is the presence
of the so-called non-eikonal form factor which makes any implementation
extremely inefficient.


Recently a reformulation of the CCFM equation has been
proposed\cite{LDC96}. In this, the Linked Dipole Chain (LDC) model,
the division between initial- and final-state radiation diagrams is
redefined using the colour dipole cascade model (CDM). After this
redefinition, the non-eikonal form factor drops out, which allows for
a more simple implementation in an event generator.


A first attempt to construct such an event generator is presented in
this paper. Although the LDC model is well suited for implementation
in a Monte Carlo program there are a number of problems to be
resolved. One problem is that both the CCFM and LDC models only deal
with purely gluonic ladders while, to make a complete event generator,
all types of ladders should be included. It is also important to
handle energy-momentum conservation in a sensible way. Another
important issue is the dependence on the input parton densities and
the Sudakov form factors needed to regularize the poles in the
splitting functions and to conserve the total momentum in the evolved
parton density functions.

In section \ref{sec:ldc} we first recall the main ideas of the LDC
model, then in section \ref{sec:MC} we present the different issues
involved in implementing the model in a Monte Carlo program. To obtain
predictions for the hadronic final state we must first, as described
in section \ref{sec:pdf}, obtain input parton densities which together
with the LDC evolution will give a satisfactory description of the
inclusive cross section. Some results for the hadronic final state are
then presented in section \ref{sec:res}. Finally, our conclusions can
be found in section \ref{sec:sum}.


\section{The Linked Dipole Chain model}
\label{sec:ldc}

It is well known that the cross section for DIS events is not
describable only by the lowest order perturbative terms. Still, it is
not necessary to consider all possible emissions. A large set of them
can be summed over and do, in principle, not affect the cross section.
In the LDC model \cite{LDC96}, the emissions that are considered to
contribute to the cross section are regarded as Initial State
Bremsstrahlung (ISB). The description of these emissions is based on
the CCFM \cite{CCFM} model which is a leading-log approximation 
of the structure function evolution in DIS. The CCFM model has been
modified by redefining which emissions should be counted as ISB,
resulting in a much simplified description.

To describe final state properties of DIS events, one must also
consider Final State Bremsstrahlung (FSB). In the LDC model this is
done within the framework of the Colour Dipole cascade Model (CDM)
\cite{kharrDIPOLE} which has previously proved to give a good
description of parton cascades in hadronic $e^+e^-$ events and DIS.
The general picture is that the initial parton ladder builds a chain
of linked colour dipoles and that the FSB is radiated from these
dipoles.

\begin{figure}
  \begin{center}
    \epsfig{figure=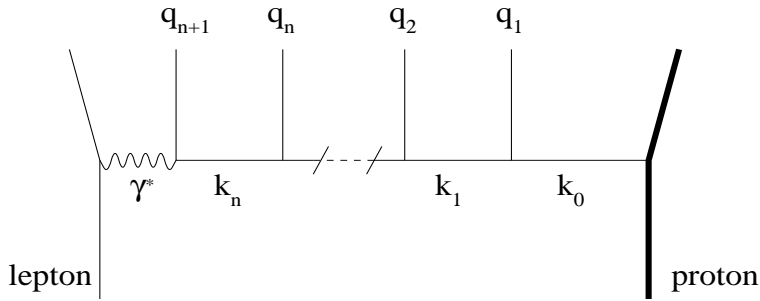,width=10cm,height=4cm}
  \end{center}
  \caption[dummy]{{\it Lepton proton scattering with $n$ perturbative
      ISB emissions.  The emitted ISB partons are denoted $\{ q_i \}$
      and the propagators are denoted $\{ k_i \}$.}}
  \label{fan}
\end{figure}

Let $\{q_i\}$ denote the momenta of emitted partons, $\{k_i\}$ denote
the momenta of the propagators (see \fig{fan})
and $z_{+i}$ be the positive light-cone momentum
fraction\footnote{Note that throughout this paper we work in the
$\gamma^*p$ centre of mass system with the proton along the positive
$z$-axis, except in section \ref{sec:res} when comparing with
experimental data, where the $\gamma^*$ is along the positive
$z$-axis.} of $k_i$ in each emission:
$k_{+i}=z_{+i}k_{+(i-1)}$. According to the CCFM model, the emissions
that contribute to the cross section (ISB) are ordered in rapidity and
energy. Furthermore, there is a restriction on the transverse momenta
of the continuing propagator in each emission:

\begin{equation}
k^2_{\perp i} > z_{+i}q^2_{\perp i}.
\end{equation}
The weight distribution, $dw$, of the initial chains 
factorizes, with the factors $dw_i$ given by the following expression
(${\bar \alpha}=3\alpha_s/\pi$):

\begin{eqnarray}
dw&=&dw_1dw_2\cdots dw_n, \nonumber \\
dw_i&=&{\bar \alpha} 
\frac{dz_{+i}}{z_{+i}} \frac{d^2q_{\perp i}}{\pi q^2_{\perp i}}
\Delta _{ne} \left(z_{+i},k_{\perp i},q_{\perp i}\right).
\end{eqnarray}
$\Delta _{ne}$ is the so called
non-eikonal form factor, given by the expression:
 
\begin{equation}
\Delta _{ne}\left(z,k_\perp,q_\perp\right)=
\exp \left[-{\bar \alpha} \log \left(\frac{1}{z}\right) 
\log \left(\frac{k^2_\perp}{zq^2_\perp}\right)\right].
\end{equation}

In the LDC model, the definition of the ISB is more restricted. Consequently,
more emissions are summed over and the expression for the weight distribution,
$dw_i$, is changed. The new restriction is that in each emission, the 
transverse momentum ($q_{\perp i}$) of the emitted parton must be larger
than the lower one of the transverse momenta of the surrounding propagators
\begin{equation}
q_{\perp i} > \min \left(k_{\perp i},k_{\perp (i-1)} \right).
\label{qtrest}
\end{equation}

\begin{figure}[t]
  \vbox{
    \hbox{
      \epsfig{figure=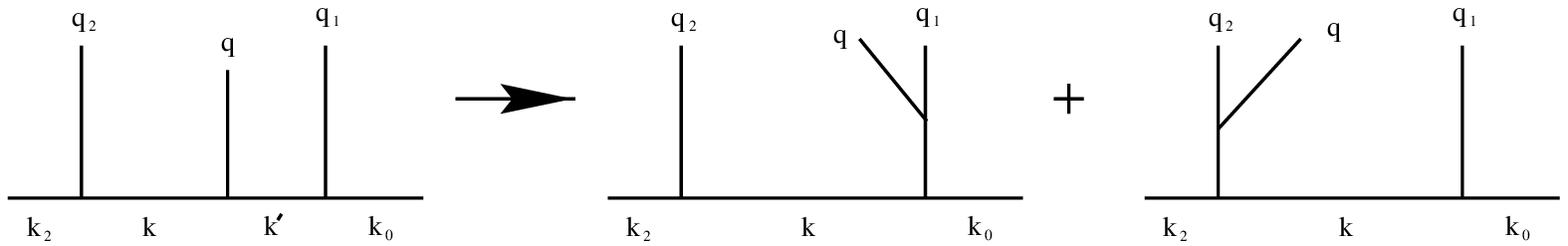,width=14cm,height=3cm}
      }
    \vspace{1.5cm}
    \hbox{
      \epsfig{figure=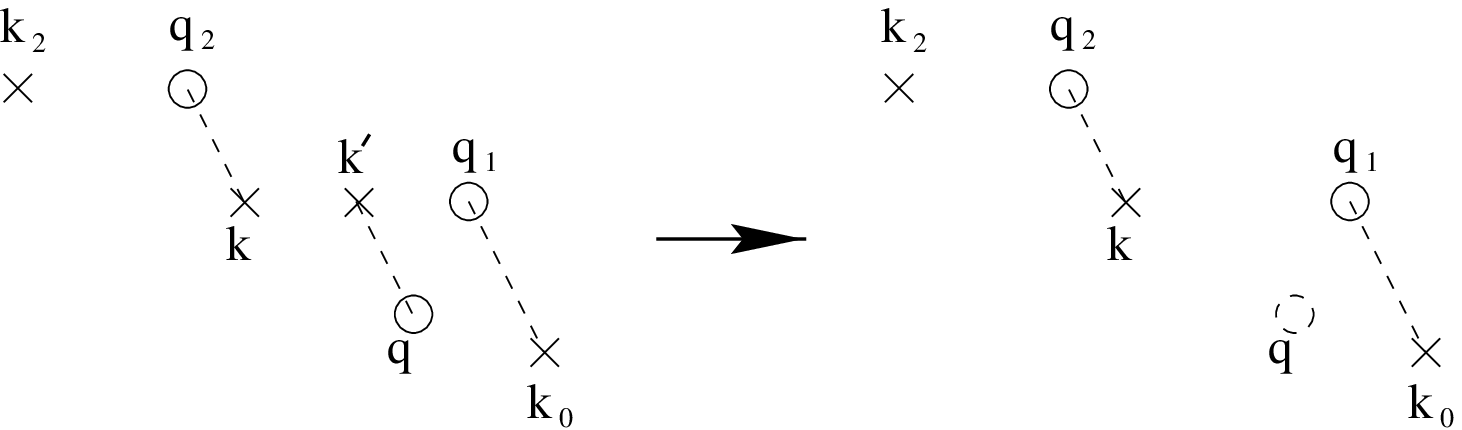,width=11cm,height=4cm}
      \hspace{0.5cm}
      \psfig{figure=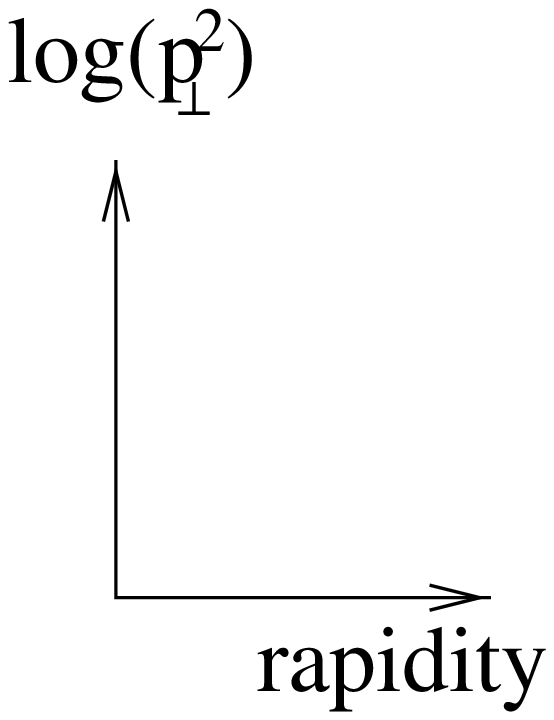,width=2cm,height=4cm}
      }
    }
  \caption[dummy]{{\it CCFM $\rightarrow$ LDC: The emission $q$ is
      regarded as ISB in the CCFM model but in LDC model it is
      regarded as FSB.  This is illustrated (above) with fan diagrams
      and (below) in a $\log p_\perp^2-$rapidity diagram.  Rapidity
      here is defined as $\log (p_+/p_\perp )$ and the dashed lines
      indicate equal $p_+$.}}
  \label{ccfmvsldc}
\end{figure}

In \fig{ccfmvsldc} we show an example of an emission which
belongs to the ISB according to the CCFM model but violates the
restriction in \eq{qtrest} and is regarded as FSB in the LDC model.

Summing over these emissions, the weight distribution of each allowed
emission now becomes
\begin{equation}
  dw_i={\bar \alpha} \frac{dz_{+i}}{z_{+i}} \frac{d^2q_{\perp i}}{\pi
    q^2_{\perp i}}.
\end{equation}
This simplification of the expression for $dw_i$ is due to the fact
that one can interpret the non-eikonal form factor, $\Delta _{ne}$, as
a Sudakov form factor, that is, it is equal to the probability of not
violating the restriction in \eq{qtrest}.

By changing variables to the propagator momenta and integrating $dw_i$ 
over the azimuthal angle (in the transverse plane) it can be written
approximately as:

\begin{equation}
dw_i=\bar{\alpha} 
\frac{dk_{\perp i}^2}{k^2_{\perp i}}
\frac{dz_{+i}}{z_{+i}}
{\rm min} \left(1,\frac{k^2_{\perp i}}{k^2_{\perp (i-1)}} \right).
\end{equation}

It is instructive to look at three different possibilities for two
subsequent emissions, specified by different orderings of the
transverse momenta of the propagators, numbered $k_1$, $k_2$ and $k_3$
(with the same order as in \fig{fan}):

\begin{itemize}
\item $k_{\perp 1}<k_{\perp 2}<k_{\perp 3}$: This is included in DGLAP
  and gives the same weight as there: $dw_2dw_3 \propto
  \frac{1}{k^2_{\perp 2}} \frac{1}{k^2_{\perp 3}}$.
\item $k_{\perp 1}<k_{\perp 2}>k_{\perp 3}$: This gives the weight
  $dw_2dw_3 \propto \frac{1}{k^4_{\perp 2}}$ and resembles a hard
  sub-collision with a momentum transfer $\hat{t} \simeq k^2_{\perp
  2}$.
\item $k_{\perp 1}>k_{\perp 2}<k_{\perp 3}$: Here $dw_2dw_3 \propto
  \frac{1}{k^2_{\perp 1}} \frac{1}{k^2_{\perp 3}}$, it has no factor
  $\frac {1}{k^2_{\perp 2}}$ and is thus infrared safe!
\end{itemize} 

The LDC model, without corrections to the leading log approximation,
has previously been studied and some qualitative results for the
structure functions and final state properties in DIS have been
presented in \refs{LDC96,kharrLDC2,HamidDIS97,HamidNext}. It is found
that the LDC model, just as the CCFM model, interpolates smoothly
between DGLAP and BFKL. The emissions along the rapidity axis can be
separated in two phases: For rapidities closest to the proton
(forward) direction, the ISB chain performs a BFKL like motion with a
constant mean $E_\perp$-flow. At a certain distance from the photon
end, the transverse momentum of the emissions begins to rise to the
photon virtuality, as expected by DGLAP. For the structure functions,
a DGLAP behaviour ($\exp(\mbox{const}\sqrt{-\log x})$) is shown for
moderate values of $x$, but for small $x$-values it has the BFKL
$x^{-\lambda}$ behaviour.

\begin{figure}
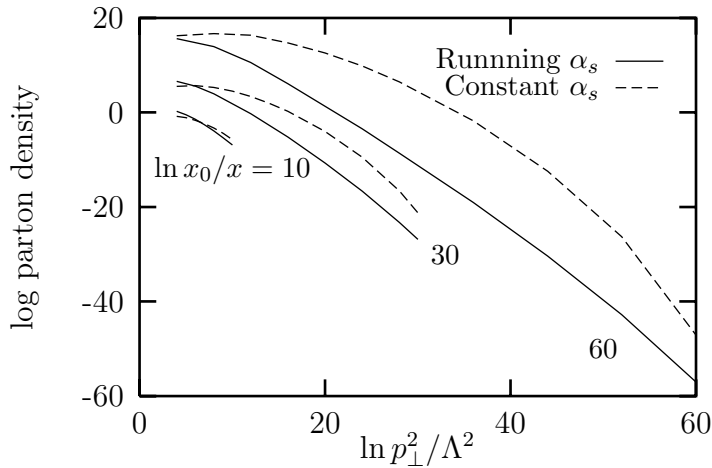

  \begin{center}
    \input st3.tex
  \end{center}
  \caption[dummy]{{\it The decrease of log of the parton density with
      $\ln p_\perp^2$ at a certain rapidity, for different $x$-values,
      for running (solid line) and constant (dashed line) coupling.}}
  \label{kharrET}
\end{figure}

A prediction from the BFKL model is that $\log{p_\perp}$ of the
emitted partons along the chain would be described by a Gaussian
distribution with a growing width $\propto \log{1/x}$. From the result
of the LDC model, one can clearly see that this BFKL behaviour is
indeed present for a constant coupling but not for a running coupling.
This is illustrated in \fig{kharrET} where the parton density at a
certain rapidity is plotted as a function of $\log{p_\perp}$ for
different values of $x$-Bjorken, for constant and running coupling.
The Gaussian behaviour is observed for constant coupling, while for a
running coupling it appears to decay exponentially.  It seems though
that the exponential decay is significant only for events with very
small $x$-values and will probably not be visible in currently
available data.

\begin{figure}
    \hbox{
      \psfig{figure=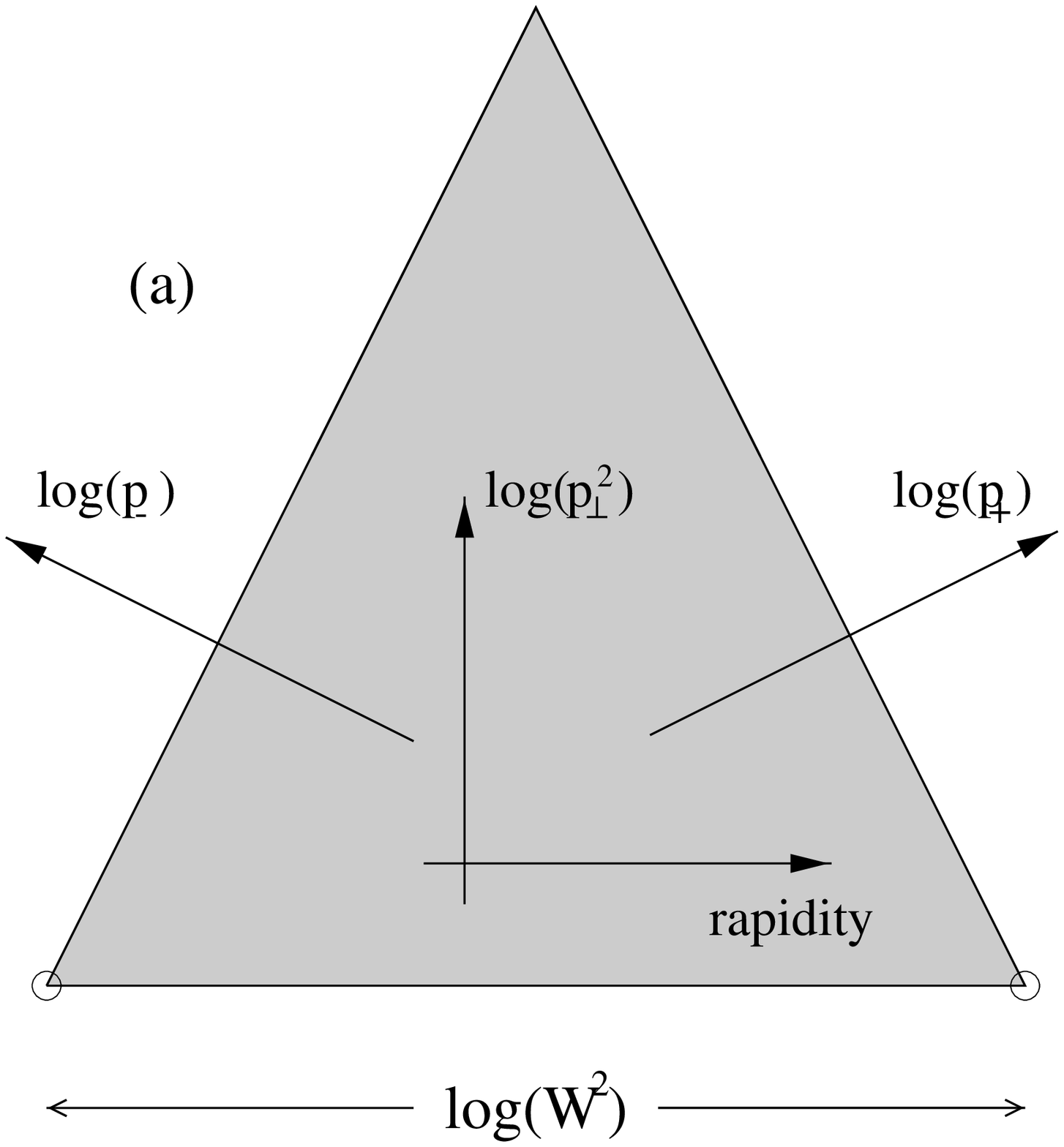,width=4.5cm,height=6cm}
      \hspace{0.2cm}
      \vbox{
        \hbox{
        \psfig{figure=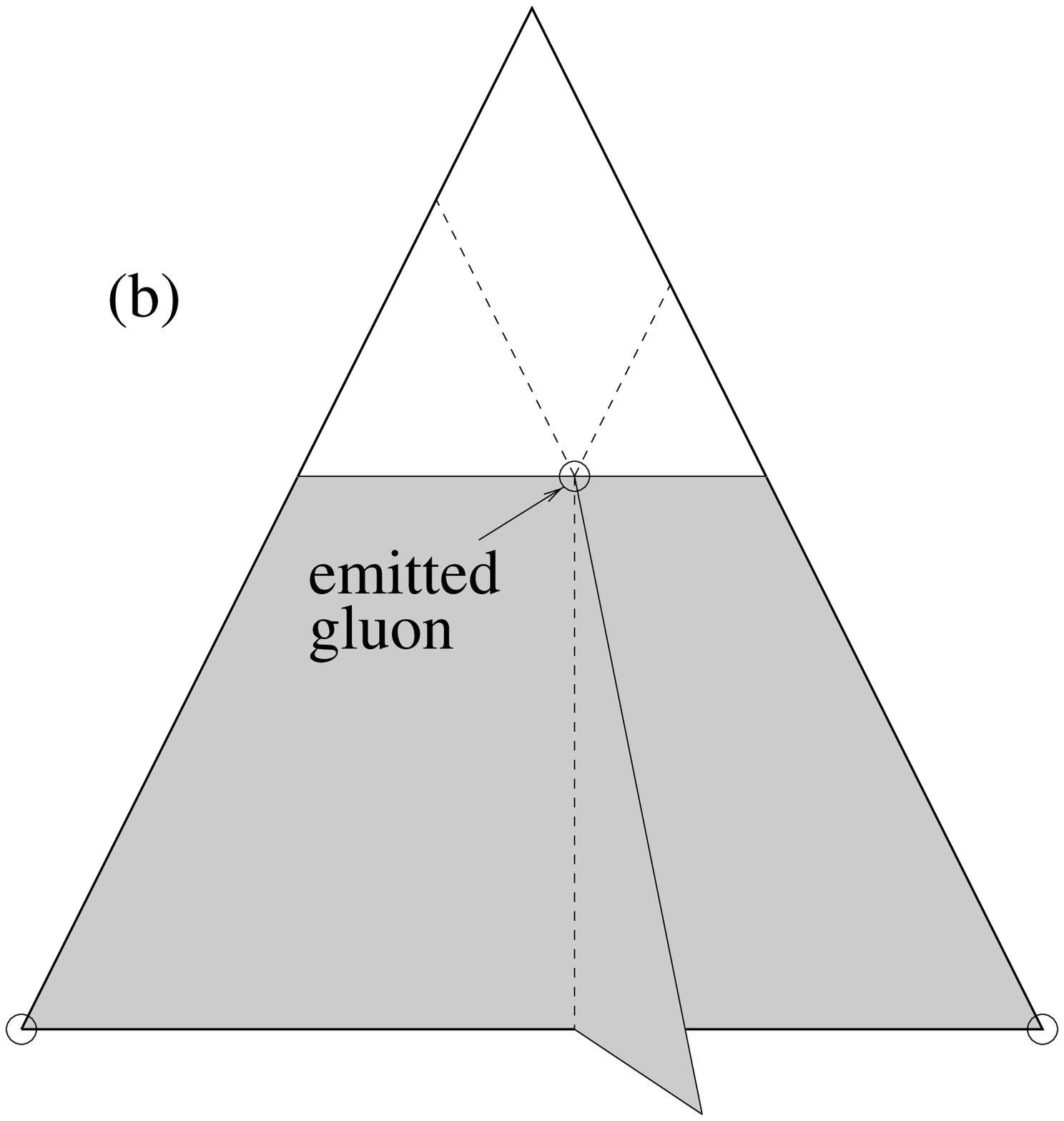,width=4.3cm,height=5.6cm}
        \hspace{0.2cm}
        \psfig{figure=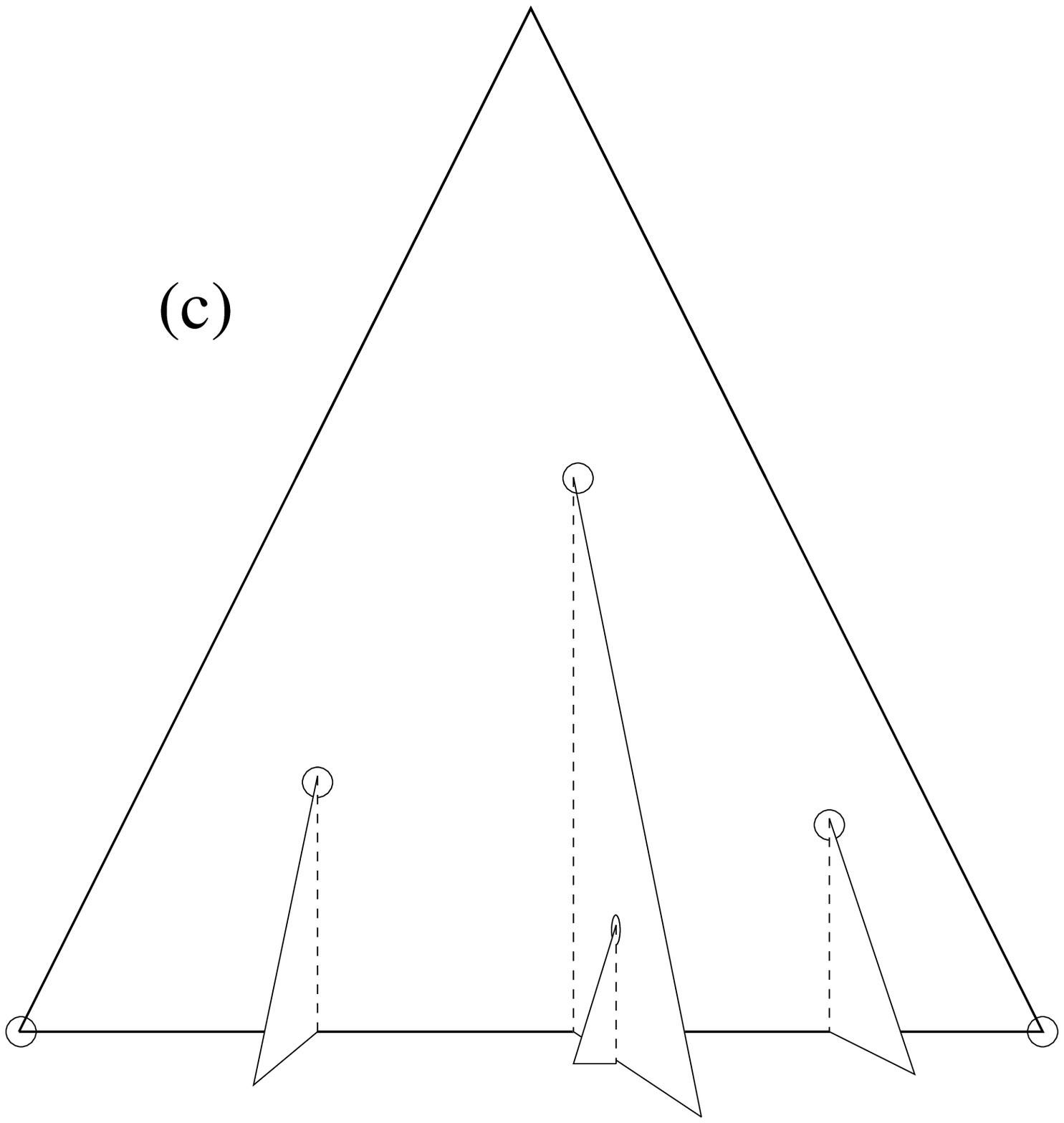,width=4.3cm,height=5.6cm}
        }
      \vspace{0.4cm}
      }
    }

  \caption[dummy]{{\it (a) The phase space of parton emission from a
      $q-{\bar q}$ pair.  (b) After the first emission, the phase
      space is split into two triangle which radiate gluons
      independently, but under a $p_\perp$-ordering condition.  (c) An
      event with four emissions.}}
  \label{kharrTRI1}
\end{figure}

The dipole model \cite{kharrDIPOLE} was originally developed for
final-state parton cascades from a quark-anti quark system.  The phase
space for gluon emission from a $q-{\bar q}$ pair is approximately
given by the triangular area in \fig{kharrTRI1}a. The gluons are
assumed to be radiated from a $q-{\bar q}$ colour dipole and after
each emission, the dipole is split into smaller dipoles
(\fig{kharrTRI1}b), which continue to radiate independently under a
$p_\perp $-ordering condition (shaded area). Also $g\rightarrow
q+{\bar q}$ splittings have been included in this model.  The size of
the dipole triangle is determined by the total $q-{\bar q}$ invariant
mass $W$.

\begin{figure}

  \hbox{
    \epsfig{figure=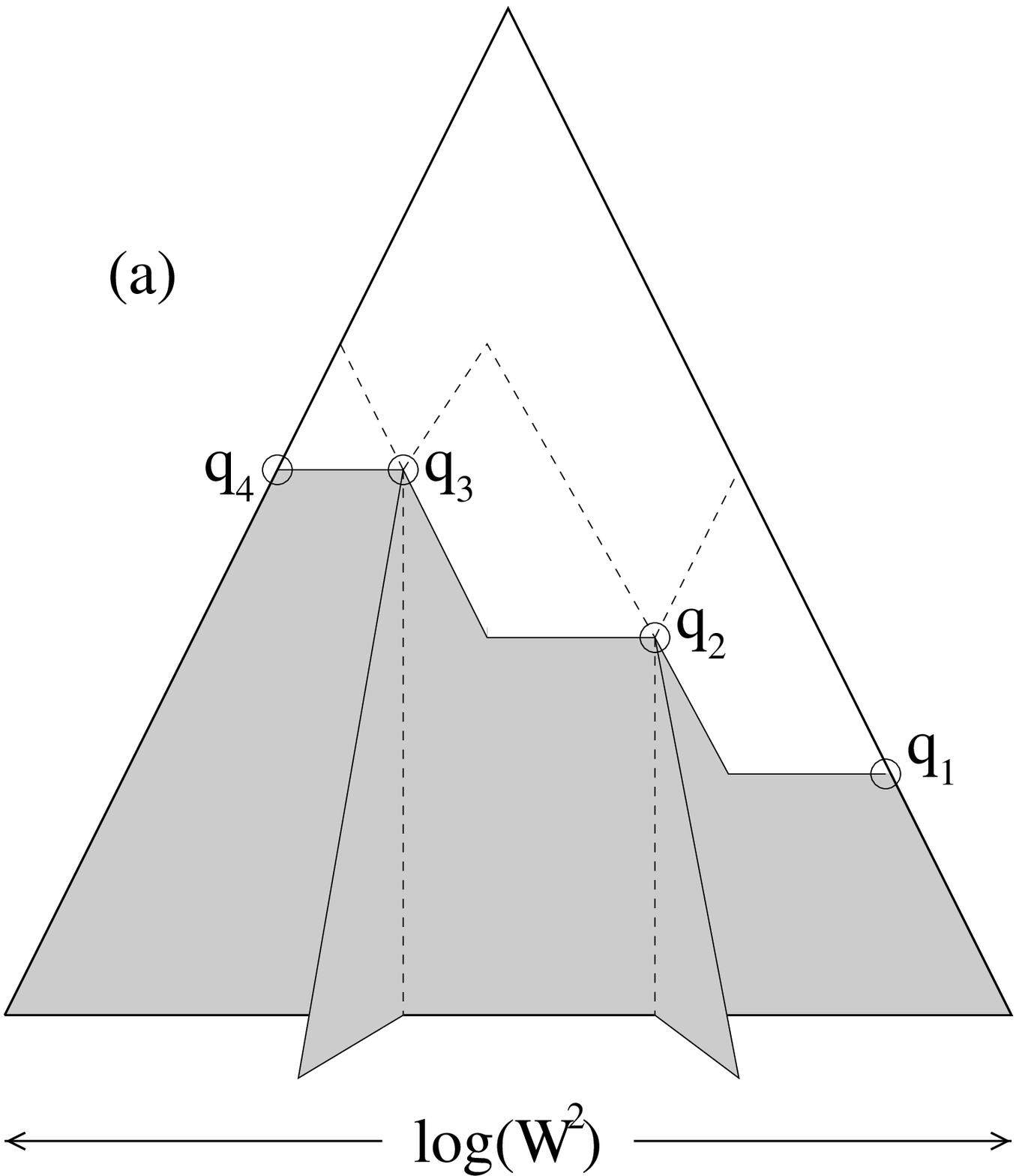,width=6cm,height=7cm}
    \hspace{2cm}
    \vbox{
      \epsfig{figure=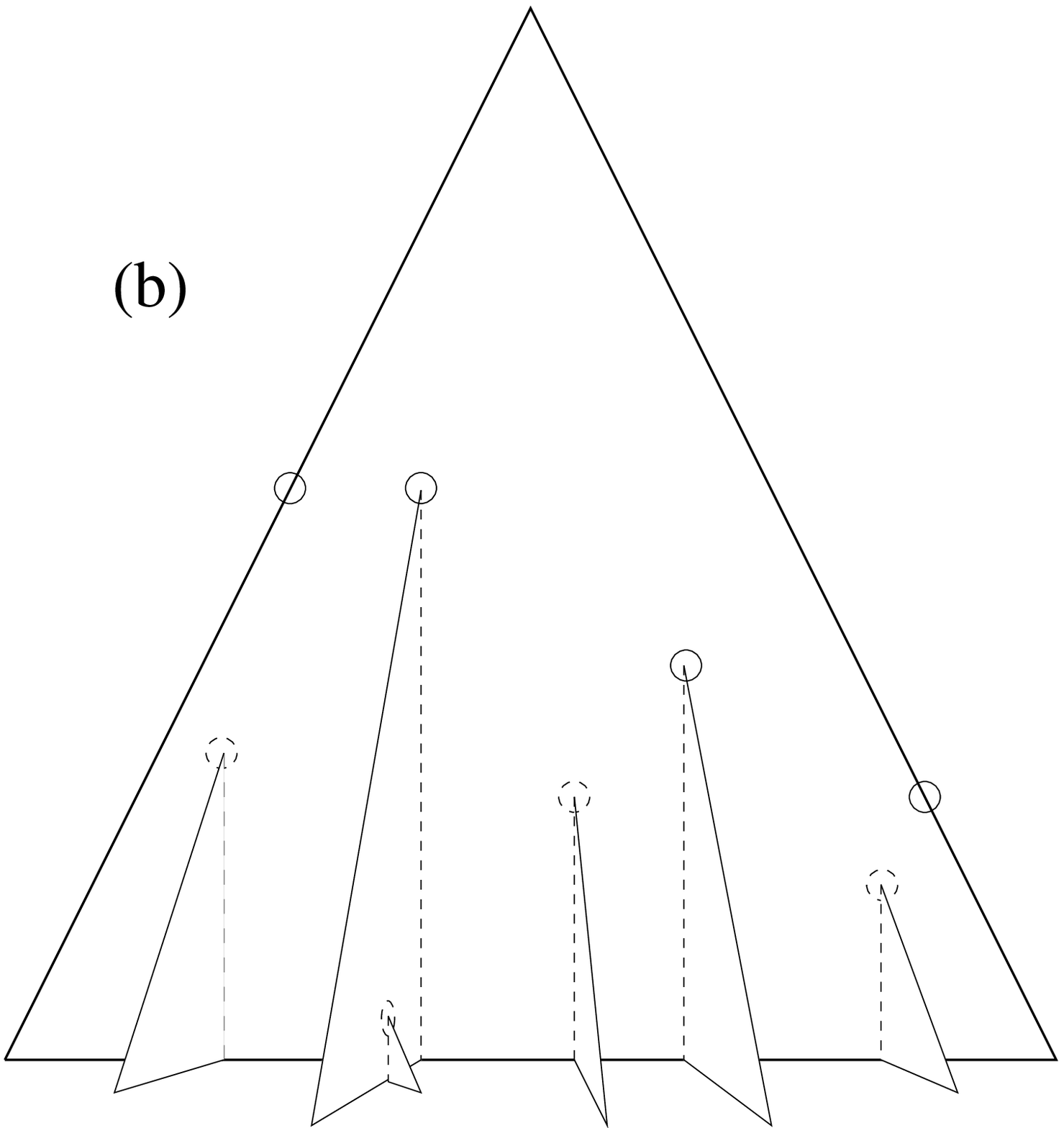,width=5.7cm,height=6.5cm}
      \vspace{0.5cm}
      }
    }

  \caption[dummy]{{\it (a) ISB emissions plotted in a dipole triangle.
      (b) The solid circles are ISB emissions and the dashed circles are
      FSB emissions.}}
  \label{LDCTRI}
\end{figure}

The momenta $\{ q_i \}$ of the emitted ISB partons in DIS are plotted in
\fig{LDCTRI}a. Due to
the ordering in positive and negative light cone momenta, one can
insert the ISB into a dipole triangle with a size determined by the photon
negative light cone momentum (left edge) and the positive light cone
momentum of the incoming (non-perturbative) gluon (right edge). After
doing this, the FSB partons can be emitted in a similar way as for the
$q-{\bar q}$ parton shower (\fig{LDCTRI}b). The phase space of the FSB
is the shaded area in \fig{LDCTRI}a.

\section{The Monte Carlo implementation}
\label{sec:MC}


The LDC model has been implemented in a Monte Carlo program and some
results within the leading log approximation have already been
presented. Here we present a more complete implementation taking into
account some non-leading corrections and generating complete events
all the way down to the final state hadron level. What follows is a
step-by-step description of the procedure.

The basic formula for the evolution of the parton densities is
\begin{eqnarray}
  x f_i(x,Q^2) = \sum_j\int_x^1 \frac{dx_0}{x_0}
  \left[\right.\begin{array}[t]{l} G_{ij}(x,Q^2,x_0,k_{\perp 0}^2)+\\~\\
  \left. S_j(Q^2,k_{\perp
    0}^2)\delta_{ij}(\ln{x}-\ln{x_0})\right] x_0f_{0j}(x_0,k_{\perp 0}^2),\end{array}
  \label{eq:evolved}
\end{eqnarray}
Where $G_{ij}(x,Q^2,x_0,k_{\perp 0}^2)$ is the sum of the weights of
all chains starting with a parton $j$ at some low scale $k_{\perp
  0}^2$ carrying a momentum fraction $x_0$, and ending up with a
parton $i$ carrying a momentum fraction $x$ being hit by a photon with
virtuality $Q^2$. The delta function corresponds to the case of no
emissions. $G$ is positive definite, and to conserve the total
momentum, $\sum_j\int_0^1xf_j(\q2,x)dx = 1$, the delta function is
multiplied by a Sudakov form factor, $S_j(Q^2,k_{\perp 0}^2)$,
representing the probability that the parton $j$ with with momentum
fraction $x_0$ at $k_{\perp 0}^2$ has not split, and thus reduced its
momentum, when probed at a higher scale \tq2. This form factor will be
discussed in detail below.

The analytic approximate upper limiting function for $G$
\cite{kharrLDC2}, given by
\begin{eqnarray}
  \label{eq:Ganalytic}
  &G_{ij}(x,Q^2,x_0,k_{\perp 0}^2) \laeq G(Q^2/k_{\perp 0}^2,x/x_0) = 
  \sqrt{\frac{a}{b}}I_1(2\sqrt{ab})&\\
  &a=\sqrt{\alb}(\ln{Q^2/k_{\perp 0}^2}+\ln{x_0/x}),\mbox{~~~}
  b=\sqrt{\alb}\ln{x_0/x}&\nonumber
\end{eqnarray}
provides us with the starting point, and for each generated chain $c$
a number of multiplicative weights are calculated
$\omega_c=\Pi_l\omega^{(l)}_c$ so that the correct form of $G_{ij}$ is
obtained as
\begin{equation}
  G_{ij}(x,Q^2,x_0,k_{\perp 0}^2) =
  \bar{\omega}_{ij}(x,Q^2,x_0,k_{\perp 0}^2)G(Q^2/k_{\perp 0}^2,x_0/x),
  \label{eq:Gexact}
\end{equation}
with the average weight
\begin{equation}
  \bar{\omega}=\frac{1}{N}\sum_{c=1}^{N}\omega_c
\end{equation}

This is how it is done:
\begin{enumerate}
\item First the $x$, $Q^2$ and the flavour $i$ of the struck quark is
  chosen using evolved parton densities and the standard Born-level
  electro-weak matrix elements. This is currently done within the
  \lepto\ program \cite{LEPTO}.
\item Then the $x_0$ and flavour of the incoming parton is chosen
  according to \eqs{eq:evolved} and (\ref{eq:Ganalytic}) with
  $k_{\perp 0}$ as a given parameter.
\item The number of emissions is chosen from
  \label{enum:wbegin}
  \begin{equation}
    \sqrt{\frac{a}{b}}I_1(2\sqrt{ab}) =
    \sum_{n=1}^{\infty}\frac{a^{n}b^{n-1}}{n!(n-1)!}
  \end{equation}
\item The positive and negative light-cone momentum fractions,
  $z_{j+}$ and $z_{j-}$ which enters in each emission $j$ is generated
  according to the ordered integral
  \begin{equation}
    \frac{a^{n}b^{n-1}}{n!(n-1)!} =
    \int\alb^n\Pi_j\frac{dz_{j+}}{z_{j+}}\frac{dz_{j-}}{z_{j-}}
    \delta(\ln{x_0}+\sum_j \ln{z_{j+}}-\ln{x})
    \label{eq:zgen}
  \end{equation}
\item At this point we need to choose the
  flavours of each link. To do this we introduce the standard
  Altarelli-Parisi splitting functions and preliminary we use
their approximations, $\tilde{P}_{i\rightarrow j}(z)$, in the 
$z\rightarrow 0$ limit
  \begin{eqnarray}
    P_{q\rightarrow q}(z) &=& C_F\frac{1+z^2}{1-z}
\approx C_F,\nonumber\\
    P_{g\rightarrow g}(z) &=& 2N_C\frac{(1-z(1-z))^2}{z(1-z)}
\approx 2N_C\frac{1}{z},\nonumber\\
    P_{g\rightarrow q}(z) &=& T_R(z^2+(1-z)^2)
\approx T_R,\\
    P_{q\rightarrow g}(z) &=& C_F\frac{1+(1-z)^2}{z}
\approx C_F\frac{2}{z},\nonumber
  \end{eqnarray}
  We get the first weight factor as the approximated splitting
  functions summed over all possible flavour combinations,
  \begin{equation}
    \label{eq:omega1}
    \omega^{(0)} =
    \sum\Pi 
\frac{\tilde{P}_{i\rightarrow j}(z_+)}{\tilde{P}_{g\rightarrow g}(z_+)}.
  \end{equation}
\item We can then use \eq{eq:omega1} to generate a specific flavour
  combination according to their individual weights.
\item Next, we generate the azimuthal angles of each emission and
  construct the exact kinematics. The delta function in \eq{eq:zgen},
  which handles the conservation of positive light-cone momenta, does
  not take into account the transverse degrees of freedom. In
  particular it would give zero positive light-cone momentum for the
  struck quark, $q_{n+1}$, in the final-state. We therefore modify
  this delta function to exactly conserve the total energy and
  momentum, effectively setting $z_{n+}$ by hand to the value
  needed. However, for some values of the azimuth angles this is not
  possible and we get a weight factor corresponding to the allowed
  integration area $\Delta \phi_j$:
  \begin{equation}
    \label{eq:genzphi}
    \omega^{(1)}=\Pi_j\frac{1}{2\pi}\int_{\Delta \phi_j}d\phi_j.
  \end{equation}
\item Then we implement the condition that the transverse momenta
  (which is generalized to the transverse mass $m_\perp$ for massive
  partons) of an emitted parton must be larger than the smallest
  virtuality $v_{j\min}$ of the connecting links $j$ and $j-1$, and
  that all virtualities must be above $k_{\perp 0}^2$, giving us the
  second weight factor
  \begin{equation}
    \label{eq:omega2}
    \omega^{(2)}=\Pi_j\Theta(m_{\perp j}^2-v_{j\min})
    \Theta(v_{j\min}-k_{\perp 0}^2).
  \end{equation}
  We note that the weights are finite even if one of $v_j$ and
  $v_{j-1}$ goes to zero, and one could imagine replacing the second
  theta function in \eq{eq:omega2} with $\Theta(v_{j\max}-k_{\perp
    0}^2)$. This would reduce the dependency on $k_{\perp 0}$ and
  would allow for more unordered chains as discussed below.
  
\item Now we introduce the running of \tas, giving a fourth weight
  \begin{equation}
    \label{eq:omega4}
    \omega^{(3)} = \Pi_j \frac{1}{\ln(m_{\perp j}^2/\Lambda^2)}
  \end{equation}
\item 
\label{enum:splitfn} 
Having obtained the virtualities of the links, we can now
  correct the splitting functions in \eq{eq:omega1}. We get the
  following cases:
  \begin{itemize}
  \item $v_{j+1}>v_{j}>v_{j-1}$: Going
    upwards from the proton side we use $P_{f_{j-1}\rightarrow f_{j}}(z_{j+})$
  \item $v_{j+1}<v_{j}<v_{j-1}$: Going
    upwards from the photon side we use $P_{f_{j+1}\rightarrow f_{j}}(z_{j-})$
  \item $v_{j+1}<v_{j}>v_{j-1}$: Corresponds to a Rutherford
    scattering and $z_{j+}\approx z_{j-}\equiv z$. Here we use
    $2\rightarrow 2$ matrix elements taking into account colour
    connections as explained in Appendix A.
  \item $v_{j+1}>v_{j}<v_{j-1}$: Here we use the $z\rightarrow 0$
    limit, $\tilde{P}_{i\rightarrow j}(z)$ of the splitting functions,
    where $z=z_+$ if $v_{j+1}>v_{j-1}$ or else $z=z_-$.
  \end{itemize}
  Note that the colour factor is independent of the ordering of the
  virtualities. Therefor we have to correct for only the kinematical
  part of the splitting function by using reduced splitting functions
  \tPslash\ where the colour factor is divided out.  Here we also
  introduce the Sudakov form factor, to be discussed below and we can
  write the fourth weight factor
  \begin{equation}
    \label{eq:omega5}
    \omega^{(4)}=\Pi_j
    \frac{S_j(v_{j-1},v_{j},v_{j+1})\Pslash^{v_{j-1}v_{j}v_{j+1}}
                      _{f_{j-1}f_{j}f_{j+1}}(z_{j+},z_{j-})}
    {\tilde{\Pslash}_{f_{j-1}\rightarrow f_{j}}(z_{j+})},
  \end{equation}
  where $S_j(v_{j-1},v_{j},v_{j+1})$ is $S_{f_{j-1}}(v_j,v_{j-1})$ or
  $S_{f_{j+1}}(v_j,v_{j+1})$ depending on whether the virtuality is
  going up or down.

  $P_{q\rightarrow q}$ and $P_{g\rightarrow g}$ both have poles as
  $z\rightarrow 1$, corresponding to emission of low-energy
  gluons. Typically these should be counted as final-state emissions,
  but to be sure to avoid divergences we introduce a cutoff
  $z_{\mbox{\tiny cut}}=0.5$. See also the discussion of double
  counting below.
  
  Note also that we use mass less splitting functions, and the
  production of heavy quarks is only suppressed by the phase space.
  This should be improved in the future.

\item \label{enum:wend} The final weight factor is introduced to correct
  the emission closest to the photon, in the cases where $v_n>v_{n-1}$,
  to reproduce the exact ${\cal O}(\alpha\as)$ matrix element as given
  eg.\ in \refc{Kramer}:

  \begin{equation}
    \label{eq:omega6}
    \omega^{(5)}=\frac{{\cal M}(Q^2,x,z_{n+},z_{n-})}
    {P^{v_{n-1}v_{n}Q^2}_{f_{n-1}f_{n}\gamma}(z_{n+},z_{n-})}
  \end{equation}
\item The generated chain is now kept with a probability
  $\omega_c=\frac{1}{W}\Pi_l\omega^{(l)}_c$, where $W$ is a scale
  factor to avoid probabilities larger than one. There is in principle
  nothing preventing weight larger than one, but they turn out to be
  very rare. Nevertheless, it may happen, and it is important to check
  that $W$ is large enough so that the results are not influenced by
  this. Chains with $\omega>1$ may optionally be saved and retrieved
  again when an event with the same flavour and similar $x$ and \tq2\
  is requested. A chain will then be used on the average $\omega_c$
  times, each time with different final-state cascade and
  hadronization. Below we have used $W=1$ giving less than 0.1\%
  events with weight larger than one.
\item To prepare for the final-state dipole radiation the emitted
  partons must be connected together and form dipoles. In the case of
  quark links, this is straight forward, the incoming quark is simply
  connected to the first emitted gluon and so on until the struck
  quark. In the case of gluon links, there are two colour lines, and a
  radiated gluon can belong to either of these. This choice is done
  completely at random. The connection between the colour line of the
  incoming parton and the proton remnant is handled in the same way as
  in the default soft radiation model of \ariadne\cite{ARIADNE}.
  
  One could imagine using other methods for determining the
  colour-flow. One suggestion is to use the colour-flow which minimize
  the total string length\footnote{As defined eg.\ by the $\lambda$
  measure of \refc{lambda}}. One could also consider colour-{\bf
  re}connections, following eg.\ the model already implemented in
  \ariadne\ \cite{colrec}, possibly giving rise to large rapidity gaps
  among the final-state hadrons.
\item The constructed dipoles can then radiate more final-state gluons
  in the phase-space limited by the virtuality and the positive and
  negative light-cone momenta of the links in the chain as in
  \fig{LDCTRI}a. Note that also in the no-emission case corresponding
  to the delta function in \eq{eq:evolved}, some radiation is allowed
  within the triangular area defined by $q_+<-Q_+$ and $q_-<Q_-$. In
  the Breit frame, this is just the area of allowed FSB in a
  $e^+e^-\rightarrow q\bar{q}$ event with centre of mass energy $Q$.
\item Finally the final state dipole chains are hadronized according
  to the Lund string fragmentation model as implemented in \jetset\
  \cite{JETSET}.
  
\end{enumerate}

This concludes the description of the actual implementation. But
before we can start producing events we have to fix the parameters
involved. These are $\Lambda$, $k_{\perp 0}$, $W$, $z_{\mbox{\tiny
cut}}$ and the input parton densities $x_0f_{0j}(x_0,k_{\perp 0}^2)$.
$W$ is not really a physical parameter, and should be set large enough
so that the result no longer depend on it. It would be natural to take
$\Lambda$ and $k_{\perp 0}$ to be the values which have been tuned for
the final state dipole cascade to reproduce LEP data,
$\Lambda_{\mbox{\tiny LEP}}$ and $k_{\perp 0\mbox{\tiny LEP}}$. One
could, of course, use an increased cutoff $k_{\perp 0}>k_{\perp
0\mbox{\tiny LEP}}$ in the ISB. This would mean that more emissions
would be moved from the initial to the final state, which would
continue emitting down to $k_{\perp 0\mbox{\tiny LEP}}$. We therefore
expect the final result to be fairly stable w.r.t.\ such variations as
long as $k_{\perp 0}$ is not too large.

There is an additional complication with a large $k_{\perp 0}$ for the
cases where the virtuality drops below $k_{\perp 0}$ somewhere along
the chain, causing a zero weight in \eq{eq:omega2}. For the total
cross section, this does not matter, as such fluctuations are included
in the input parton densities at a lower $x_0$, corresponding to the
momentum fraction of the link closest to the photon which is below the
cutoff. For the final state, however, it means that we are excluding
some radiation close to the direction of the incoming hadron.

As discussed above, one could replace the second theta function in
\eq{eq:omega2} with $\Theta(v_{j\max}-k_{\perp 0}^2)$ since the
$m_\perp^2$ then would still be in the perturbative region. In this
way the result would be less sensitive to variations of $k_{\perp 0}$,
as the perturbative system on the proton side of the sub-cutoff link
would still be generated. This would not, of course, solve the problem
altogether as one can imagine chains where two or more consecutive
links are below the cutoff.

In step \ref{enum:splitfn} we have replaced the $N_c/z$ pole, which is
used in the emissions of the original leading log LDC model, with the
standard Altarelli-Parisi splitting functions. This should be a
sensible way of including some sub-leading effects, as long as the
splitting functions are regularized in a correct way. In each
emission, two particles are produced and one vanishes. For the parton
distributions, this means we must subtract and add partons
correspondingly. 

A simple way of treating this double counting problem is to add only
one of the produced partons assuming that the mother parton is not
affected by the emission.  The choice of which of the two partons to
add is not trivial. One way is to introduce a cut-off $z_{cut}=0.5$
allowing only $z<z_{cut}$. This is a good choice in the $g\rightarrow
g$ splitting since the gluon with $z>0.5$ is more similar to the
mother gluon. For the other emissions, which involve both quarks and
gluons, it is more important to make the choice which leads to a better
approximation of the quark distributions. This can be done by allowing
all $q \rightarrow g$ splittings, forbidding all $q \rightarrow q$
splittings and allowing $g \rightarrow q$ splittings only if the quark
(or anti-quark) interacts directly with the photon.

In a more sophisticated treatment, the subtraction of a parton is done
with Sudakov form factors corresponding to the probability for the
partons not to vanish before the splitting can occur. In this way, we
can take into account that e.g. the possibility for a gluon to split
into a quark anti-quark pair with low virtuality reduces its
contribution to emissions with higher virtualities. The suppression
factor becomes an exponential of an integral over the splitting
\begin{equation}
S_g=\exp\left[ -\int_0^1 P_{g\rightarrow q}(z)dz \int
\frac{\alpha_s(q_\perp^2)}{2\pi} \frac{dq_\perp^2}{q_\perp^2} \right].
\end{equation}
The region of integration corresponds to the region of allowed
emissions. Here we only use an approximate form. A more thorough
investigation will be presented in a future paper.

The lower limit on $q_\perp^2$ is given by the lowest virtuality of an
emission step multiplied by a fudge factor, $e^\delta$, to account
for the suppression of the emission probability for small
$q_\perp$-values (\eq{qtrest}). The value of $\delta=0.4$ that is used
has proven to be an effective cut-off in the leading log treatment of
the LDC model \cite{kharrLDC2,HamidNext}. The upper limit is set
to the highest virtuality of an emission step. One could imagine
having a higher, or a $z$ dependent limit, but for simplicity we only
use the virtuality in this publication.

The Sudakov factor due to the $g\rightarrow g$ splitting depends on
the choice of $z_{cut}$. For $z_{cut}=0.5$, there is no double
counting to correct for and for larger $z_{cut}$ the integration of
the splitting function is in the region $0.5<z<z_{cut}$. If $q
\rightarrow q$ splittings are allowed, quarks are suppressed with a
Sudakov factor where the integration region is given by $0<z<z_{cut}$.

\begin{figure}[t]
  \begin{center}
    \hbox{
      \epsfig{figure=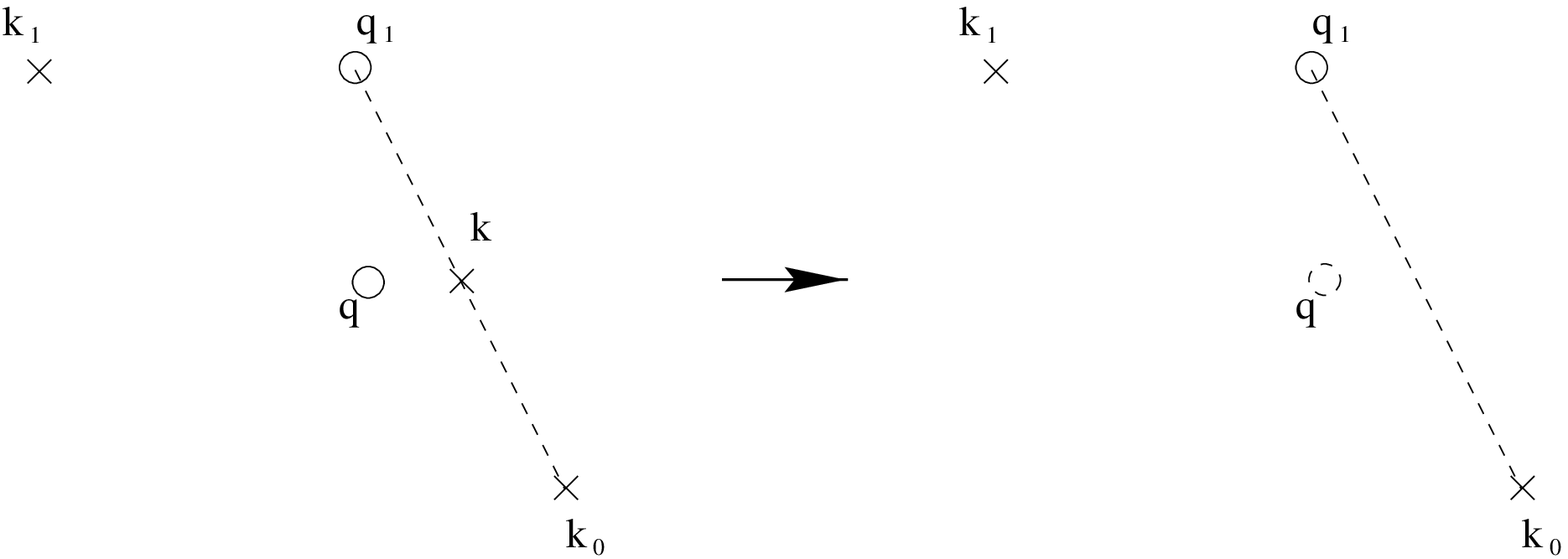,width=12cm,height=4cm}
      \hspace{0.5cm}
      \psfig{figure=koord.ps,width=2cm,height=4cm}
      }
    \caption[dummy]{{\it In the case where the virtuality is increasing,
        a splitting with $z$ larger than 0.5 corresponds to an
        emission of a gluon $q$ which is inside the phase space where
        subsequent FSB emissions is allowed. Such splittings should
        therefore be removed from the ISB chains to avoid double
        counting.}}
    \label{fig:largez}
  \end{center}
\end{figure}

The situation is quite different when we are interested in the final
state properties. Here, the Sudakov form factors are not as important
since they for most events roughly give an overall factor
$\prod_{j=1}^nS(v_{j-1},v_j,v_{j+1})\sim S(k_{\perp 0}^2,Q^2)$ which
does not have an influence on the relative contributions of different
final states. On the other hand, disallowing some of the initial state
emissions to reduce double counting has a large effect on the final
state properties since it reduces high $p_\perp$ emissions. This is a
problem except for gluon emissions with $z>0.5$ in regions with ordered
virtuality, since these gluons can be treated as final state emissions
as shown in \fig{fig:largez}.  Consequently, we would expect it to be
a good approximation for the final state generation to skip the
Sudakov form factors and to allow some of the splittings which lead to
double counting.

It is clear that the result is very dependent on the non-perturbative
input parton densities, which are basically unknown. If eg.\ the gluon
density is very divergent at small $x$, the $x_0$ chosen from
\eq{eq:evolved} will tend to be small, limiting the total phase space
available for radiation $\Delta y\propto\ln{x_0/x}$. The input parton
densities can, however, be constrained somewhat from the total cross
section, and we can parametrize them and make a fit of the parameters
to eg.\ \tf2\ data at different $x$ and $Q^2$.


\section{Fitting the input parton densities}
\label{sec:pdf}

From \eq{eq:evolved} we can write the leading order expression for
\tf2\ as
\begin{eqnarray}
  \label{eq:F2}
  \f2=\sum_{i\ne 0} e_i^2 \sum_j\int_x^1\frac{dx_0}{x_0}
  \left[\right.\begin{array}[t]{l}G_{ij}(x,Q^2,x_0,k_{\perp 0}^2)+\\~\\
    S_j(Q^2,k_{\perp
    0}^2)\delta_{ij}(\ln{x}-\ln{x_0})\left.\right]
  x_0f_{0j}(x_0,k_{\perp 0}^2).\end{array}
\end{eqnarray}

For given values of $x$, $Q^2$, $x_0$, $k_{\perp 0}^2$, $i$ and $j$,
we can calculate $G_{ij}(x,Q^2,x_0,k_{\perp 0}^2)$ from steps
\ref{enum:wbegin} through \ref{enum:wend} in the previous section
using \eq{eq:Gexact}. For any given parametrization of the input
densities it is then possible to calculate \tf2\ and compare with
experimental data.

In principle one could also fit to other data, such as prompt photon
and jet production in hadron-hadron collisions. This is not possible
in our current implementation which only gives the evolved densities
for quarks and anti-quarks. This means that the input gluon
distribution is only constrained indirectly from the \tq2\ dependence
of \tf2, and in this paper we only make a very crude fit using only
four different parameters for $x_0f_{0j}(x_0,k_{\perp 0}^2)$.

The input densities are parametrized as
\begin{equation}
  \label{eq:f0param}
  xf_{0j}(x) = A_j x^{\alpha_j}(1-x)^{\beta_j}.
\end{equation}
For the valence distributions $u_v(x)$ and $d_v(x)$ we use the same
form, with $\beta_v=3$ leaving $\alpha_v$ free and using the normalization
\begin{equation}
  \label{eq:valnorm}
  \int_0^1u_v(x)dx=2,\mbox{~~~}\int_0^1d_v(x)dx=1
\end{equation}
to fix $A_{uv}$ and $A_{dv}$. For the gluon distribution, $\beta_g=4$
while $\alpha_g$ and $A_g$ are left free. All the sea-quarks
distributions have the same form with $\beta_S=4$, leaving $\alpha_S$
free and setting $2A_s=2A_{\bar{s}}=A_u=A_{\bar{u}}=A_d=A_{\bar{d}}$
so that the total momentum
\begin{equation}
  \label{eq:allnorm}
  \int_0^1dx\sum_jxf_{0j}(x)=1
\end{equation}
is conserved.

We only use data from proton \tf2\ measurements from H1\cite{H1F2},
ZEUS\cite{ZEUSF2}, NMC\cite{NMCF2} and E665\cite{E665F2} without
allowing for any normalization uncertainty factors. We use only data
for $\q2>1.5$ GeV and $x<0.5$ to ensure a reasonable length of the
evolution. We then make six sets of fits using different options in
the generation of the $G$ function:
\begin{Aenumerate}
\item LDC default: $k_{\perp 0}=0.6$ GeV,
   $\Lambda=0.22$ GeV.
\item \label{enum:DGLAP} DGLAP: As for A but only allow chains with
   monotonically increasing virtualities of the links from the proton
   side.
\item DGLAP': As for B, but chains where the virtuality of the link
   closest to the virtual photon is larger than $Q^2$ are
   permitted. We use this as a kind of higher-order corrected DGLAP
   evolution although, of course, not equivalent to NLO evolution.
\item As for A but $k_{\perp 0}=1$ GeV, to check the
  sensitivity to this cutoff.
\item As for A but without the Sudakov form factor. Instead
  $P_{q\rightarrow q}(z)$ is set to zero and $P_{g\rightarrow q}(z)$
  is nonzero only in the splitting closest to the photon.
\item As for A but $\beta_g = \beta_S = 5$ to check the
  sensitivity to the fit parameters.
\item As for A but only fitting to \tf2\ data with $x<0.1$, to reduce
  the sensitivity to the step size $\delta \ln{x_0/x}=0.2$ used when
  integrating \eq{eq:F2}, and to the high-$x$ form of the input
  parametrization.
\item As for A but allow the virtuality of some links to be below
  $k_{\perp 0}$ as long as the largest virtuality of two consecutive
  links always is above $k_{\perp 0}$.
\end{Aenumerate}

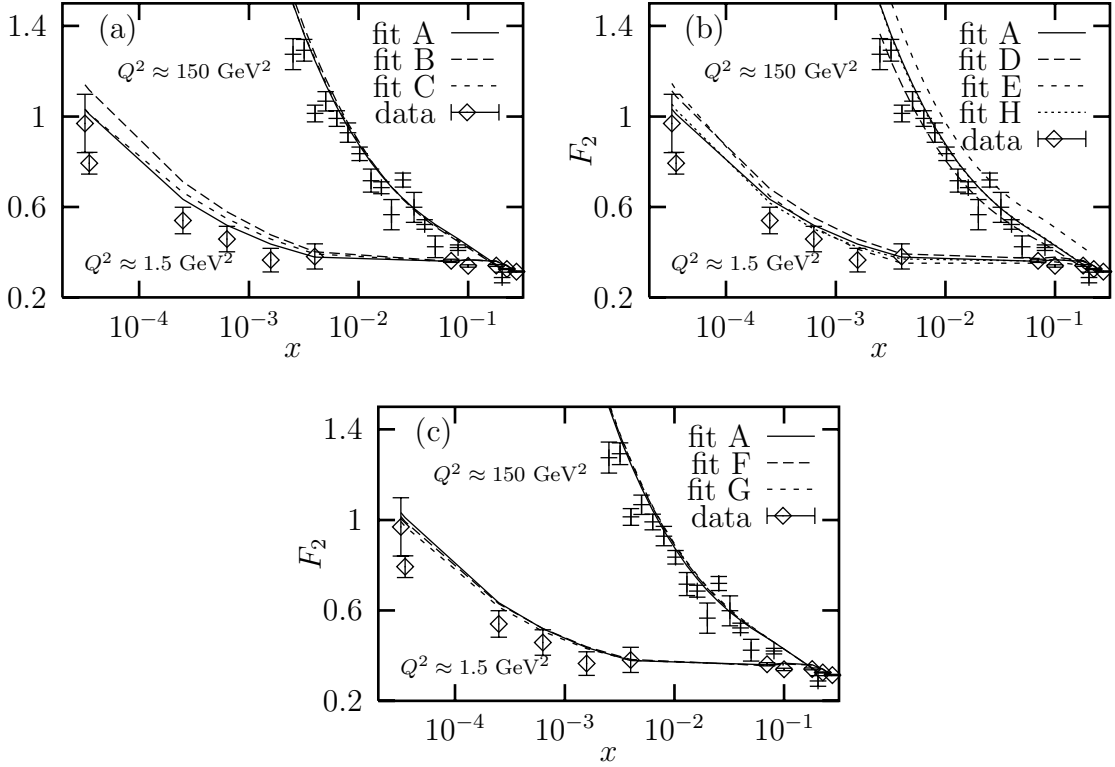
\begin{figure}[t]
  \vskip -1cm
  \hbox{
    \hskip -1.5cm
\setlength{\unitlength}{0.1bp}
\special{!
/gnudict 40 dict def
gnudict begin
/Color false def
/Solid false def
/gnulinewidth 5.000 def
/vshift -33 def
/dl {10 mul} def
/hpt 31.5 def
/vpt 31.5 def
/M {moveto} bind def
/L {lineto} bind def
/R {rmoveto} bind def
/V {rlineto} bind def
/vpt2 vpt 2 mul def
/hpt2 hpt 2 mul def
/Lshow { currentpoint stroke M
  0 vshift R show } def
/Rshow { currentpoint stroke M
  dup stringwidth pop neg vshift R show } def
/Cshow { currentpoint stroke M
  dup stringwidth pop -2 div vshift R show } def
/DL { Color {setrgbcolor Solid {pop []} if 0 setdash }
 {pop pop pop Solid {pop []} if 0 setdash} ifelse } def
/BL { stroke gnulinewidth 2 mul setlinewidth } def
/AL { stroke gnulinewidth 2 div setlinewidth } def
/PL { stroke gnulinewidth setlinewidth } def
/LTb { BL [] 0 0 0 DL } def
/LTa { AL [1 dl 2 dl] 0 setdash 0 0 0 setrgbcolor } def
/LT0 { PL [] 0 1 0 DL } def
/LT1 { PL [4 dl 2 dl] 0 0 1 DL } def
/LT2 { PL [2 dl 3 dl] 1 0 0 DL } def
/LT3 { PL [1 dl 1.5 dl] 1 0 1 DL } def
/LT4 { PL [5 dl 2 dl 1 dl 2 dl] 0 1 1 DL } def
/LT5 { PL [4 dl 3 dl 1 dl 3 dl] 1 1 0 DL } def
/LT6 { PL [2 dl 2 dl 2 dl 4 dl] 0 0 0 DL } def
/LT7 { PL [2 dl 2 dl 2 dl 2 dl 2 dl 4 dl] 1 0.3 0 DL } def
/LT8 { PL [2 dl 2 dl 2 dl 2 dl 2 dl 2 dl 2 dl 4 dl] 0.5 0.5 0.5 DL } def
/P { stroke [] 0 setdash
  currentlinewidth 2 div sub M
  0 currentlinewidth V stroke } def
/D { stroke [] 0 setdash 2 copy vpt add M
  hpt neg vpt neg V hpt vpt neg V
  hpt vpt V hpt neg vpt V closepath stroke
  P } def
/A { stroke [] 0 setdash vpt sub M 0 vpt2 V
  currentpoint stroke M
  hpt neg vpt neg R hpt2 0 V stroke
  } def
/B { stroke [] 0 setdash 2 copy exch hpt sub exch vpt add M
  0 vpt2 neg V hpt2 0 V 0 vpt2 V
  hpt2 neg 0 V closepath stroke
  P } def
/C { stroke [] 0 setdash exch hpt sub exch vpt add M
  hpt2 vpt2 neg V currentpoint stroke M
  hpt2 neg 0 R hpt2 vpt2 V stroke } def
/T { stroke [] 0 setdash 2 copy vpt 1.12 mul add M
  hpt neg vpt -1.62 mul V
  hpt 2 mul 0 V
  hpt neg vpt 1.62 mul V closepath stroke
  P  } def
/S { 2 copy A C} def
end
}
\begin{picture}(2519,1511)(0,0)
\special{"
gnudict begin
gsave
50 50 translate
0.100 0.100 scale
0 setgray
/Helvetica findfont 100 scalefont setfont
newpath
-500.000000 -500.000000 translate
LTa
LTb
600 251 M
63 0 V
1673 0 R
-63 0 V
600 592 M
63 0 V
1673 0 R
-63 0 V
600 933 M
63 0 V
1673 0 R
-63 0 V
600 1275 M
63 0 V
1673 0 R
-63 0 V
889 251 M
0 63 V
0 1046 R
0 -63 V
1303 251 M
0 63 V
0 1046 R
0 -63 V
1716 251 M
0 63 V
0 1046 R
0 -63 V
2129 251 M
0 63 V
0 1046 R
0 -63 V
600 251 M
1736 0 V
0 1109 V
-1736 0 V
600 251 L
LT0
2065 1249 M
180 0 V
685 961 M
16 -19 V
1054 621 L
166 -98 V
165 -71 V
166 -49 V
514 -16 V
64 6 V
106 -6 V
40 -17 V
36 -25 V
LT1
2065 1149 M
180 0 V
685 1053 M
16 -20 V
1054 687 L
1220 574 L
165 -87 V
166 -64 V
514 -36 V
64 6 V
106 -5 V
40 -16 V
36 -25 V
LT2
2065 1049 M
180 0 V
685 960 M
16 -17 V
1054 647 L
166 -99 V
165 -77 V
166 -56 V
514 -30 V
64 7 V
106 -5 V
40 -16 V
36 -24 V
LT0
1469 1358 M
42 -116 V
41 -99 V
40 -89 V
42 -85 V
42 -76 V
44 -70 V
43 -63 V
39 -49 V
38 -44 V
43 -43 V
42 -38 V
40 -31 V
40 -29 V
86 -54 V
2257 360 L
LT1
1478 1360 M
33 -91 V
41 -101 V
40 -92 V
42 -88 V
42 -80 V
44 -74 V
43 -67 V
39 -52 V
38 -48 V
43 -46 V
42 -40 V
40 -34 V
40 -30 V
86 -55 V
2257 356 L
LT2
1469 1356 M
42 -111 V
41 -96 V
40 -88 V
42 -84 V
42 -77 V
44 -71 V
43 -64 V
39 -50 V
38 -46 V
43 -45 V
42 -39 V
40 -33 V
40 -30 V
86 -55 V
2257 358 L
LT0
2125 949 D
685 907 D
701 757 D
1054 541 D
1220 471 D
1385 392 D
1551 405 D
2065 389 D
2129 370 D
2235 371 D
2275 358 D
2311 348 D
2065 949 M
180 0 V
-180 31 R
0 -62 V
180 62 R
0 -62 V
685 797 M
0 220 V
654 797 M
62 0 V
-62 220 R
62 0 V
701 716 M
0 82 V
670 716 M
62 0 V
-62 82 R
62 0 V
1054 491 M
0 100 V
1023 491 M
62 0 V
-62 100 R
62 0 V
1220 423 M
0 96 V
-31 -96 R
62 0 V
-62 96 R
62 0 V
1385 347 M
0 89 V
-31 -89 R
62 0 V
-62 89 R
62 0 V
135 -78 R
0 95 V
-31 -95 R
62 0 V
-62 95 R
62 0 V
483 -70 R
0 12 V
-31 -12 R
62 0 V
-62 12 R
62 0 V
33 -30 R
0 9 V
-31 -9 R
62 0 V
-62 9 R
62 0 V
75 -6 R
0 7 V
-31 -7 R
62 0 V
-62 7 R
62 0 V
9 -19 R
0 4 V
-31 -4 R
62 0 V
-62 4 R
62 0 V
5 -14 R
0 5 V
-31 -5 R
62 0 V
-62 5 R
62 0 V
LT0
1469 1168 A
1511 1183 A
1552 945 A
1592 991 A
1634 926 A
1676 872 A
1720 793 A
1763 691 A
1802 665 A
1840 563 A
1883 694 A
1925 591 A
1965 526 A
2005 442 A
2091 439 A
2257 326 A
1469 1110 M
0 117 V
-31 -117 R
62 0 V
-62 117 R
62 0 V
11 -85 R
0 82 V
-31 -82 R
62 0 V
-62 82 R
62 0 V
10 -311 R
0 63 V
-31 -63 R
62 0 V
-62 63 R
62 0 V
9 -22 R
0 73 V
-31 -73 R
62 0 V
-62 73 R
62 0 V
11 -131 R
0 59 V
-31 -59 R
62 0 V
-62 59 R
62 0 V
11 -119 R
0 73 V
-31 -73 R
62 0 V
-62 73 R
62 0 V
13 -142 R
0 51 V
-31 -51 R
62 0 V
-62 51 R
62 0 V
12 -170 R
0 87 V
-31 -87 R
62 0 V
-62 87 R
62 0 V
8 -93 R
0 45 V
-31 -45 R
62 0 V
-62 45 R
62 0 V
7 -181 R
0 114 V
1809 506 M
62 0 V
-62 114 R
62 0 V
12 47 R
0 53 V
-31 -53 R
62 0 V
-62 53 R
62 0 V
11 -186 R
0 113 V
1894 534 M
62 0 V
-62 113 R
62 0 V
9 -139 R
0 36 V
-31 -36 R
62 0 V
-62 36 R
62 0 V
9 -143 R
0 82 V
-31 -82 R
62 0 V
-62 82 R
62 0 V
55 -55 R
0 21 V
-31 -21 R
62 0 V
-62 21 R
62 0 V
2257 306 M
0 40 V
-31 -40 R
62 0 V
-62 40 R
62 0 V
stroke
grestore
end
showpage
}
\put(2005,949){\makebox(0,0)[r]{data}}
\put(2005,1049){\makebox(0,0)[r]{fit C}}
\put(2005,1149){\makebox(0,0)[r]{fit B}}
\put(2005,1249){\makebox(0,0)[r]{fit A}}
\put(683,379){\makebox(0,0)[l]{{\scriptsize $Q^2\approx 1.5$ GeV$^2$}}}
\put(807,1104){\makebox(0,0)[l]{{\scriptsize $Q^2\approx 150$ GeV$^2$}}}
\put(808,1260){\makebox(0,0){(a)}}
\put(1468,51){\makebox(0,0){$x$}}
\put(400,805){%
\special{ps: gsave currentpoint currentpoint translate
270 rotate neg exch neg exch translate}%
\makebox(0,0)[b]{\shortstack{$ $}}%
\special{ps: currentpoint grestore moveto}%
}
\put(2129,151){\makebox(0,0){$10^{-1}$}}
\put(1716,151){\makebox(0,0){$10^{-2}$}}
\put(1303,151){\makebox(0,0){$10^{-3}$}}
\put(889,151){\makebox(0,0){$10^{-4}$}}
\put(540,1275){\makebox(0,0)[r]{1.4}}
\put(540,933){\makebox(0,0)[r]{1}}
\put(540,592){\makebox(0,0)[r]{0.6}}
\put(540,251){\makebox(0,0)[r]{0.2}}
\end{picture}
    \hskip -1.5cm
\setlength{\unitlength}{0.1bp}
\special{!
/gnudict 40 dict def
gnudict begin
/Color false def
/Solid false def
/gnulinewidth 5.000 def
/vshift -33 def
/dl {10 mul} def
/hpt 31.5 def
/vpt 31.5 def
/M {moveto} bind def
/L {lineto} bind def
/R {rmoveto} bind def
/V {rlineto} bind def
/vpt2 vpt 2 mul def
/hpt2 hpt 2 mul def
/Lshow { currentpoint stroke M
  0 vshift R show } def
/Rshow { currentpoint stroke M
  dup stringwidth pop neg vshift R show } def
/Cshow { currentpoint stroke M
  dup stringwidth pop -2 div vshift R show } def
/DL { Color {setrgbcolor Solid {pop []} if 0 setdash }
 {pop pop pop Solid {pop []} if 0 setdash} ifelse } def
/BL { stroke gnulinewidth 2 mul setlinewidth } def
/AL { stroke gnulinewidth 2 div setlinewidth } def
/PL { stroke gnulinewidth setlinewidth } def
/LTb { BL [] 0 0 0 DL } def
/LTa { AL [1 dl 2 dl] 0 setdash 0 0 0 setrgbcolor } def
/LT0 { PL [] 0 1 0 DL } def
/LT1 { PL [4 dl 2 dl] 0 0 1 DL } def
/LT2 { PL [2 dl 3 dl] 1 0 0 DL } def
/LT3 { PL [1 dl 1.5 dl] 1 0 1 DL } def
/LT4 { PL [5 dl 2 dl 1 dl 2 dl] 0 1 1 DL } def
/LT5 { PL [4 dl 3 dl 1 dl 3 dl] 1 1 0 DL } def
/LT6 { PL [2 dl 2 dl 2 dl 4 dl] 0 0 0 DL } def
/LT7 { PL [2 dl 2 dl 2 dl 2 dl 2 dl 4 dl] 1 0.3 0 DL } def
/LT8 { PL [2 dl 2 dl 2 dl 2 dl 2 dl 2 dl 2 dl 4 dl] 0.5 0.5 0.5 DL } def
/P { stroke [] 0 setdash
  currentlinewidth 2 div sub M
  0 currentlinewidth V stroke } def
/D { stroke [] 0 setdash 2 copy vpt add M
  hpt neg vpt neg V hpt vpt neg V
  hpt vpt V hpt neg vpt V closepath stroke
  P } def
/A { stroke [] 0 setdash vpt sub M 0 vpt2 V
  currentpoint stroke M
  hpt neg vpt neg R hpt2 0 V stroke
  } def
/B { stroke [] 0 setdash 2 copy exch hpt sub exch vpt add M
  0 vpt2 neg V hpt2 0 V 0 vpt2 V
  hpt2 neg 0 V closepath stroke
  P } def
/C { stroke [] 0 setdash exch hpt sub exch vpt add M
  hpt2 vpt2 neg V currentpoint stroke M
  hpt2 neg 0 R hpt2 vpt2 V stroke } def
/T { stroke [] 0 setdash 2 copy vpt 1.12 mul add M
  hpt neg vpt -1.62 mul V
  hpt 2 mul 0 V
  hpt neg vpt 1.62 mul V closepath stroke
  P  } def
/S { 2 copy A C} def
end
}
\begin{picture}(2519,1511)(0,0)
\special{"
gnudict begin
gsave
50 50 translate
0.100 0.100 scale
0 setgray
/Helvetica findfont 100 scalefont setfont
newpath
-500.000000 -500.000000 translate
LTa
LTb
600 251 M
63 0 V
1673 0 R
-63 0 V
600 592 M
63 0 V
1673 0 R
-63 0 V
600 933 M
63 0 V
1673 0 R
-63 0 V
600 1275 M
63 0 V
1673 0 R
-63 0 V
889 251 M
0 63 V
0 1046 R
0 -63 V
1303 251 M
0 63 V
0 1046 R
0 -63 V
1716 251 M
0 63 V
0 1046 R
0 -63 V
2129 251 M
0 63 V
0 1046 R
0 -63 V
600 251 M
1736 0 V
0 1109 V
-1736 0 V
600 251 L
LT0
2065 1249 M
180 0 V
685 961 M
16 -19 V
1054 621 L
166 -98 V
165 -71 V
166 -49 V
514 -16 V
64 6 V
106 -6 V
40 -17 V
36 -25 V
LT1
2065 1149 M
180 0 V
685 1030 M
16 -21 V
1054 661 L
1220 552 L
165 -81 V
166 -56 V
514 -17 V
64 5 V
106 -13 V
40 -20 V
36 -29 V
LT2
2065 1049 M
180 0 V
-1560 9 R
16 -25 V
1054 633 L
1220 516 L
165 -83 V
166 -53 V
514 0 V
64 5 V
106 -14 V
40 -18 V
36 -27 V
LT3
2065 949 M
180 0 V
685 980 M
16 -22 V
1054 607 L
166 -98 V
165 -68 V
166 -45 V
514 -8 V
64 6 V
106 -7 V
40 -17 V
36 -25 V
LT0
1469 1358 M
42 -116 V
41 -99 V
40 -89 V
42 -85 V
42 -76 V
44 -70 V
43 -63 V
39 -49 V
38 -44 V
43 -43 V
42 -38 V
40 -31 V
40 -29 V
86 -54 V
2257 360 L
LT1
1469 1245 M
42 -104 V
41 -89 V
40 -81 V
42 -76 V
42 -69 V
44 -63 V
43 -56 V
39 -45 V
38 -39 V
43 -38 V
42 -33 V
40 -28 V
40 -25 V
86 -48 V
2257 350 L
LT2
1513 1360 M
39 -104 V
40 -99 V
42 -93 V
42 -83 V
44 -77 V
43 -67 V
39 -53 V
38 -47 V
43 -45 V
42 -39 V
40 -32 V
40 -28 V
86 -52 V
2257 423 L
LT3
1469 1354 M
42 -118 V
41 -100 V
40 -89 V
42 -84 V
42 -76 V
44 -69 V
43 -61 V
39 -48 V
38 -43 V
43 -42 V
42 -37 V
40 -31 V
40 -28 V
86 -54 V
2257 361 L
LT0
2125 849 D
685 907 D
701 757 D
1054 541 D
1220 471 D
1385 392 D
1551 405 D
2065 389 D
2129 370 D
2235 371 D
2275 358 D
2311 348 D
2065 849 M
180 0 V
-180 31 R
0 -62 V
180 62 R
0 -62 V
685 797 M
0 220 V
654 797 M
62 0 V
-62 220 R
62 0 V
701 716 M
0 82 V
670 716 M
62 0 V
-62 82 R
62 0 V
1054 491 M
0 100 V
1023 491 M
62 0 V
-62 100 R
62 0 V
1220 423 M
0 96 V
-31 -96 R
62 0 V
-62 96 R
62 0 V
1385 347 M
0 89 V
-31 -89 R
62 0 V
-62 89 R
62 0 V
135 -78 R
0 95 V
-31 -95 R
62 0 V
-62 95 R
62 0 V
483 -70 R
0 12 V
-31 -12 R
62 0 V
-62 12 R
62 0 V
33 -30 R
0 9 V
-31 -9 R
62 0 V
-62 9 R
62 0 V
75 -6 R
0 7 V
-31 -7 R
62 0 V
-62 7 R
62 0 V
9 -19 R
0 4 V
-31 -4 R
62 0 V
-62 4 R
62 0 V
5 -14 R
0 5 V
-31 -5 R
62 0 V
-62 5 R
62 0 V
LT0
1469 1168 A
1511 1183 A
1552 945 A
1592 991 A
1634 926 A
1676 872 A
1720 793 A
1763 691 A
1802 665 A
1840 563 A
1883 694 A
1925 591 A
1965 526 A
2005 442 A
2091 439 A
2257 326 A
1469 1110 M
0 117 V
-31 -117 R
62 0 V
-62 117 R
62 0 V
11 -85 R
0 82 V
-31 -82 R
62 0 V
-62 82 R
62 0 V
10 -311 R
0 63 V
-31 -63 R
62 0 V
-62 63 R
62 0 V
9 -22 R
0 73 V
-31 -73 R
62 0 V
-62 73 R
62 0 V
11 -131 R
0 59 V
-31 -59 R
62 0 V
-62 59 R
62 0 V
11 -119 R
0 73 V
-31 -73 R
62 0 V
-62 73 R
62 0 V
13 -142 R
0 51 V
-31 -51 R
62 0 V
-62 51 R
62 0 V
12 -170 R
0 87 V
-31 -87 R
62 0 V
-62 87 R
62 0 V
8 -93 R
0 45 V
-31 -45 R
62 0 V
-62 45 R
62 0 V
7 -181 R
0 114 V
1809 506 M
62 0 V
-62 114 R
62 0 V
12 47 R
0 53 V
-31 -53 R
62 0 V
-62 53 R
62 0 V
11 -186 R
0 113 V
1894 534 M
62 0 V
-62 113 R
62 0 V
9 -139 R
0 36 V
-31 -36 R
62 0 V
-62 36 R
62 0 V
9 -143 R
0 82 V
-31 -82 R
62 0 V
-62 82 R
62 0 V
55 -55 R
0 21 V
-31 -21 R
62 0 V
-62 21 R
62 0 V
2257 306 M
0 40 V
-31 -40 R
62 0 V
-62 40 R
62 0 V
stroke
grestore
end
showpage
}
\put(2005,849){\makebox(0,0)[r]{data}}
\put(2005,949){\makebox(0,0)[r]{fit H}}
\put(2005,1049){\makebox(0,0)[r]{fit E}}
\put(2005,1149){\makebox(0,0)[r]{fit D}}
\put(2005,1249){\makebox(0,0)[r]{fit A}}
\put(683,379){\makebox(0,0)[l]{{\scriptsize $Q^2\approx 1.5$ GeV$^2$}}}
\put(807,1104){\makebox(0,0)[l]{{\scriptsize $Q^2\approx 150$ GeV$^2$}}}
\put(808,1260){\makebox(0,0){(b)}}
\put(1468,51){\makebox(0,0){$x$}}
\put(400,805){%
\special{ps: gsave currentpoint currentpoint translate
270 rotate neg exch neg exch translate}%
\makebox(0,0)[b]{\shortstack{$F_2$}}%
\special{ps: currentpoint grestore moveto}%
}
\put(2129,151){\makebox(0,0){$10^{-1}$}}
\put(1716,151){\makebox(0,0){$10^{-2}$}}
\put(1303,151){\makebox(0,0){$10^{-3}$}}
\put(889,151){\makebox(0,0){$10^{-4}$}}
\put(540,1275){\makebox(0,0)[r]{1.4}}
\put(540,933){\makebox(0,0)[r]{1}}
\put(540,592){\makebox(0,0)[r]{0.6}}
\put(540,251){\makebox(0,0)[r]{0.2}}
\end{picture}
    }
  \begin{center}
\setlength{\unitlength}{0.1bp}
\special{!
/gnudict 40 dict def
gnudict begin
/Color false def
/Solid false def
/gnulinewidth 5.000 def
/vshift -33 def
/dl {10 mul} def
/hpt 31.5 def
/vpt 31.5 def
/M {moveto} bind def
/L {lineto} bind def
/R {rmoveto} bind def
/V {rlineto} bind def
/vpt2 vpt 2 mul def
/hpt2 hpt 2 mul def
/Lshow { currentpoint stroke M
  0 vshift R show } def
/Rshow { currentpoint stroke M
  dup stringwidth pop neg vshift R show } def
/Cshow { currentpoint stroke M
  dup stringwidth pop -2 div vshift R show } def
/DL { Color {setrgbcolor Solid {pop []} if 0 setdash }
 {pop pop pop Solid {pop []} if 0 setdash} ifelse } def
/BL { stroke gnulinewidth 2 mul setlinewidth } def
/AL { stroke gnulinewidth 2 div setlinewidth } def
/PL { stroke gnulinewidth setlinewidth } def
/LTb { BL [] 0 0 0 DL } def
/LTa { AL [1 dl 2 dl] 0 setdash 0 0 0 setrgbcolor } def
/LT0 { PL [] 0 1 0 DL } def
/LT1 { PL [4 dl 2 dl] 0 0 1 DL } def
/LT2 { PL [2 dl 3 dl] 1 0 0 DL } def
/LT3 { PL [1 dl 1.5 dl] 1 0 1 DL } def
/LT4 { PL [5 dl 2 dl 1 dl 2 dl] 0 1 1 DL } def
/LT5 { PL [4 dl 3 dl 1 dl 3 dl] 1 1 0 DL } def
/LT6 { PL [2 dl 2 dl 2 dl 4 dl] 0 0 0 DL } def
/LT7 { PL [2 dl 2 dl 2 dl 2 dl 2 dl 4 dl] 1 0.3 0 DL } def
/LT8 { PL [2 dl 2 dl 2 dl 2 dl 2 dl 2 dl 2 dl 4 dl] 0.5 0.5 0.5 DL } def
/P { stroke [] 0 setdash
  currentlinewidth 2 div sub M
  0 currentlinewidth V stroke } def
/D { stroke [] 0 setdash 2 copy vpt add M
  hpt neg vpt neg V hpt vpt neg V
  hpt vpt V hpt neg vpt V closepath stroke
  P } def
/A { stroke [] 0 setdash vpt sub M 0 vpt2 V
  currentpoint stroke M
  hpt neg vpt neg R hpt2 0 V stroke
  } def
/B { stroke [] 0 setdash 2 copy exch hpt sub exch vpt add M
  0 vpt2 neg V hpt2 0 V 0 vpt2 V
  hpt2 neg 0 V closepath stroke
  P } def
/C { stroke [] 0 setdash exch hpt sub exch vpt add M
  hpt2 vpt2 neg V currentpoint stroke M
  hpt2 neg 0 R hpt2 vpt2 V stroke } def
/T { stroke [] 0 setdash 2 copy vpt 1.12 mul add M
  hpt neg vpt -1.62 mul V
  hpt 2 mul 0 V
  hpt neg vpt 1.62 mul V closepath stroke
  P  } def
/S { 2 copy A C} def
end
}
\begin{picture}(2519,1511)(0,0)
\special{"
gnudict begin
gsave
50 50 translate
0.100 0.100 scale
0 setgray
/Helvetica findfont 100 scalefont setfont
newpath
-500.000000 -500.000000 translate
LTa
LTb
600 251 M
63 0 V
1673 0 R
-63 0 V
600 592 M
63 0 V
1673 0 R
-63 0 V
600 933 M
63 0 V
1673 0 R
-63 0 V
600 1275 M
63 0 V
1673 0 R
-63 0 V
889 251 M
0 63 V
0 1046 R
0 -63 V
1303 251 M
0 63 V
0 1046 R
0 -63 V
1716 251 M
0 63 V
0 1046 R
0 -63 V
2129 251 M
0 63 V
0 1046 R
0 -63 V
600 251 M
1736 0 V
0 1109 V
-1736 0 V
600 251 L
LT0
2065 1249 M
180 0 V
685 961 M
16 -19 V
1054 621 L
166 -98 V
165 -71 V
166 -49 V
514 -16 V
64 6 V
106 -6 V
40 -17 V
36 -25 V
LT1
2065 1149 M
180 0 V
685 946 M
16 -19 V
1054 617 L
166 -93 V
165 -69 V
166 -48 V
514 -23 V
64 6 V
106 -4 V
40 -15 V
36 -23 V
LT2
2065 1049 M
180 0 V
685 927 M
16 -18 V
1054 605 L
166 -92 V
165 -65 V
166 -44 V
514 -19 V
64 2 V
LT0
1469 1358 M
42 -116 V
41 -99 V
40 -89 V
42 -85 V
42 -76 V
44 -70 V
43 -63 V
39 -49 V
38 -44 V
43 -43 V
42 -38 V
40 -31 V
40 -29 V
86 -54 V
2257 360 L
LT1
1472 1360 M
39 -108 V
41 -99 V
40 -89 V
42 -84 V
42 -77 V
44 -70 V
43 -63 V
39 -50 V
38 -45 V
43 -44 V
42 -40 V
40 -33 V
40 -29 V
86 -57 V
2257 359 L
LT2
1474 1360 M
37 -103 V
41 -99 V
40 -89 V
42 -85 V
42 -76 V
44 -71 V
43 -62 V
39 -50 V
38 -45 V
43 -44 V
42 -40 V
40 -33 V
40 -31 V
86 -60 V
LT0
2125 949 D
685 907 D
701 757 D
1054 541 D
1220 471 D
1385 392 D
1551 405 D
2065 389 D
2129 370 D
2235 371 D
2275 358 D
2311 348 D
2065 949 M
180 0 V
-180 31 R
0 -62 V
180 62 R
0 -62 V
685 797 M
0 220 V
654 797 M
62 0 V
-62 220 R
62 0 V
701 716 M
0 82 V
670 716 M
62 0 V
-62 82 R
62 0 V
1054 491 M
0 100 V
1023 491 M
62 0 V
-62 100 R
62 0 V
1220 423 M
0 96 V
-31 -96 R
62 0 V
-62 96 R
62 0 V
1385 347 M
0 89 V
-31 -89 R
62 0 V
-62 89 R
62 0 V
135 -78 R
0 95 V
-31 -95 R
62 0 V
-62 95 R
62 0 V
483 -70 R
0 12 V
-31 -12 R
62 0 V
-62 12 R
62 0 V
33 -30 R
0 9 V
-31 -9 R
62 0 V
-62 9 R
62 0 V
75 -6 R
0 7 V
-31 -7 R
62 0 V
-62 7 R
62 0 V
9 -19 R
0 4 V
-31 -4 R
62 0 V
-62 4 R
62 0 V
5 -14 R
0 5 V
-31 -5 R
62 0 V
-62 5 R
62 0 V
LT0
1469 1168 A
1511 1183 A
1552 945 A
1592 991 A
1634 926 A
1676 872 A
1720 793 A
1763 691 A
1802 665 A
1840 563 A
1883 694 A
1925 591 A
1965 526 A
2005 442 A
2091 439 A
2257 326 A
1469 1110 M
0 117 V
-31 -117 R
62 0 V
-62 117 R
62 0 V
11 -85 R
0 82 V
-31 -82 R
62 0 V
-62 82 R
62 0 V
10 -311 R
0 63 V
-31 -63 R
62 0 V
-62 63 R
62 0 V
9 -22 R
0 73 V
-31 -73 R
62 0 V
-62 73 R
62 0 V
11 -131 R
0 59 V
-31 -59 R
62 0 V
-62 59 R
62 0 V
11 -119 R
0 73 V
-31 -73 R
62 0 V
-62 73 R
62 0 V
13 -142 R
0 51 V
-31 -51 R
62 0 V
-62 51 R
62 0 V
12 -170 R
0 87 V
-31 -87 R
62 0 V
-62 87 R
62 0 V
8 -93 R
0 45 V
-31 -45 R
62 0 V
-62 45 R
62 0 V
7 -181 R
0 114 V
1809 506 M
62 0 V
-62 114 R
62 0 V
12 47 R
0 53 V
-31 -53 R
62 0 V
-62 53 R
62 0 V
11 -186 R
0 113 V
1894 534 M
62 0 V
-62 113 R
62 0 V
9 -139 R
0 36 V
-31 -36 R
62 0 V
-62 36 R
62 0 V
9 -143 R
0 82 V
-31 -82 R
62 0 V
-62 82 R
62 0 V
55 -55 R
0 21 V
-31 -21 R
62 0 V
-62 21 R
62 0 V
2257 306 M
0 40 V
-31 -40 R
62 0 V
-62 40 R
62 0 V
stroke
grestore
end
showpage
}
\put(2005,949){\makebox(0,0)[r]{data}}
\put(2005,1049){\makebox(0,0)[r]{fit G}}
\put(2005,1149){\makebox(0,0)[r]{fit F}}
\put(2005,1249){\makebox(0,0)[r]{fit A}}
\put(683,379){\makebox(0,0)[l]{{\scriptsize $Q^2\approx 1.5$ GeV$^2$}}}
\put(807,1104){\makebox(0,0)[l]{{\scriptsize $Q^2\approx 150$ GeV$^2$}}}
\put(808,1260){\makebox(0,0){(c)}}
\put(1468,51){\makebox(0,0){$x$}}
\put(400,805){%
\special{ps: gsave currentpoint currentpoint translate
270 rotate neg exch neg exch translate}%
\makebox(0,0)[b]{\shortstack{$F_2$}}%
\special{ps: currentpoint grestore moveto}%
}
\put(2129,151){\makebox(0,0){$10^{-1}$}}
\put(1716,151){\makebox(0,0){$10^{-2}$}}
\put(1303,151){\makebox(0,0){$10^{-3}$}}
\put(889,151){\makebox(0,0){$10^{-4}$}}
\put(540,1275){\makebox(0,0)[r]{1.4}}
\put(540,933){\makebox(0,0)[r]{1}}
\put(540,592){\makebox(0,0)[r]{0.6}}
\put(540,251){\makebox(0,0)[r]{0.2}}
\end{picture}
  \end{center}
  \caption[dummy]{{\it The fitted \tf2\ as a
      function of $x$ for different strategies and for two different
      values of \tq2, compared with data from
      \cite{H1F2,ZEUSF2,NMCF2,E665F2}.}}
  \label{fig:F2x}
\end{figure}
\begin{figure}[t]
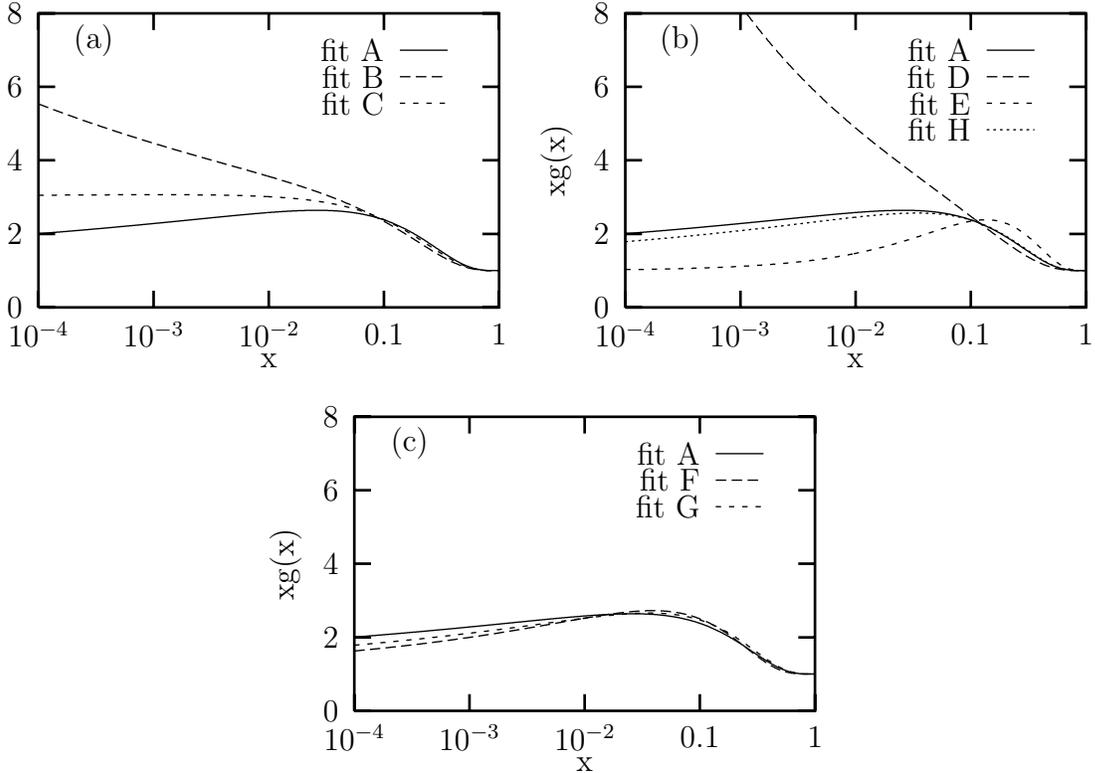

  \vskip -1cm
  \hbox{
    \hskip -1.5cm
    \input figgina.tex
    \hskip -1.5cm
    \input figginb.tex
    }
  \begin{center}
    \input figginc.tex
  \end{center}
  \caption[dummy]{{\it The fitted input gluon density as a function of
      $x$ for different strategies.}}
  \label{fig:gin}
\end{figure}

The results of the fits are presented in \figs{fig:F2x} and
\ref{fig:gin}. For the default case, the fit is quite acceptable. We
note in particular that the fitted input gluon density is slowly
decreasing with $1/x$, although we must keep in mind that the gluon
distribution is only indirectly constrained.

For the DGLAP case the fit is much worse. The number of allowed ISB
emissions is here strongly restricted, especially for small \tq2\ and
$x$. This results in much slower evolution which forces the input
densities to rise with $1/x$. In fit C, where the link closest to the
virtual photon is allowed to be above \tq2, the fit is on the other
hand again quite acceptable. The input gluon distribution is no longer
rising with $1/x$, on the other hand it is as strongly decreasing as
in fit A. It is known that \tf2\ can be fitted using conventional
DGLAP evolution with a valence-like flat input distribution at small
input scales, close to the one used here, as in the GRV
parametrizations \cite{GRV}.  Such fits can, however, not be directly
compared to this one, as we here have less parameters. But we can
conclude that the ISB chains with unordered \tkt\ do play a rôle in
our case, although most of the effect can be obtained allowing only
for one `stepping down', closest to the photon in the chain.

Increasing the input scale to $1$ GeV in fit D makes a big difference
particularly at small $x$ as seen in \figs{fig:F2x}b and
\ref{fig:gin}b. Also here the number of allowed ISB emissions is
strongly restricted and again the input gluon is forced to increase
strongly with $1/x$. This should come as no surprise. We expect,
however, that when we below study the hadronic final states, the
result should be less sensitive to the input scale used.

The importance of the Sudakov form factor is apparent in
\figs{fig:F2x}b and \ref{fig:gin}b, comparing fits A and E, especially
for the input gluon distribution. Also in \figs{fig:F2x}b and
\ref{fig:gin}b we show the fit H, where some propagators below the
cutoff is allowed. The reproduction of \tf2\ does not change much, but
we see that the input gluon decreases slightly faster with $1/x$ than
for the default fit A.

The fits F and G in \figs{fig:F2x}c and \ref{fig:gin}c show how
sensitive the fit is to the input distribution is at large $x$.
Changing $\beta_g$ and $\beta_S$ from $4$ to $5$ does not influence
the fit very much and neither does the omission of the data points at
high $x$, although in both cases the input gluon is shifted somewhat
to higher $x$.


\section{Results for the hadronic final states at HERA}
\label{sec:res}

The results for the hadronic final states will depend on the input
densities used. If eg.\ the input densities are increasing with
$1/x_0$, chains starting with low $x_0$ will be favoured and the
length of the chains will be shorter resulting in fewer emissions and
less activity in general due to the reduction of the available phase
space, in particular close to the direction of the incoming hadron.
Also a smaller input gluon density will result in fewer chains
initiated with a gluon, which has higher charge than an incoming
quark, and again the probability of emissions will become smaller. In
addition, the non-perturbative hadronization is smaller if only one
string is stretched between the perturbative system and the hadron
remnant in the case of an incoming quark.

\begin{figure}
  \vskip -1.8cm
  \hbox{
    \hskip -1.7cm
\setlength{\unitlength}{0.1bp}
\special{!
/gnudict 40 dict def
gnudict begin
/Color false def
/Solid false def
/gnulinewidth 5.000 def
/vshift -33 def
/dl {10 mul} def
/hpt 31.5 def
/vpt 31.5 def
/M {moveto} bind def
/L {lineto} bind def
/R {rmoveto} bind def
/V {rlineto} bind def
/vpt2 vpt 2 mul def
/hpt2 hpt 2 mul def
/Lshow { currentpoint stroke M
  0 vshift R show } def
/Rshow { currentpoint stroke M
  dup stringwidth pop neg vshift R show } def
/Cshow { currentpoint stroke M
  dup stringwidth pop -2 div vshift R show } def
/DL { Color {setrgbcolor Solid {pop []} if 0 setdash }
 {pop pop pop Solid {pop []} if 0 setdash} ifelse } def
/BL { stroke gnulinewidth 2 mul setlinewidth } def
/AL { stroke gnulinewidth 2 div setlinewidth } def
/PL { stroke gnulinewidth setlinewidth } def
/LTb { BL [] 0 0 0 DL } def
/LTa { AL [1 dl 2 dl] 0 setdash 0 0 0 setrgbcolor } def
/LT0 { PL [] 0 1 0 DL } def
/LT1 { PL [4 dl 2 dl] 0 0 1 DL } def
/LT2 { PL [2 dl 3 dl] 1 0 0 DL } def
/LT3 { PL [1 dl 1.5 dl] 1 0 1 DL } def
/LT4 { PL [5 dl 2 dl 1 dl 2 dl] 0 1 1 DL } def
/LT5 { PL [4 dl 3 dl 1 dl 3 dl] 1 1 0 DL } def
/LT6 { PL [2 dl 2 dl 2 dl 4 dl] 0 0 0 DL } def
/LT7 { PL [2 dl 2 dl 2 dl 2 dl 2 dl 4 dl] 1 0.3 0 DL } def
/LT8 { PL [2 dl 2 dl 2 dl 2 dl 2 dl 2 dl 2 dl 4 dl] 0.5 0.5 0.5 DL } def
/P { stroke [] 0 setdash
  currentlinewidth 2 div sub M
  0 currentlinewidth V stroke } def
/D { stroke [] 0 setdash 2 copy vpt add M
  hpt neg vpt neg V hpt vpt neg V
  hpt vpt V hpt neg vpt V closepath stroke
  P } def
/A { stroke [] 0 setdash vpt sub M 0 vpt2 V
  currentpoint stroke M
  hpt neg vpt neg R hpt2 0 V stroke
  } def
/B { stroke [] 0 setdash 2 copy exch hpt sub exch vpt add M
  0 vpt2 neg V hpt2 0 V 0 vpt2 V
  hpt2 neg 0 V closepath stroke
  P } def
/C { stroke [] 0 setdash exch hpt sub exch vpt add M
  hpt2 vpt2 neg V currentpoint stroke M
  hpt2 neg 0 R hpt2 vpt2 V stroke } def
/T { stroke [] 0 setdash 2 copy vpt 1.12 mul add M
  hpt neg vpt -1.62 mul V
  hpt 2 mul 0 V
  hpt neg vpt 1.62 mul V closepath stroke
  P  } def
/S { 2 copy A C} def
end
}
\begin{picture}(2519,1511)(0,0)
\special{"
gnudict begin
gsave
50 50 translate
0.100 0.100 scale
0 setgray
/Helvetica findfont 100 scalefont setfont
newpath
-500.000000 -500.000000 translate
LTa
LTb
600 251 M
63 0 V
1673 0 R
-63 0 V
600 695 M
63 0 V
1673 0 R
-63 0 V
600 1138 M
63 0 V
1673 0 R
-63 0 V
600 251 M
0 63 V
0 1046 R
0 -63 V
1034 251 M
0 63 V
0 1046 R
0 -63 V
1468 251 M
0 63 V
0 1046 R
0 -63 V
1902 251 M
0 63 V
0 1046 R
0 -63 V
2336 251 M
0 63 V
0 1046 R
0 -63 V
600 251 M
1736 0 V
0 1109 V
-1736 0 V
600 251 L
LT0
1962 672 M
180 0 V
600 461 M
43 67 V
87 151 V
87 136 V
87 55 V
87 43 V
86 48 V
87 17 V
87 3 V
87 27 V
87 24 V
86 5 V
87 35 V
87 28 V
87 30 V
87 48 V
86 26 V
87 -59 V
87 -149 V
87 -275 V
87 -223 V
43 -77 V
LT1
1962 572 M
180 0 V
600 424 M
43 54 V
87 142 V
87 133 V
87 136 V
87 92 V
86 32 V
87 47 V
87 32 V
87 22 V
87 47 V
86 11 V
87 3 V
87 -34 V
87 26 V
87 -30 V
86 10 V
87 -9 V
87 -148 V
87 -233 V
87 -255 V
43 -81 V
LT2
1962 472 M
180 0 V
600 461 M
43 35 V
87 27 V
87 53 V
87 47 V
87 7 V
86 26 V
87 42 V
87 -16 V
87 40 V
87 36 V
86 -19 V
87 12 V
87 46 V
87 31 V
87 1 V
86 65 V
87 31 V
87 -77 V
87 -130 V
87 -182 V
43 -76 V
LT3
1962 372 M
180 0 V
600 462 M
43 48 V
87 101 V
87 42 V
87 77 V
87 71 V
86 33 V
87 44 V
87 33 V
87 74 V
87 3 V
86 21 V
87 -15 V
87 28 V
87 47 V
87 -15 V
86 21 V
87 -43 V
87 -144 V
87 -176 V
87 -180 V
43 -82 V
LT0
1251 1026 A
1338 1135 A
1425 1143 A
1511 1236 A
1598 1132 A
1685 1154 A
1772 1231 A
1859 1226 A
1945 1104 A
2032 1148 A
2119 1091 A
2206 704 A
2293 479 A
1251 974 M
0 103 V
1220 974 M
62 0 V
-62 103 R
62 0 V
56 8 R
0 100 V
-31 -100 R
62 0 V
-62 100 R
62 0 V
56 -93 R
0 102 V
-31 -102 R
62 0 V
-62 102 R
62 0 V
55 -14 R
0 112 V
-31 -112 R
62 0 V
-62 112 R
62 0 V
56 -213 R
0 106 V
-31 -106 R
62 0 V
-62 106 R
62 0 V
56 -94 R
0 126 V
-31 -126 R
62 0 V
-62 126 R
62 0 V
56 -65 R
0 158 V
-31 -158 R
62 0 V
-62 158 R
62 0 V
56 -159 R
0 151 V
-31 -151 R
62 0 V
-62 151 R
62 0 V
55 -256 R
0 115 V
-31 -115 R
62 0 V
-62 115 R
62 0 V
56 -77 R
0 128 V
-31 -128 R
62 0 V
-62 128 R
62 0 V
56 -185 R
0 128 V
-31 -128 R
62 0 V
-62 128 R
62 0 V
56 -490 R
0 78 V
-31 -78 R
62 0 V
-62 78 R
62 0 V
56 -290 R
0 53 V
-31 -53 R
62 0 V
-62 53 R
62 0 V
stroke
grestore
end
showpage
}
\put(1902,372){\makebox(0,0)[r]{{\scriptsize \lepto\ SCI}}}
\put(1902,472){\makebox(0,0)[r]{{\scriptsize \lepto}}}
\put(1902,572){\makebox(0,0)[r]{{\scriptsize \ariadne}}}
\put(1902,672){\makebox(0,0)[r]{{\scriptsize LDC~A}}}
\put(748,1260){\makebox(0,0){(a)}}
\put(1468,51){\makebox(0,0){$\eta$}}
\put(460,805){%
\special{ps: gsave currentpoint currentpoint translate
270 rotate neg exch neg exch translate}%
\makebox(0,0)[b]{\shortstack{$dE_\perp/d\eta$ (GeV)}}%
\special{ps: currentpoint grestore moveto}%
}
\put(2336,151){\makebox(0,0){5}}
\put(1902,151){\makebox(0,0){2.5}}
\put(1468,151){\makebox(0,0){0}}
\put(1034,151){\makebox(0,0){-2.5}}
\put(600,151){\makebox(0,0){-5}}
\put(540,1138){\makebox(0,0)[r]{2}}
\put(540,695){\makebox(0,0)[r]{1}}
\put(540,251){\makebox(0,0)[r]{0}}
\end{picture}
    \hskip -1.2cm
\setlength{\unitlength}{0.1bp}
\special{!
/gnudict 40 dict def
gnudict begin
/Color false def
/Solid false def
/gnulinewidth 5.000 def
/vshift -33 def
/dl {10 mul} def
/hpt 31.5 def
/vpt 31.5 def
/M {moveto} bind def
/L {lineto} bind def
/R {rmoveto} bind def
/V {rlineto} bind def
/vpt2 vpt 2 mul def
/hpt2 hpt 2 mul def
/Lshow { currentpoint stroke M
  0 vshift R show } def
/Rshow { currentpoint stroke M
  dup stringwidth pop neg vshift R show } def
/Cshow { currentpoint stroke M
  dup stringwidth pop -2 div vshift R show } def
/DL { Color {setrgbcolor Solid {pop []} if 0 setdash }
 {pop pop pop Solid {pop []} if 0 setdash} ifelse } def
/BL { stroke gnulinewidth 2 mul setlinewidth } def
/AL { stroke gnulinewidth 2 div setlinewidth } def
/PL { stroke gnulinewidth setlinewidth } def
/LTb { BL [] 0 0 0 DL } def
/LTa { AL [1 dl 2 dl] 0 setdash 0 0 0 setrgbcolor } def
/LT0 { PL [] 0 1 0 DL } def
/LT1 { PL [4 dl 2 dl] 0 0 1 DL } def
/LT2 { PL [2 dl 3 dl] 1 0 0 DL } def
/LT3 { PL [1 dl 1.5 dl] 1 0 1 DL } def
/LT4 { PL [5 dl 2 dl 1 dl 2 dl] 0 1 1 DL } def
/LT5 { PL [4 dl 3 dl 1 dl 3 dl] 1 1 0 DL } def
/LT6 { PL [2 dl 2 dl 2 dl 4 dl] 0 0 0 DL } def
/LT7 { PL [2 dl 2 dl 2 dl 2 dl 2 dl 4 dl] 1 0.3 0 DL } def
/LT8 { PL [2 dl 2 dl 2 dl 2 dl 2 dl 2 dl 2 dl 4 dl] 0.5 0.5 0.5 DL } def
/P { stroke [] 0 setdash
  currentlinewidth 2 div sub M
  0 currentlinewidth V stroke } def
/D { stroke [] 0 setdash 2 copy vpt add M
  hpt neg vpt neg V hpt vpt neg V
  hpt vpt V hpt neg vpt V closepath stroke
  P } def
/A { stroke [] 0 setdash vpt sub M 0 vpt2 V
  currentpoint stroke M
  hpt neg vpt neg R hpt2 0 V stroke
  } def
/B { stroke [] 0 setdash 2 copy exch hpt sub exch vpt add M
  0 vpt2 neg V hpt2 0 V 0 vpt2 V
  hpt2 neg 0 V closepath stroke
  P } def
/C { stroke [] 0 setdash exch hpt sub exch vpt add M
  hpt2 vpt2 neg V currentpoint stroke M
  hpt2 neg 0 R hpt2 vpt2 V stroke } def
/T { stroke [] 0 setdash 2 copy vpt 1.12 mul add M
  hpt neg vpt -1.62 mul V
  hpt 2 mul 0 V
  hpt neg vpt 1.62 mul V closepath stroke
  P  } def
/S { 2 copy A C} def
end
}
\begin{picture}(2519,1511)(0,0)
\special{"
gnudict begin
gsave
50 50 translate
0.100 0.100 scale
0 setgray
/Helvetica findfont 100 scalefont setfont
newpath
-500.000000 -500.000000 translate
LTa
LTb
600 251 M
63 0 V
1673 0 R
-63 0 V
600 695 M
63 0 V
1673 0 R
-63 0 V
600 1138 M
63 0 V
1673 0 R
-63 0 V
600 251 M
0 63 V
0 1046 R
0 -63 V
1034 251 M
0 63 V
0 1046 R
0 -63 V
1468 251 M
0 63 V
0 1046 R
0 -63 V
1902 251 M
0 63 V
0 1046 R
0 -63 V
2336 251 M
0 63 V
0 1046 R
0 -63 V
600 251 M
1736 0 V
0 1109 V
-1736 0 V
600 251 L
LT0
1788 672 M
180 0 V
600 294 M
43 24 V
87 102 V
87 147 V
87 137 V
87 129 V
86 60 V
87 24 V
87 12 V
87 34 V
87 60 V
86 67 V
87 41 V
87 55 V
87 41 V
87 -37 V
86 -126 V
87 -244 V
87 -279 V
87 -176 V
87 -81 V
43 -12 V
LT1
1788 572 M
180 0 V
600 299 M
43 17 V
87 73 V
87 120 V
87 168 V
87 97 V
86 104 V
87 75 V
87 19 V
87 72 V
87 46 V
86 27 V
87 83 V
87 21 V
87 54 V
87 -13 V
86 -181 V
87 -308 V
87 -255 V
87 -159 V
87 -74 V
43 -12 V
LT2
1788 472 M
180 0 V
600 319 M
43 28 V
87 80 V
87 79 V
87 69 V
87 80 V
86 75 V
87 64 V
87 59 V
87 48 V
87 41 V
86 68 V
87 70 V
87 23 V
87 79 V
87 -9 V
86 -168 V
87 -236 V
87 -227 V
87 -170 V
87 -79 V
43 -15 V
LT3
1788 372 M
180 0 V
600 313 M
43 26 V
87 74 V
87 93 V
87 76 V
87 91 V
86 80 V
87 97 V
87 100 V
87 82 V
87 91 V
86 56 V
87 59 V
87 -1 V
87 24 V
87 -52 V
86 -212 V
87 -215 V
87 -249 V
87 -166 V
87 -81 V
43 -12 V
LT0
1425 910 A
1511 1123 A
1598 1219 A
1685 1213 A
1772 1179 A
1859 1336 A
1945 1242 A
2032 874 A
2119 566 A
2206 342 A
2293 287 A
1425 862 M
0 95 V
-31 -95 R
62 0 V
-62 95 R
62 0 V
55 107 R
0 118 V
-31 -118 R
62 0 V
-62 118 R
62 0 V
56 -32 R
0 139 V
-31 -139 R
62 0 V
-62 139 R
62 0 V
56 -141 R
0 130 V
-31 -130 R
62 0 V
-62 130 R
62 0 V
56 -168 R
0 137 V
-31 -137 R
62 0 V
-62 137 R
62 0 V
56 8 R
0 105 V
-31 -105 R
62 0 V
-62 105 R
62 0 V
55 -197 R
0 157 V
-31 -157 R
62 0 V
-62 157 R
62 0 V
56 -492 R
0 93 V
-31 -93 R
62 0 V
-62 93 R
62 0 V
56 -383 R
0 55 V
-31 -55 R
62 0 V
-62 55 R
62 0 V
56 -261 R
0 20 V
-31 -20 R
62 0 V
-62 20 R
62 0 V
56 -71 R
0 13 V
-31 -13 R
62 0 V
-62 13 R
62 0 V
stroke
grestore
end
showpage
}
\put(1728,372){\makebox(0,0)[r]{{\scriptsize \lepto\ SCI}}}
\put(1728,472){\makebox(0,0)[r]{{\scriptsize \lepto}}}
\put(1728,572){\makebox(0,0)[r]{{\scriptsize \ariadne}}}
\put(1728,672){\makebox(0,0)[r]{{\scriptsize LDC~A}}}
\put(748,1260){\makebox(0,0){(b)}}
\put(1468,51){\makebox(0,0){$\eta$}}
\put(460,805){%
\special{ps: gsave currentpoint currentpoint translate
270 rotate neg exch neg exch translate}%
\makebox(0,0)[b]{\shortstack{$dE_\perp/d\eta$ (GeV)}}%
\special{ps: currentpoint grestore moveto}%
}
\put(2336,151){\makebox(0,0){5}}
\put(1902,151){\makebox(0,0){2.5}}
\put(1468,151){\makebox(0,0){0}}
\put(1034,151){\makebox(0,0){-2.5}}
\put(600,151){\makebox(0,0){-5}}
\put(540,1138){\makebox(0,0)[r]{2}}
\put(540,695){\makebox(0,0)[r]{1}}
\put(540,251){\makebox(0,0)[r]{0}}
\end{picture}
    }
  \vskip -0.2cm
  \hbox{
    \hskip -1.7cm
\setlength{\unitlength}{0.1bp}
\special{!
/gnudict 40 dict def
gnudict begin
/Color false def
/Solid false def
/gnulinewidth 5.000 def
/vshift -33 def
/dl {10 mul} def
/hpt 31.5 def
/vpt 31.5 def
/M {moveto} bind def
/L {lineto} bind def
/R {rmoveto} bind def
/V {rlineto} bind def
/vpt2 vpt 2 mul def
/hpt2 hpt 2 mul def
/Lshow { currentpoint stroke M
  0 vshift R show } def
/Rshow { currentpoint stroke M
  dup stringwidth pop neg vshift R show } def
/Cshow { currentpoint stroke M
  dup stringwidth pop -2 div vshift R show } def
/DL { Color {setrgbcolor Solid {pop []} if 0 setdash }
 {pop pop pop Solid {pop []} if 0 setdash} ifelse } def
/BL { stroke gnulinewidth 2 mul setlinewidth } def
/AL { stroke gnulinewidth 2 div setlinewidth } def
/PL { stroke gnulinewidth setlinewidth } def
/LTb { BL [] 0 0 0 DL } def
/LTa { AL [1 dl 2 dl] 0 setdash 0 0 0 setrgbcolor } def
/LT0 { PL [] 0 1 0 DL } def
/LT1 { PL [4 dl 2 dl] 0 0 1 DL } def
/LT2 { PL [2 dl 3 dl] 1 0 0 DL } def
/LT3 { PL [1 dl 1.5 dl] 1 0 1 DL } def
/LT4 { PL [5 dl 2 dl 1 dl 2 dl] 0 1 1 DL } def
/LT5 { PL [4 dl 3 dl 1 dl 3 dl] 1 1 0 DL } def
/LT6 { PL [2 dl 2 dl 2 dl 4 dl] 0 0 0 DL } def
/LT7 { PL [2 dl 2 dl 2 dl 2 dl 2 dl 4 dl] 1 0.3 0 DL } def
/LT8 { PL [2 dl 2 dl 2 dl 2 dl 2 dl 2 dl 2 dl 4 dl] 0.5 0.5 0.5 DL } def
/P { stroke [] 0 setdash
  currentlinewidth 2 div sub M
  0 currentlinewidth V stroke } def
/D { stroke [] 0 setdash 2 copy vpt add M
  hpt neg vpt neg V hpt vpt neg V
  hpt vpt V hpt neg vpt V closepath stroke
  P } def
/A { stroke [] 0 setdash vpt sub M 0 vpt2 V
  currentpoint stroke M
  hpt neg vpt neg R hpt2 0 V stroke
  } def
/B { stroke [] 0 setdash 2 copy exch hpt sub exch vpt add M
  0 vpt2 neg V hpt2 0 V 0 vpt2 V
  hpt2 neg 0 V closepath stroke
  P } def
/C { stroke [] 0 setdash exch hpt sub exch vpt add M
  hpt2 vpt2 neg V currentpoint stroke M
  hpt2 neg 0 R hpt2 vpt2 V stroke } def
/T { stroke [] 0 setdash 2 copy vpt 1.12 mul add M
  hpt neg vpt -1.62 mul V
  hpt 2 mul 0 V
  hpt neg vpt 1.62 mul V closepath stroke
  P  } def
/S { 2 copy A C} def
end
}
\begin{picture}(2519,1511)(0,0)
\special{"
gnudict begin
gsave
50 50 translate
0.100 0.100 scale
0 setgray
/Helvetica findfont 100 scalefont setfont
newpath
-500.000000 -500.000000 translate
LTa
LTb
600 251 M
63 0 V
1673 0 R
-63 0 V
600 621 M
63 0 V
1673 0 R
-63 0 V
600 990 M
63 0 V
1673 0 R
-63 0 V
600 1360 M
63 0 V
1673 0 R
-63 0 V
600 251 M
0 63 V
0 1046 R
0 -63 V
986 251 M
0 63 V
0 1046 R
0 -63 V
1372 251 M
0 63 V
0 1046 R
0 -63 V
1757 251 M
0 63 V
0 1046 R
0 -63 V
2143 251 M
0 63 V
0 1046 R
0 -63 V
600 251 M
1736 0 V
0 1109 V
-1736 0 V
600 251 L
LT0
1142 1249 M
180 0 V
745 432 M
96 70 V
116 81 V
164 116 V
222 140 V
318 112 V
434 56 V
LT1
1142 1149 M
180 0 V
745 461 M
96 81 V
116 83 V
164 111 V
222 160 V
318 110 V
434 97 V
LT2
1142 1049 M
180 0 V
745 397 M
96 58 V
116 82 V
164 27 V
222 103 V
318 23 V
434 -86 V
LT3
1142 949 M
180 0 V
745 399 M
96 47 V
116 71 V
164 114 V
222 48 V
318 37 V
434 119 V
LT0
745 425 A
841 484 A
957 565 A
1121 691 A
1343 806 A
1661 1101 A
2095 1024 A
745 417 M
0 15 V
714 417 M
62 0 V
-62 15 R
62 0 V
65 37 R
0 30 V
810 469 M
62 0 V
-62 30 R
62 0 V
85 48 R
0 37 V
926 547 M
62 0 V
-62 37 R
62 0 V
133 77 R
0 59 V
-31 -59 R
62 0 V
-62 59 R
62 0 V
191 26 R
0 119 V
1312 746 M
62 0 V
-62 119 R
62 0 V
287 96 R
0 281 V
1630 961 M
62 0 V
-62 281 R
62 0 V
2095 750 M
0 547 V
2064 750 M
62 0 V
-62 547 R
62 0 V
stroke
grestore
end
showpage
}
\put(1082,949){\makebox(0,0)[r]{{\scriptsize \lepto\ SCI}}}
\put(1082,1049){\makebox(0,0)[r]{{\scriptsize \lepto}}}
\put(1082,1149){\makebox(0,0)[r]{{\scriptsize \ariadne}}}
\put(1082,1249){\makebox(0,0)[r]{{\scriptsize LDC~A}}}
\put(688,1260){\makebox(0,0){(c)}}
\put(1468,51){\makebox(0,0){$x_F$}}
\put(460,805){%
\special{ps: gsave currentpoint currentpoint translate
270 rotate neg exch neg exch translate}%
\makebox(0,0)[b]{\shortstack{$\langle p_\perp^2 \rangle$ (Gev$^2$)}}%
\special{ps: currentpoint grestore moveto}%
}
\put(2143,151){\makebox(0,0){0.8}}
\put(1757,151){\makebox(0,0){0.6}}
\put(1372,151){\makebox(0,0){0.4}}
\put(986,151){\makebox(0,0){0.2}}
\put(600,151){\makebox(0,0){0}}
\put(540,1360){\makebox(0,0)[r]{3}}
\put(540,990){\makebox(0,0)[r]{2}}
\put(540,621){\makebox(0,0)[r]{1}}
\put(540,251){\makebox(0,0)[r]{0}}
\end{picture}
    \hskip -1.2cm
\setlength{\unitlength}{0.1bp}
\special{!
/gnudict 40 dict def
gnudict begin
/Color false def
/Solid false def
/gnulinewidth 5.000 def
/vshift -33 def
/dl {10 mul} def
/hpt 31.5 def
/vpt 31.5 def
/M {moveto} bind def
/L {lineto} bind def
/R {rmoveto} bind def
/V {rlineto} bind def
/vpt2 vpt 2 mul def
/hpt2 hpt 2 mul def
/Lshow { currentpoint stroke M
  0 vshift R show } def
/Rshow { currentpoint stroke M
  dup stringwidth pop neg vshift R show } def
/Cshow { currentpoint stroke M
  dup stringwidth pop -2 div vshift R show } def
/DL { Color {setrgbcolor Solid {pop []} if 0 setdash }
 {pop pop pop Solid {pop []} if 0 setdash} ifelse } def
/BL { stroke gnulinewidth 2 mul setlinewidth } def
/AL { stroke gnulinewidth 2 div setlinewidth } def
/PL { stroke gnulinewidth setlinewidth } def
/LTb { BL [] 0 0 0 DL } def
/LTa { AL [1 dl 2 dl] 0 setdash 0 0 0 setrgbcolor } def
/LT0 { PL [] 0 1 0 DL } def
/LT1 { PL [4 dl 2 dl] 0 0 1 DL } def
/LT2 { PL [2 dl 3 dl] 1 0 0 DL } def
/LT3 { PL [1 dl 1.5 dl] 1 0 1 DL } def
/LT4 { PL [5 dl 2 dl 1 dl 2 dl] 0 1 1 DL } def
/LT5 { PL [4 dl 3 dl 1 dl 3 dl] 1 1 0 DL } def
/LT6 { PL [2 dl 2 dl 2 dl 4 dl] 0 0 0 DL } def
/LT7 { PL [2 dl 2 dl 2 dl 2 dl 2 dl 4 dl] 1 0.3 0 DL } def
/LT8 { PL [2 dl 2 dl 2 dl 2 dl 2 dl 2 dl 2 dl 4 dl] 0.5 0.5 0.5 DL } def
/P { stroke [] 0 setdash
  currentlinewidth 2 div sub M
  0 currentlinewidth V stroke } def
/D { stroke [] 0 setdash 2 copy vpt add M
  hpt neg vpt neg V hpt vpt neg V
  hpt vpt V hpt neg vpt V closepath stroke
  P } def
/A { stroke [] 0 setdash vpt sub M 0 vpt2 V
  currentpoint stroke M
  hpt neg vpt neg R hpt2 0 V stroke
  } def
/B { stroke [] 0 setdash 2 copy exch hpt sub exch vpt add M
  0 vpt2 neg V hpt2 0 V 0 vpt2 V
  hpt2 neg 0 V closepath stroke
  P } def
/C { stroke [] 0 setdash exch hpt sub exch vpt add M
  hpt2 vpt2 neg V currentpoint stroke M
  hpt2 neg 0 R hpt2 vpt2 V stroke } def
/T { stroke [] 0 setdash 2 copy vpt 1.12 mul add M
  hpt neg vpt -1.62 mul V
  hpt 2 mul 0 V
  hpt neg vpt 1.62 mul V closepath stroke
  P  } def
/S { 2 copy A C} def
end
}
\begin{picture}(2519,1511)(0,0)
\special{"
gnudict begin
gsave
50 50 translate
0.100 0.100 scale
0 setgray
/Helvetica findfont 100 scalefont setfont
newpath
-500.000000 -500.000000 translate
LTa
LTb
600 251 M
63 0 V
1673 0 R
-63 0 V
600 551 M
63 0 V
1673 0 R
-63 0 V
600 851 M
63 0 V
1673 0 R
-63 0 V
600 1150 M
63 0 V
1673 0 R
-63 0 V
600 251 M
0 63 V
0 1046 R
0 -63 V
1179 251 M
0 63 V
0 1046 R
0 -63 V
1757 251 M
0 63 V
0 1046 R
0 -63 V
2336 251 M
0 63 V
0 1046 R
0 -63 V
600 251 M
1736 0 V
0 1109 V
-1736 0 V
600 251 L
LT0
1962 1241 M
180 0 V
629 1283 M
58 57 V
58 -60 V
58 -89 V
57 -94 V
918 997 L
58 -94 V
58 -74 V
58 -68 V
72 -85 V
1265 563 L
131 -82 V
1613 378 L
1788 251 L
LT1
1962 1141 M
180 0 V
629 1267 M
58 55 V
58 -60 V
58 -79 V
57 -76 V
58 -88 V
58 -83 V
58 -69 V
58 -44 V
72 -58 V
101 -71 V
1396 583 L
217 -64 V
1902 397 L
2191 265 L
103 -14 V
LT2
1962 1041 M
180 0 V
629 1254 M
58 61 V
58 -80 V
58 -106 V
57 -118 V
918 874 L
58 -96 V
58 -143 V
58 11 V
72 -77 V
1265 442 L
131 -58 V
217 0 V
LT3
1962 941 M
180 0 V
629 1298 M
54 62 V
7 0 R
55 -72 V
58 -101 V
57 -123 V
918 932 L
976 800 L
58 -64 V
58 -90 V
72 -95 V
101 -55 V
1396 357 L
98 -106 V
LT0
687 1327 A
745 1264 A
803 1168 A
860 1108 A
918 989 A
976 955 A
1034 890 A
1092 789 A
1164 785 A
1265 659 A
1396 569 A
1613 521 A
687 1311 M
0 30 V
-31 -30 R
62 0 V
-62 30 R
62 0 V
27 -89 R
0 24 V
-31 -24 R
62 0 V
-62 24 R
62 0 V
27 -116 R
0 16 V
-31 -16 R
62 0 V
-62 16 R
62 0 V
26 -81 R
0 25 V
-31 -25 R
62 0 V
-62 25 R
62 0 V
918 957 M
0 58 V
887 957 M
62 0 V
-62 58 R
62 0 V
27 -81 R
0 39 V
945 934 M
62 0 V
-62 39 R
62 0 V
27 -107 R
0 44 V
-31 -44 R
62 0 V
-62 44 R
62 0 V
27 -147 R
0 48 V
-31 -48 R
62 0 V
-62 48 R
62 0 V
41 -52 R
0 48 V
-31 -48 R
62 0 V
-62 48 R
62 0 V
70 -204 R
0 96 V
-31 -96 R
62 0 V
-62 96 R
62 0 V
1396 512 M
0 96 V
-31 -96 R
62 0 V
-62 96 R
62 0 V
1613 251 M
0 352 V
1582 251 M
62 0 V
-62 352 R
62 0 V
LT0
687 1327 A
745 1264 A
803 1168 A
860 1108 A
918 989 A
976 955 A
1034 890 A
1092 789 A
1164 785 A
1265 659 A
1396 569 A
1613 521 A
687 1324 M
0 6 V
-31 -6 R
62 0 V
-62 6 R
62 0 V
27 -69 R
0 7 V
-31 -7 R
62 0 V
-62 7 R
62 0 V
27 -104 R
0 8 V
-31 -8 R
62 0 V
-62 8 R
62 0 V
26 -70 R
0 12 V
-31 -12 R
62 0 V
-62 12 R
62 0 V
918 980 M
0 18 V
887 980 M
62 0 V
-62 18 R
62 0 V
27 -55 R
0 23 V
945 943 M
62 0 V
-62 23 R
62 0 V
27 -92 R
0 30 V
-31 -30 R
62 0 V
-62 30 R
62 0 V
27 -136 R
0 40 V
-31 -40 R
62 0 V
-62 40 R
62 0 V
41 -43 R
0 37 V
-31 -37 R
62 0 V
-62 37 R
62 0 V
70 -169 R
0 48 V
-31 -48 R
62 0 V
-62 48 R
62 0 V
1396 534 M
0 62 V
-31 -62 R
62 0 V
-62 62 R
62 0 V
1613 482 M
0 69 V
-31 -69 R
62 0 V
-62 69 R
62 0 V
stroke
grestore
end
showpage
}
\put(1902,941){\makebox(0,0)[r]{{\scriptsize \lepto\ SCI}}}
\put(1902,1041){\makebox(0,0)[r]{{\scriptsize \lepto}}}
\put(1902,1141){\makebox(0,0)[r]{{\scriptsize \ariadne}}}
\put(1902,1241){\makebox(0,0)[r]{{\scriptsize LDC~A}}}
\put(988,1260){\makebox(0,0){(d)}}
\put(1468,51){\makebox(0,0){$\kt$ (GeV)}}
\put(280,805){%
\special{ps: gsave currentpoint currentpoint translate
270 rotate neg exch neg exch translate}%
\makebox(0,0)[b]{\shortstack{$1/N dn/d\kt$}}%
\special{ps: currentpoint grestore moveto}%
}
\put(2336,151){\makebox(0,0){6}}
\put(1757,151){\makebox(0,0){4}}
\put(1179,151){\makebox(0,0){2}}
\put(600,151){\makebox(0,0){0}}
\put(540,1150){\makebox(0,0)[r]{$1$}}
\put(540,851){\makebox(0,0)[r]{$10^{-1}$}}
\put(540,551){\makebox(0,0)[r]{$10^{-2}$}}
\put(540,251){\makebox(0,0)[r]{$10^{-3}$}}
\end{picture}
    }
  \vskip -0.2cm
  \hbox{
    \hskip -1.7cm
\setlength{\unitlength}{0.1bp}
\special{!
/gnudict 40 dict def
gnudict begin
/Color false def
/Solid false def
/gnulinewidth 5.000 def
/vshift -33 def
/dl {10 mul} def
/hpt 31.5 def
/vpt 31.5 def
/M {moveto} bind def
/L {lineto} bind def
/R {rmoveto} bind def
/V {rlineto} bind def
/vpt2 vpt 2 mul def
/hpt2 hpt 2 mul def
/Lshow { currentpoint stroke M
  0 vshift R show } def
/Rshow { currentpoint stroke M
  dup stringwidth pop neg vshift R show } def
/Cshow { currentpoint stroke M
  dup stringwidth pop -2 div vshift R show } def
/DL { Color {setrgbcolor Solid {pop []} if 0 setdash }
 {pop pop pop Solid {pop []} if 0 setdash} ifelse } def
/BL { stroke gnulinewidth 2 mul setlinewidth } def
/AL { stroke gnulinewidth 2 div setlinewidth } def
/PL { stroke gnulinewidth setlinewidth } def
/LTb { BL [] 0 0 0 DL } def
/LTa { AL [1 dl 2 dl] 0 setdash 0 0 0 setrgbcolor } def
/LT0 { PL [] 0 1 0 DL } def
/LT1 { PL [4 dl 2 dl] 0 0 1 DL } def
/LT2 { PL [2 dl 3 dl] 1 0 0 DL } def
/LT3 { PL [1 dl 1.5 dl] 1 0 1 DL } def
/LT4 { PL [5 dl 2 dl 1 dl 2 dl] 0 1 1 DL } def
/LT5 { PL [4 dl 3 dl 1 dl 3 dl] 1 1 0 DL } def
/LT6 { PL [2 dl 2 dl 2 dl 4 dl] 0 0 0 DL } def
/LT7 { PL [2 dl 2 dl 2 dl 2 dl 2 dl 4 dl] 1 0.3 0 DL } def
/LT8 { PL [2 dl 2 dl 2 dl 2 dl 2 dl 2 dl 2 dl 4 dl] 0.5 0.5 0.5 DL } def
/P { stroke [] 0 setdash
  currentlinewidth 2 div sub M
  0 currentlinewidth V stroke } def
/D { stroke [] 0 setdash 2 copy vpt add M
  hpt neg vpt neg V hpt vpt neg V
  hpt vpt V hpt neg vpt V closepath stroke
  P } def
/A { stroke [] 0 setdash vpt sub M 0 vpt2 V
  currentpoint stroke M
  hpt neg vpt neg R hpt2 0 V stroke
  } def
/B { stroke [] 0 setdash 2 copy exch hpt sub exch vpt add M
  0 vpt2 neg V hpt2 0 V 0 vpt2 V
  hpt2 neg 0 V closepath stroke
  P } def
/C { stroke [] 0 setdash exch hpt sub exch vpt add M
  hpt2 vpt2 neg V currentpoint stroke M
  hpt2 neg 0 R hpt2 vpt2 V stroke } def
/T { stroke [] 0 setdash 2 copy vpt 1.12 mul add M
  hpt neg vpt -1.62 mul V
  hpt 2 mul 0 V
  hpt neg vpt 1.62 mul V closepath stroke
  P  } def
/S { 2 copy A C} def
end
}
\begin{picture}(2519,1511)(0,0)
\special{"
gnudict begin
gsave
50 50 translate
0.100 0.100 scale
0 setgray
/Helvetica findfont 100 scalefont setfont
newpath
-500.000000 -500.000000 translate
LTa
LTb
600 251 M
63 0 V
1673 0 R
-63 0 V
600 568 M
63 0 V
1673 0 R
-63 0 V
600 885 M
63 0 V
1673 0 R
-63 0 V
600 1202 M
63 0 V
1673 0 R
-63 0 V
817 251 M
0 63 V
0 1046 R
0 -63 V
1251 251 M
0 63 V
0 1046 R
0 -63 V
1685 251 M
0 63 V
0 1046 R
0 -63 V
2119 251 M
0 63 V
0 1046 R
0 -63 V
600 251 M
1736 0 V
0 1109 V
-1736 0 V
600 251 L
LT0
1311 1265 M
180 0 V
654 489 M
763 460 L
108 25 V
109 16 V
108 4 V
109 13 V
108 18 V
109 80 V
108 142 V
109 46 V
108 102 V
109 84 V
108 -39 V
2065 739 L
2173 450 L
2282 290 L
LT1
1311 1165 M
180 0 V
654 662 M
109 38 V
871 682 L
980 669 L
108 25 V
109 62 V
108 -7 V
109 78 V
108 109 V
109 -22 V
108 62 V
109 55 V
108 -1 V
2065 790 L
2173 476 L
2282 288 L
LT2
1311 1065 M
180 0 V
654 312 M
109 14 V
108 30 V
109 24 V
108 -35 V
109 33 V
108 11 V
109 35 V
108 -4 V
109 138 V
108 5 V
109 114 V
108 33 V
2065 580 L
2173 389 L
2282 280 L
LT3
1311 965 M
180 0 V
654 315 M
109 8 V
108 50 V
109 12 V
108 54 V
109 -11 V
108 4 V
109 53 V
108 -17 V
109 43 V
108 110 V
109 74 V
108 14 V
2065 544 L
2173 416 L
2282 282 L
LT0
1414 821 A
1522 850 A
1631 1037 A
1739 1014 A
1848 1112 A
1956 919 A
1414 761 M
0 121 V
1383 761 M
62 0 V
-62 121 R
62 0 V
77 -145 R
0 226 V
1491 737 M
62 0 V
-62 226 R
62 0 V
78 -61 R
0 270 V
1600 902 M
62 0 V
-62 270 R
62 0 V
77 -232 R
0 147 V
1708 940 M
62 0 V
-62 147 R
62 0 V
78 -43 R
0 135 V
-31 -135 R
62 0 V
-62 135 R
62 0 V
77 -304 R
0 88 V
-31 -88 R
62 0 V
-62 88 R
62 0 V
LT0
1414 821 A
1522 850 A
1631 1037 A
1739 1014 A
1848 1112 A
1956 919 A
1414 784 M
0 75 V
-31 -75 R
62 0 V
-62 75 R
62 0 V
77 -46 R
0 74 V
-31 -74 R
62 0 V
-62 74 R
62 0 V
78 102 R
0 96 V
-31 -96 R
62 0 V
-62 96 R
62 0 V
77 -115 R
0 87 V
-31 -87 R
62 0 V
-62 87 R
62 0 V
78 11 R
0 87 V
-31 -87 R
62 0 V
-62 87 R
62 0 V
77 -271 R
0 70 V
-31 -70 R
62 0 V
-62 70 R
62 0 V
stroke
grestore
end
showpage
}
\put(1251,965){\makebox(0,0)[r]{{\scriptsize \lepto\ SCI}}}
\put(1251,1065){\makebox(0,0)[r]{{\scriptsize \lepto}}}
\put(1251,1165){\makebox(0,0)[r]{{\scriptsize \ariadne}}}
\put(1251,1265){\makebox(0,0)[r]{{\scriptsize LDC~A}}}
\put(748,1260){\makebox(0,0){(e)}}
\put(1468,51){\makebox(0,0){$\eta$}}
\put(340,805){%
\special{ps: gsave currentpoint currentpoint translate
270 rotate neg exch neg exch translate}%
\makebox(0,0)[b]{\shortstack{$1/N dn/d\eta$}}%
\special{ps: currentpoint grestore moveto}%
}
\put(2119,151){\makebox(0,0){4}}
\put(1685,151){\makebox(0,0){2}}
\put(1251,151){\makebox(0,0){0}}
\put(817,151){\makebox(0,0){-2}}
\put(540,1202){\makebox(0,0)[r]{0.3}}
\put(540,885){\makebox(0,0)[r]{0.2}}
\put(540,568){\makebox(0,0)[r]{0.1}}
\put(540,251){\makebox(0,0)[r]{0}}
\end{picture}
    \hskip -1.2cm
\setlength{\unitlength}{0.1bp}
\special{!
/gnudict 40 dict def
gnudict begin
/Color false def
/Solid false def
/gnulinewidth 5.000 def
/vshift -33 def
/dl {10 mul} def
/hpt 31.5 def
/vpt 31.5 def
/M {moveto} bind def
/L {lineto} bind def
/R {rmoveto} bind def
/V {rlineto} bind def
/vpt2 vpt 2 mul def
/hpt2 hpt 2 mul def
/Lshow { currentpoint stroke M
  0 vshift R show } def
/Rshow { currentpoint stroke M
  dup stringwidth pop neg vshift R show } def
/Cshow { currentpoint stroke M
  dup stringwidth pop -2 div vshift R show } def
/DL { Color {setrgbcolor Solid {pop []} if 0 setdash }
 {pop pop pop Solid {pop []} if 0 setdash} ifelse } def
/BL { stroke gnulinewidth 2 mul setlinewidth } def
/AL { stroke gnulinewidth 2 div setlinewidth } def
/PL { stroke gnulinewidth setlinewidth } def
/LTb { BL [] 0 0 0 DL } def
/LTa { AL [1 dl 2 dl] 0 setdash 0 0 0 setrgbcolor } def
/LT0 { PL [] 0 1 0 DL } def
/LT1 { PL [4 dl 2 dl] 0 0 1 DL } def
/LT2 { PL [2 dl 3 dl] 1 0 0 DL } def
/LT3 { PL [1 dl 1.5 dl] 1 0 1 DL } def
/LT4 { PL [5 dl 2 dl 1 dl 2 dl] 0 1 1 DL } def
/LT5 { PL [4 dl 3 dl 1 dl 3 dl] 1 1 0 DL } def
/LT6 { PL [2 dl 2 dl 2 dl 4 dl] 0 0 0 DL } def
/LT7 { PL [2 dl 2 dl 2 dl 2 dl 2 dl 4 dl] 1 0.3 0 DL } def
/LT8 { PL [2 dl 2 dl 2 dl 2 dl 2 dl 2 dl 2 dl 4 dl] 0.5 0.5 0.5 DL } def
/P { stroke [] 0 setdash
  currentlinewidth 2 div sub M
  0 currentlinewidth V stroke } def
/D { stroke [] 0 setdash 2 copy vpt add M
  hpt neg vpt neg V hpt vpt neg V
  hpt vpt V hpt neg vpt V closepath stroke
  P } def
/A { stroke [] 0 setdash vpt sub M 0 vpt2 V
  currentpoint stroke M
  hpt neg vpt neg R hpt2 0 V stroke
  } def
/B { stroke [] 0 setdash 2 copy exch hpt sub exch vpt add M
  0 vpt2 neg V hpt2 0 V 0 vpt2 V
  hpt2 neg 0 V closepath stroke
  P } def
/C { stroke [] 0 setdash exch hpt sub exch vpt add M
  hpt2 vpt2 neg V currentpoint stroke M
  hpt2 neg 0 R hpt2 vpt2 V stroke } def
/T { stroke [] 0 setdash 2 copy vpt 1.12 mul add M
  hpt neg vpt -1.62 mul V
  hpt 2 mul 0 V
  hpt neg vpt 1.62 mul V closepath stroke
  P  } def
/S { 2 copy A C} def
end
}
\begin{picture}(2519,1511)(0,0)
\special{"
gnudict begin
gsave
50 50 translate
0.100 0.100 scale
0 setgray
/Helvetica findfont 100 scalefont setfont
newpath
-500.000000 -500.000000 translate
LTa
LTb
600 251 M
63 0 V
1673 0 R
-63 0 V
600 806 M
63 0 V
1673 0 R
-63 0 V
600 1360 M
63 0 V
1673 0 R
-63 0 V
600 251 M
0 63 V
0 1046 R
0 -63 V
1468 251 M
0 63 V
0 1046 R
0 -63 V
2336 251 M
0 63 V
0 1046 R
0 -63 V
600 251 M
1736 0 V
0 1109 V
-1736 0 V
600 251 L
LT0
1789 1249 M
180 0 V
811 519 M
287 71 V
262 41 V
319 -10 V
287 58 V
262 45 V
LT1
1789 1149 M
180 0 V
811 729 M
287 21 V
262 16 V
319 -24 V
287 91 V
262 112 V
LT2
1789 1049 M
180 0 V
811 353 M
287 74 V
262 22 V
319 76 V
287 11 V
262 267 V
LT3
1789 949 M
180 0 V
811 328 M
287 75 V
262 86 V
319 39 V
287 54 V
262 225 V
LT0
811 783 A
1098 772 A
1360 794 A
1679 817 A
1966 828 A
2228 1049 A
811 728 M
0 111 V
780 728 M
62 0 V
780 839 M
62 0 V
1098 702 M
0 140 V
1067 702 M
62 0 V
-62 140 R
62 0 V
1360 735 M
0 119 V
1329 735 M
62 0 V
-62 119 R
62 0 V
288 -87 R
0 99 V
-31 -99 R
62 0 V
-62 99 R
62 0 V
256 -88 R
0 100 V
1935 778 M
62 0 V
-62 100 R
62 0 V
231 5 R
0 333 V
2197 883 M
62 0 V
-62 333 R
62 0 V
LT0
811 783 A
1098 772 A
1360 794 A
1679 817 A
1966 828 A
2228 1049 A
811 750 M
0 67 V
780 750 M
62 0 V
-62 67 R
62 0 V
256 -67 R
0 44 V
-31 -44 R
62 0 V
-62 44 R
62 0 V
231 -22 R
0 45 V
-31 -45 R
62 0 V
-62 45 R
62 0 V
288 -23 R
0 45 V
-31 -45 R
62 0 V
-62 45 R
62 0 V
256 -33 R
0 44 V
-31 -44 R
62 0 V
-62 44 R
62 0 V
231 144 R
0 111 V
2197 994 M
62 0 V
-62 111 R
62 0 V
stroke
grestore
end
showpage
}
\put(1729,949){\makebox(0,0)[r]{{\scriptsize \lepto\ SCI}}}
\put(1729,1049){\makebox(0,0)[r]{{\scriptsize \lepto}}}
\put(1729,1149){\makebox(0,0)[r]{{\scriptsize \ariadne}}}
\put(1729,1249){\makebox(0,0)[r]{{\scriptsize LDC~A}}}
\put(748,1260){\makebox(0,0){(f)}}
\put(1468,51){\makebox(0,0){$x$}}
\put(280,805){%
\special{ps: gsave currentpoint currentpoint translate
270 rotate neg exch neg exch translate}%
\makebox(0,0)[b]{\shortstack{$dR_2/dx$}}%
\special{ps: currentpoint grestore moveto}%
}
\put(2336,151){\makebox(0,0){$10^{-2}$}}
\put(1468,151){\makebox(0,0){$10^{-3}$}}
\put(600,151){\makebox(0,0){$10^{-4}$}}
\put(540,1360){\makebox(0,0)[r]{0.1}}
\put(540,806){\makebox(0,0)[r]{0.05}}
\put(540,251){\makebox(0,0)[r]{0}}
\end{picture}
    }
  \vskip -3mm
  \caption[dummy]{{\it Comparison of the default LDC~A model with HERA
      data as given in \cite{HZTOOL} and other event generators. The
      distributions are as follows: (a) The transverse energy flow as
      a function of pseudo rapidity for events with $0.0001<x<0.0002$,
      $5<Q^2/\mbox{GeV}^2<10$ \cite{ForwardJets}.  (b) As (a) but for
      events with $0.003<x<0.01$, $20<Q^2/\mbox{GeV}^2<50$. (c) The
      average squared transverse momentum of charged particles as a
      function of $x_F$ \cite{HZ95221}. (d) The transverse momentum
      distribution of charged particles in the pseudo rapidity bin
      $0.5<\eta<1.5$ for events with $0.0002<x<0.0005$,
      $6<Q^2/\mbox{GeV}^2<10$ \cite{ForwardParticles}. (e) The
      pseudo rapidity distribution of charged particles with a
      transverse momentum larger than 1 GeV for the same kinematical
      bin as in (d). (f) The two-jet ratio $R_2(x)$ as a function of
      $x$ \cite{R2H1}. All measurements were made in the hadronic
      centre of mass system and only events without a large rapidity
      gap were included.  The full line is LDC~A, long-dashed is
      \ariadne\ 4.08 with default parameter settings, dotted is
      \lepto\ 6.4 with default parameter settings and short-dashed is
      the same but with SCI and the special sea-quark remnant
      treatment \cite{LeptoSCI} switched off.}}
  \label{fig:comp}
\end{figure}
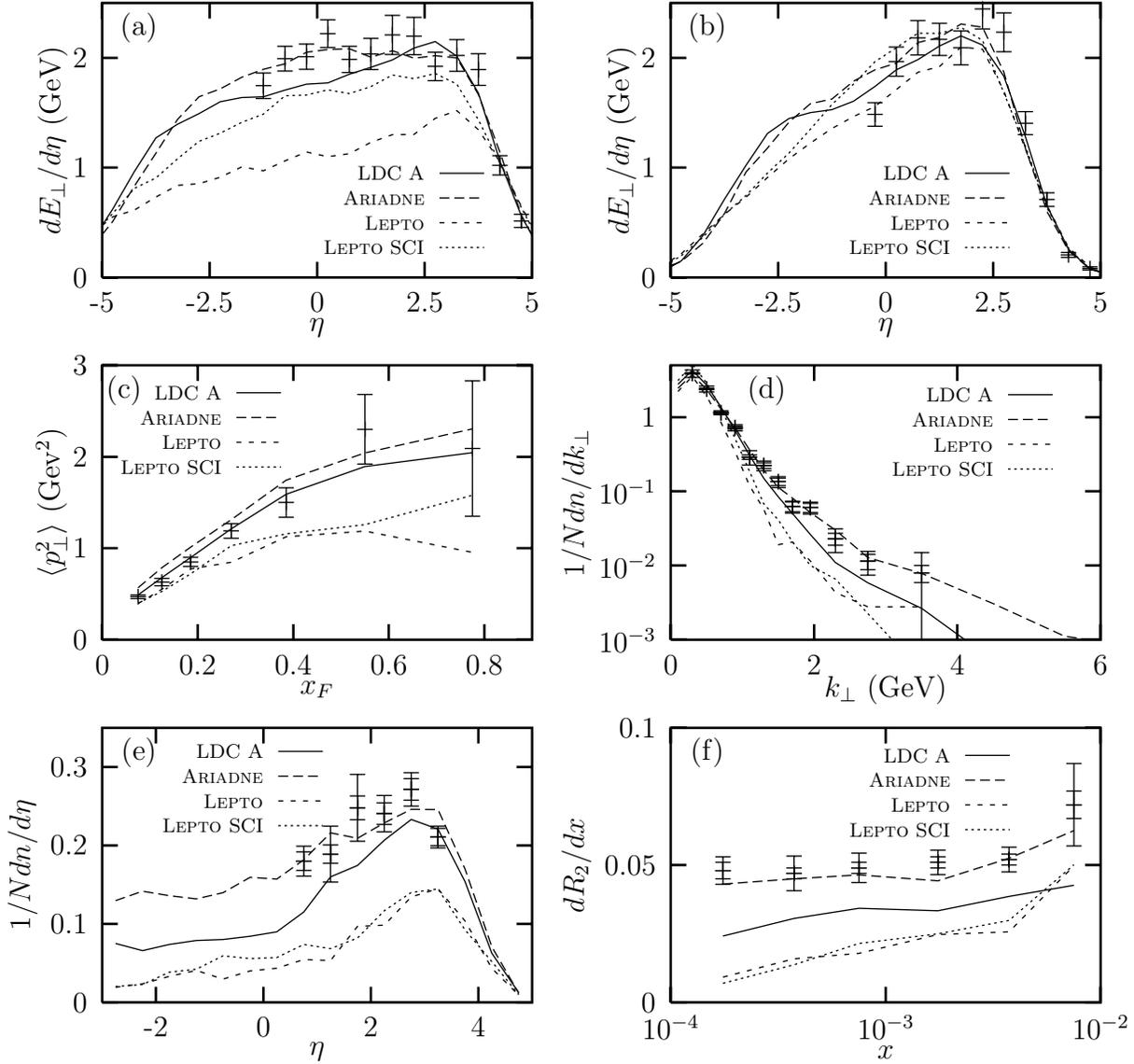

\begin{figure}
  \vskip -1cm
  \hbox{
    \hskip -2.0cm
\setlength{\unitlength}{0.1bp}
\special{!
/gnudict 40 dict def
gnudict begin
/Color false def
/Solid false def
/gnulinewidth 5.000 def
/vshift -33 def
/dl {10 mul} def
/hpt 31.5 def
/vpt 31.5 def
/M {moveto} bind def
/L {lineto} bind def
/R {rmoveto} bind def
/V {rlineto} bind def
/vpt2 vpt 2 mul def
/hpt2 hpt 2 mul def
/Lshow { currentpoint stroke M
  0 vshift R show } def
/Rshow { currentpoint stroke M
  dup stringwidth pop neg vshift R show } def
/Cshow { currentpoint stroke M
  dup stringwidth pop -2 div vshift R show } def
/DL { Color {setrgbcolor Solid {pop []} if 0 setdash }
 {pop pop pop Solid {pop []} if 0 setdash} ifelse } def
/BL { stroke gnulinewidth 2 mul setlinewidth } def
/AL { stroke gnulinewidth 2 div setlinewidth } def
/PL { stroke gnulinewidth setlinewidth } def
/LTb { BL [] 0 0 0 DL } def
/LTa { AL [1 dl 2 dl] 0 setdash 0 0 0 setrgbcolor } def
/LT0 { PL [] 0 1 0 DL } def
/LT1 { PL [4 dl 2 dl] 0 0 1 DL } def
/LT2 { PL [2 dl 3 dl] 1 0 0 DL } def
/LT3 { PL [1 dl 1.5 dl] 1 0 1 DL } def
/LT4 { PL [5 dl 2 dl 1 dl 2 dl] 0 1 1 DL } def
/LT5 { PL [4 dl 3 dl 1 dl 3 dl] 1 1 0 DL } def
/LT6 { PL [2 dl 2 dl 2 dl 4 dl] 0 0 0 DL } def
/LT7 { PL [2 dl 2 dl 2 dl 2 dl 2 dl 4 dl] 1 0.3 0 DL } def
/LT8 { PL [2 dl 2 dl 2 dl 2 dl 2 dl 2 dl 2 dl 4 dl] 0.5 0.5 0.5 DL } def
/P { stroke [] 0 setdash
  currentlinewidth 2 div sub M
  0 currentlinewidth V stroke } def
/D { stroke [] 0 setdash 2 copy vpt add M
  hpt neg vpt neg V hpt vpt neg V
  hpt vpt V hpt neg vpt V closepath stroke
  P } def
/A { stroke [] 0 setdash vpt sub M 0 vpt2 V
  currentpoint stroke M
  hpt neg vpt neg R hpt2 0 V stroke
  } def
/B { stroke [] 0 setdash 2 copy exch hpt sub exch vpt add M
  0 vpt2 neg V hpt2 0 V 0 vpt2 V
  hpt2 neg 0 V closepath stroke
  P } def
/C { stroke [] 0 setdash exch hpt sub exch vpt add M
  hpt2 vpt2 neg V currentpoint stroke M
  hpt2 neg 0 R hpt2 vpt2 V stroke } def
/T { stroke [] 0 setdash 2 copy vpt 1.12 mul add M
  hpt neg vpt -1.62 mul V
  hpt 2 mul 0 V
  hpt neg vpt 1.62 mul V closepath stroke
  P  } def
/S { 2 copy A C} def
end
}
\begin{picture}(2519,1511)(0,0)
\special{"
gnudict begin
gsave
50 50 translate
0.100 0.100 scale
0 setgray
/Helvetica findfont 100 scalefont setfont
newpath
-500.000000 -500.000000 translate
LTa
LTb
600 251 M
63 0 V
1673 0 R
-63 0 V
600 568 M
63 0 V
1673 0 R
-63 0 V
600 885 M
63 0 V
1673 0 R
-63 0 V
600 1202 M
63 0 V
1673 0 R
-63 0 V
817 251 M
0 63 V
0 1046 R
0 -63 V
1251 251 M
0 63 V
0 1046 R
0 -63 V
1685 251 M
0 63 V
0 1046 R
0 -63 V
2119 251 M
0 63 V
0 1046 R
0 -63 V
600 251 M
1736 0 V
0 1109 V
-1736 0 V
600 251 L
LT0
1311 1265 M
180 0 V
654 489 M
763 460 L
108 25 V
109 16 V
108 4 V
109 13 V
108 18 V
109 80 V
108 142 V
109 46 V
108 102 V
109 84 V
108 -39 V
2065 739 L
2173 450 L
2282 290 L
LT1
1311 1165 M
180 0 V
654 388 M
109 0 V
871 362 L
109 51 V
108 -7 V
109 -15 V
108 11 V
109 24 V
108 5 V
109 28 V
108 -17 V
109 4 V
108 51 V
109 42 V
2173 439 L
2282 298 L
LT3
1311 1065 M
180 0 V
654 465 M
763 445 L
108 28 V
980 456 L
108 82 V
109 -28 V
108 39 V
109 32 V
108 87 V
109 103 V
108 17 V
109 117 V
108 47 V
2065 681 L
2173 471 L
2282 283 L
LT0
1414 821 A
1522 850 A
1631 1037 A
1739 1014 A
1848 1112 A
1956 919 A
1414 761 M
0 121 V
1383 761 M
62 0 V
-62 121 R
62 0 V
77 -145 R
0 226 V
1491 737 M
62 0 V
-62 226 R
62 0 V
78 -61 R
0 270 V
1600 902 M
62 0 V
-62 270 R
62 0 V
77 -232 R
0 147 V
1708 940 M
62 0 V
-62 147 R
62 0 V
78 -43 R
0 135 V
-31 -135 R
62 0 V
-62 135 R
62 0 V
77 -304 R
0 88 V
-31 -88 R
62 0 V
-62 88 R
62 0 V
LT0
1414 821 A
1522 850 A
1631 1037 A
1739 1014 A
1848 1112 A
1956 919 A
1414 784 M
0 75 V
-31 -75 R
62 0 V
-62 75 R
62 0 V
77 -46 R
0 74 V
-31 -74 R
62 0 V
-62 74 R
62 0 V
78 102 R
0 96 V
-31 -96 R
62 0 V
-62 96 R
62 0 V
77 -115 R
0 87 V
-31 -87 R
62 0 V
-62 87 R
62 0 V
78 11 R
0 87 V
-31 -87 R
62 0 V
-62 87 R
62 0 V
77 -271 R
0 70 V
-31 -70 R
62 0 V
-62 70 R
62 0 V
stroke
grestore
end
showpage
}
\put(1251,1065){\makebox(0,0)[r]{{\scriptsize LDC~C}}}
\put(1251,1165){\makebox(0,0)[r]{{\scriptsize LDC~B}}}
\put(1251,1265){\makebox(0,0)[r]{{\scriptsize LDC~A}}}
\put(808,1260){\makebox(0,0){(a)}}
\put(1468,51){\makebox(0,0){$\eta$}}
\put(220,805){%
\special{ps: gsave currentpoint currentpoint translate
270 rotate neg exch neg exch translate}%
\makebox(0,0)[b]{\shortstack{ }}%
\special{ps: currentpoint grestore moveto}%
}
\put(2119,151){\makebox(0,0){4}}
\put(1685,151){\makebox(0,0){2}}
\put(1251,151){\makebox(0,0){0}}
\put(817,151){\makebox(0,0){-2}}
\put(540,1202){\makebox(0,0)[r]{0.3}}
\put(540,885){\makebox(0,0)[r]{0.2}}
\put(540,568){\makebox(0,0)[r]{0.1}}
\put(540,251){\makebox(0,0)[r]{0}}
\end{picture}
    \hskip -1.0cm
\setlength{\unitlength}{0.1bp}
\special{!
/gnudict 40 dict def
gnudict begin
/Color false def
/Solid false def
/gnulinewidth 5.000 def
/vshift -33 def
/dl {10 mul} def
/hpt 31.5 def
/vpt 31.5 def
/M {moveto} bind def
/L {lineto} bind def
/R {rmoveto} bind def
/V {rlineto} bind def
/vpt2 vpt 2 mul def
/hpt2 hpt 2 mul def
/Lshow { currentpoint stroke M
  0 vshift R show } def
/Rshow { currentpoint stroke M
  dup stringwidth pop neg vshift R show } def
/Cshow { currentpoint stroke M
  dup stringwidth pop -2 div vshift R show } def
/DL { Color {setrgbcolor Solid {pop []} if 0 setdash }
 {pop pop pop Solid {pop []} if 0 setdash} ifelse } def
/BL { stroke gnulinewidth 2 mul setlinewidth } def
/AL { stroke gnulinewidth 2 div setlinewidth } def
/PL { stroke gnulinewidth setlinewidth } def
/LTb { BL [] 0 0 0 DL } def
/LTa { AL [1 dl 2 dl] 0 setdash 0 0 0 setrgbcolor } def
/LT0 { PL [] 0 1 0 DL } def
/LT1 { PL [4 dl 2 dl] 0 0 1 DL } def
/LT2 { PL [2 dl 3 dl] 1 0 0 DL } def
/LT3 { PL [1 dl 1.5 dl] 1 0 1 DL } def
/LT4 { PL [5 dl 2 dl 1 dl 2 dl] 0 1 1 DL } def
/LT5 { PL [4 dl 3 dl 1 dl 3 dl] 1 1 0 DL } def
/LT6 { PL [2 dl 2 dl 2 dl 4 dl] 0 0 0 DL } def
/LT7 { PL [2 dl 2 dl 2 dl 2 dl 2 dl 4 dl] 1 0.3 0 DL } def
/LT8 { PL [2 dl 2 dl 2 dl 2 dl 2 dl 2 dl 2 dl 4 dl] 0.5 0.5 0.5 DL } def
/P { stroke [] 0 setdash
  currentlinewidth 2 div sub M
  0 currentlinewidth V stroke } def
/D { stroke [] 0 setdash 2 copy vpt add M
  hpt neg vpt neg V hpt vpt neg V
  hpt vpt V hpt neg vpt V closepath stroke
  P } def
/A { stroke [] 0 setdash vpt sub M 0 vpt2 V
  currentpoint stroke M
  hpt neg vpt neg R hpt2 0 V stroke
  } def
/B { stroke [] 0 setdash 2 copy exch hpt sub exch vpt add M
  0 vpt2 neg V hpt2 0 V 0 vpt2 V
  hpt2 neg 0 V closepath stroke
  P } def
/C { stroke [] 0 setdash exch hpt sub exch vpt add M
  hpt2 vpt2 neg V currentpoint stroke M
  hpt2 neg 0 R hpt2 vpt2 V stroke } def
/T { stroke [] 0 setdash 2 copy vpt 1.12 mul add M
  hpt neg vpt -1.62 mul V
  hpt 2 mul 0 V
  hpt neg vpt 1.62 mul V closepath stroke
  P  } def
/S { 2 copy A C} def
end
}
\begin{picture}(2519,1511)(0,0)
\special{"
gnudict begin
gsave
50 50 translate
0.100 0.100 scale
0 setgray
/Helvetica findfont 100 scalefont setfont
newpath
-500.000000 -500.000000 translate
LTa
LTb
600 251 M
63 0 V
1673 0 R
-63 0 V
600 568 M
63 0 V
1673 0 R
-63 0 V
600 885 M
63 0 V
1673 0 R
-63 0 V
600 1202 M
63 0 V
1673 0 R
-63 0 V
817 251 M
0 63 V
0 1046 R
0 -63 V
1251 251 M
0 63 V
0 1046 R
0 -63 V
1685 251 M
0 63 V
0 1046 R
0 -63 V
2119 251 M
0 63 V
0 1046 R
0 -63 V
600 251 M
1736 0 V
0 1109 V
-1736 0 V
600 251 L
LT0
1311 1265 M
180 0 V
654 489 M
763 460 L
108 25 V
109 16 V
108 4 V
109 13 V
108 18 V
109 80 V
108 142 V
109 46 V
108 102 V
109 84 V
108 -39 V
2065 739 L
2173 450 L
2282 290 L
LT1
1311 1165 M
180 0 V
654 473 M
763 453 L
108 20 V
980 459 L
108 24 V
109 29 V
108 73 V
109 60 V
108 79 V
109 78 V
108 116 V
109 28 V
108 -14 V
2065 781 L
2173 451 L
2282 312 L
LT2
1311 1065 M
180 0 V
654 495 M
109 7 V
871 479 L
980 465 L
108 30 V
109 16 V
108 78 V
109 28 V
108 107 V
109 96 V
108 65 V
109 105 V
108 -63 V
2065 715 L
2173 450 L
2282 293 L
LT3
1311 965 M
180 0 V
654 486 M
109 15 V
871 473 L
109 8 V
108 5 V
109 41 V
108 46 V
109 50 V
108 70 V
109 115 V
108 117 V
109 101 V
108 -91 V
2065 763 L
2173 445 L
2282 280 L
LT0
1414 821 A
1522 850 A
1631 1037 A
1739 1014 A
1848 1112 A
1956 919 A
1414 761 M
0 121 V
1383 761 M
62 0 V
-62 121 R
62 0 V
77 -145 R
0 226 V
1491 737 M
62 0 V
-62 226 R
62 0 V
78 -61 R
0 270 V
1600 902 M
62 0 V
-62 270 R
62 0 V
77 -232 R
0 147 V
1708 940 M
62 0 V
-62 147 R
62 0 V
78 -43 R
0 135 V
-31 -135 R
62 0 V
-62 135 R
62 0 V
77 -304 R
0 88 V
-31 -88 R
62 0 V
-62 88 R
62 0 V
LT0
1414 821 A
1522 850 A
1631 1037 A
1739 1014 A
1848 1112 A
1956 919 A
1414 784 M
0 75 V
-31 -75 R
62 0 V
-62 75 R
62 0 V
77 -46 R
0 74 V
-31 -74 R
62 0 V
-62 74 R
62 0 V
78 102 R
0 96 V
-31 -96 R
62 0 V
-62 96 R
62 0 V
77 -115 R
0 87 V
-31 -87 R
62 0 V
-62 87 R
62 0 V
78 11 R
0 87 V
-31 -87 R
62 0 V
-62 87 R
62 0 V
77 -271 R
0 70 V
-31 -70 R
62 0 V
-62 70 R
62 0 V
stroke
grestore
end
showpage
}
\put(1251,965){\makebox(0,0)[r]{{\scriptsize LDC~H}}}
\put(1251,1065){\makebox(0,0)[r]{{\scriptsize LDC~A$_z$}}}
\put(1251,1165){\makebox(0,0)[r]{{\scriptsize LDC~D}}}
\put(1251,1265){\makebox(0,0)[r]{{\scriptsize LDC~A}}}
\put(808,1260){\makebox(0,0){(b)}}
\put(1468,51){\makebox(0,0){$\eta$}}
\put(220,805){%
\special{ps: gsave currentpoint currentpoint translate
270 rotate neg exch neg exch translate}%
\makebox(0,0)[b]{\shortstack{$1/N dn/d\eta$}}%
\special{ps: currentpoint grestore moveto}%
}
\put(2119,151){\makebox(0,0){4}}
\put(1685,151){\makebox(0,0){2}}
\put(1251,151){\makebox(0,0){0}}
\put(817,151){\makebox(0,0){-2}}
\put(540,1202){\makebox(0,0)[r]{0.3}}
\put(540,885){\makebox(0,0)[r]{0.2}}
\put(540,568){\makebox(0,0)[r]{0.1}}
\put(540,251){\makebox(0,0)[r]{0}}
\end{picture}
    }
  \vskip 0.5cm
  \hbox{
    \hskip -2.0cm
\setlength{\unitlength}{0.1bp}
\special{!
/gnudict 40 dict def
gnudict begin
/Color false def
/Solid false def
/gnulinewidth 5.000 def
/vshift -33 def
/dl {10 mul} def
/hpt 31.5 def
/vpt 31.5 def
/M {moveto} bind def
/L {lineto} bind def
/R {rmoveto} bind def
/V {rlineto} bind def
/vpt2 vpt 2 mul def
/hpt2 hpt 2 mul def
/Lshow { currentpoint stroke M
  0 vshift R show } def
/Rshow { currentpoint stroke M
  dup stringwidth pop neg vshift R show } def
/Cshow { currentpoint stroke M
  dup stringwidth pop -2 div vshift R show } def
/DL { Color {setrgbcolor Solid {pop []} if 0 setdash }
 {pop pop pop Solid {pop []} if 0 setdash} ifelse } def
/BL { stroke gnulinewidth 2 mul setlinewidth } def
/AL { stroke gnulinewidth 2 div setlinewidth } def
/PL { stroke gnulinewidth setlinewidth } def
/LTb { BL [] 0 0 0 DL } def
/LTa { AL [1 dl 2 dl] 0 setdash 0 0 0 setrgbcolor } def
/LT0 { PL [] 0 1 0 DL } def
/LT1 { PL [4 dl 2 dl] 0 0 1 DL } def
/LT2 { PL [2 dl 3 dl] 1 0 0 DL } def
/LT3 { PL [1 dl 1.5 dl] 1 0 1 DL } def
/LT4 { PL [5 dl 2 dl 1 dl 2 dl] 0 1 1 DL } def
/LT5 { PL [4 dl 3 dl 1 dl 3 dl] 1 1 0 DL } def
/LT6 { PL [2 dl 2 dl 2 dl 4 dl] 0 0 0 DL } def
/LT7 { PL [2 dl 2 dl 2 dl 2 dl 2 dl 4 dl] 1 0.3 0 DL } def
/LT8 { PL [2 dl 2 dl 2 dl 2 dl 2 dl 2 dl 2 dl 4 dl] 0.5 0.5 0.5 DL } def
/P { stroke [] 0 setdash
  currentlinewidth 2 div sub M
  0 currentlinewidth V stroke } def
/D { stroke [] 0 setdash 2 copy vpt add M
  hpt neg vpt neg V hpt vpt neg V
  hpt vpt V hpt neg vpt V closepath stroke
  P } def
/A { stroke [] 0 setdash vpt sub M 0 vpt2 V
  currentpoint stroke M
  hpt neg vpt neg R hpt2 0 V stroke
  } def
/B { stroke [] 0 setdash 2 copy exch hpt sub exch vpt add M
  0 vpt2 neg V hpt2 0 V 0 vpt2 V
  hpt2 neg 0 V closepath stroke
  P } def
/C { stroke [] 0 setdash exch hpt sub exch vpt add M
  hpt2 vpt2 neg V currentpoint stroke M
  hpt2 neg 0 R hpt2 vpt2 V stroke } def
/T { stroke [] 0 setdash 2 copy vpt 1.12 mul add M
  hpt neg vpt -1.62 mul V
  hpt 2 mul 0 V
  hpt neg vpt 1.62 mul V closepath stroke
  P  } def
/S { 2 copy A C} def
end
}
\begin{picture}(2519,1511)(0,0)
\special{"
gnudict begin
gsave
50 50 translate
0.100 0.100 scale
0 setgray
/Helvetica findfont 100 scalefont setfont
newpath
-500.000000 -500.000000 translate
LTa
LTb
600 251 M
63 0 V
1673 0 R
-63 0 V
600 568 M
63 0 V
1673 0 R
-63 0 V
600 885 M
63 0 V
1673 0 R
-63 0 V
600 1202 M
63 0 V
1673 0 R
-63 0 V
817 251 M
0 63 V
0 1046 R
0 -63 V
1251 251 M
0 63 V
0 1046 R
0 -63 V
1685 251 M
0 63 V
0 1046 R
0 -63 V
2119 251 M
0 63 V
0 1046 R
0 -63 V
600 251 M
1736 0 V
0 1109 V
-1736 0 V
600 251 L
LT0
1311 1265 M
180 0 V
654 489 M
763 460 L
108 25 V
109 16 V
108 4 V
109 13 V
108 18 V
109 80 V
108 142 V
109 46 V
108 102 V
109 84 V
108 -39 V
2065 739 L
2173 450 L
2282 290 L
LT1
1311 1165 M
180 0 V
654 510 M
763 490 L
108 21 V
109 10 V
108 7 V
109 10 V
108 26 V
109 119 V
108 91 V
109 106 V
108 104 V
109 56 V
108 -74 V
2065 699 L
2173 425 L
2282 285 L
LT3
1311 1065 M
180 0 V
654 460 M
109 19 V
108 15 V
980 479 L
108 39 V
109 15 V
108 46 V
109 89 V
108 74 V
109 61 V
108 136 V
109 108 V
108 -72 V
2065 752 L
2173 432 L
2282 276 L
LT0
1414 821 A
1522 850 A
1631 1037 A
1739 1014 A
1848 1112 A
1956 919 A
1414 761 M
0 121 V
1383 761 M
62 0 V
-62 121 R
62 0 V
77 -145 R
0 226 V
1491 737 M
62 0 V
-62 226 R
62 0 V
78 -61 R
0 270 V
1600 902 M
62 0 V
-62 270 R
62 0 V
77 -232 R
0 147 V
1708 940 M
62 0 V
-62 147 R
62 0 V
78 -43 R
0 135 V
-31 -135 R
62 0 V
-62 135 R
62 0 V
77 -304 R
0 88 V
-31 -88 R
62 0 V
-62 88 R
62 0 V
LT0
1414 821 A
1522 850 A
1631 1037 A
1739 1014 A
1848 1112 A
1956 919 A
1414 784 M
0 75 V
-31 -75 R
62 0 V
-62 75 R
62 0 V
77 -46 R
0 74 V
-31 -74 R
62 0 V
-62 74 R
62 0 V
78 102 R
0 96 V
-31 -96 R
62 0 V
-62 96 R
62 0 V
77 -115 R
0 87 V
-31 -87 R
62 0 V
-62 87 R
62 0 V
78 11 R
0 87 V
-31 -87 R
62 0 V
-62 87 R
62 0 V
77 -271 R
0 70 V
-31 -70 R
62 0 V
-62 70 R
62 0 V
stroke
grestore
end
showpage
}
\put(1251,1065){\makebox(0,0)[r]{{\scriptsize LDC~A$_0$}}}
\put(1251,1165){\makebox(0,0)[r]{{\scriptsize LDC~E}}}
\put(1251,1265){\makebox(0,0)[r]{{\scriptsize LDC~A}}}
\put(808,1260){\makebox(0,0){(c)}}
\put(1468,51){\makebox(0,0){$\eta$}}
\put(220,805){%
\special{ps: gsave currentpoint currentpoint translate
270 rotate neg exch neg exch translate}%
\makebox(0,0)[b]{\shortstack{ }}%
\special{ps: currentpoint grestore moveto}%
}
\put(2119,151){\makebox(0,0){4}}
\put(1685,151){\makebox(0,0){2}}
\put(1251,151){\makebox(0,0){0}}
\put(817,151){\makebox(0,0){-2}}
\put(540,1202){\makebox(0,0)[r]{0.3}}
\put(540,885){\makebox(0,0)[r]{0.2}}
\put(540,568){\makebox(0,0)[r]{0.1}}
\put(540,251){\makebox(0,0)[r]{0}}
\end{picture}
    \hskip -1.0cm
\setlength{\unitlength}{0.1bp}
\special{!
/gnudict 40 dict def
gnudict begin
/Color false def
/Solid false def
/gnulinewidth 5.000 def
/vshift -33 def
/dl {10 mul} def
/hpt 31.5 def
/vpt 31.5 def
/M {moveto} bind def
/L {lineto} bind def
/R {rmoveto} bind def
/V {rlineto} bind def
/vpt2 vpt 2 mul def
/hpt2 hpt 2 mul def
/Lshow { currentpoint stroke M
  0 vshift R show } def
/Rshow { currentpoint stroke M
  dup stringwidth pop neg vshift R show } def
/Cshow { currentpoint stroke M
  dup stringwidth pop -2 div vshift R show } def
/DL { Color {setrgbcolor Solid {pop []} if 0 setdash }
 {pop pop pop Solid {pop []} if 0 setdash} ifelse } def
/BL { stroke gnulinewidth 2 mul setlinewidth } def
/AL { stroke gnulinewidth 2 div setlinewidth } def
/PL { stroke gnulinewidth setlinewidth } def
/LTb { BL [] 0 0 0 DL } def
/LTa { AL [1 dl 2 dl] 0 setdash 0 0 0 setrgbcolor } def
/LT0 { PL [] 0 1 0 DL } def
/LT1 { PL [4 dl 2 dl] 0 0 1 DL } def
/LT2 { PL [2 dl 3 dl] 1 0 0 DL } def
/LT3 { PL [1 dl 1.5 dl] 1 0 1 DL } def
/LT4 { PL [5 dl 2 dl 1 dl 2 dl] 0 1 1 DL } def
/LT5 { PL [4 dl 3 dl 1 dl 3 dl] 1 1 0 DL } def
/LT6 { PL [2 dl 2 dl 2 dl 4 dl] 0 0 0 DL } def
/LT7 { PL [2 dl 2 dl 2 dl 2 dl 2 dl 4 dl] 1 0.3 0 DL } def
/LT8 { PL [2 dl 2 dl 2 dl 2 dl 2 dl 2 dl 2 dl 4 dl] 0.5 0.5 0.5 DL } def
/P { stroke [] 0 setdash
  currentlinewidth 2 div sub M
  0 currentlinewidth V stroke } def
/D { stroke [] 0 setdash 2 copy vpt add M
  hpt neg vpt neg V hpt vpt neg V
  hpt vpt V hpt neg vpt V closepath stroke
  P } def
/A { stroke [] 0 setdash vpt sub M 0 vpt2 V
  currentpoint stroke M
  hpt neg vpt neg R hpt2 0 V stroke
  } def
/B { stroke [] 0 setdash 2 copy exch hpt sub exch vpt add M
  0 vpt2 neg V hpt2 0 V 0 vpt2 V
  hpt2 neg 0 V closepath stroke
  P } def
/C { stroke [] 0 setdash exch hpt sub exch vpt add M
  hpt2 vpt2 neg V currentpoint stroke M
  hpt2 neg 0 R hpt2 vpt2 V stroke } def
/T { stroke [] 0 setdash 2 copy vpt 1.12 mul add M
  hpt neg vpt -1.62 mul V
  hpt 2 mul 0 V
  hpt neg vpt 1.62 mul V closepath stroke
  P  } def
/S { 2 copy A C} def
end
}
\begin{picture}(2519,1511)(0,0)
\special{"
gnudict begin
gsave
50 50 translate
0.100 0.100 scale
0 setgray
/Helvetica findfont 100 scalefont setfont
newpath
-500.000000 -500.000000 translate
LTa
LTb
600 251 M
63 0 V
1673 0 R
-63 0 V
600 568 M
63 0 V
1673 0 R
-63 0 V
600 885 M
63 0 V
1673 0 R
-63 0 V
600 1202 M
63 0 V
1673 0 R
-63 0 V
817 251 M
0 63 V
0 1046 R
0 -63 V
1251 251 M
0 63 V
0 1046 R
0 -63 V
1685 251 M
0 63 V
0 1046 R
0 -63 V
2119 251 M
0 63 V
0 1046 R
0 -63 V
600 251 M
1736 0 V
0 1109 V
-1736 0 V
600 251 L
LT0
1311 1265 M
180 0 V
654 489 M
763 460 L
108 25 V
109 16 V
108 4 V
109 13 V
108 18 V
109 80 V
108 142 V
109 46 V
108 102 V
109 84 V
108 -39 V
2065 739 L
2173 450 L
2282 290 L
LT1
1311 1165 M
180 0 V
654 460 M
109 27 V
871 475 L
109 32 V
108 12 V
109 14 V
108 16 V
109 79 V
108 94 V
109 104 V
108 79 V
109 72 V
108 -14 V
2065 758 L
2173 470 L
2282 296 L
LT3
1311 1065 M
180 0 V
654 490 M
763 473 L
108 0 V
109 15 V
108 -2 V
109 30 V
108 51 V
109 57 V
108 78 V
109 138 V
108 67 V
109 84 V
108 -22 V
2065 783 L
2173 463 L
2282 281 L
LT0
1414 821 A
1522 850 A
1631 1037 A
1739 1014 A
1848 1112 A
1956 919 A
1414 761 M
0 121 V
1383 761 M
62 0 V
-62 121 R
62 0 V
77 -145 R
0 226 V
1491 737 M
62 0 V
-62 226 R
62 0 V
78 -61 R
0 270 V
1600 902 M
62 0 V
-62 270 R
62 0 V
77 -232 R
0 147 V
1708 940 M
62 0 V
-62 147 R
62 0 V
78 -43 R
0 135 V
-31 -135 R
62 0 V
-62 135 R
62 0 V
77 -304 R
0 88 V
-31 -88 R
62 0 V
-62 88 R
62 0 V
LT0
1414 821 A
1522 850 A
1631 1037 A
1739 1014 A
1848 1112 A
1956 919 A
1414 784 M
0 75 V
-31 -75 R
62 0 V
-62 75 R
62 0 V
77 -46 R
0 74 V
-31 -74 R
62 0 V
-62 74 R
62 0 V
78 102 R
0 96 V
-31 -96 R
62 0 V
-62 96 R
62 0 V
77 -115 R
0 87 V
-31 -87 R
62 0 V
-62 87 R
62 0 V
78 11 R
0 87 V
-31 -87 R
62 0 V
-62 87 R
62 0 V
77 -271 R
0 70 V
-31 -70 R
62 0 V
-62 70 R
62 0 V
stroke
grestore
end
showpage
}
\put(1251,1065){\makebox(0,0)[r]{{\scriptsize LDC~G}}}
\put(1251,1165){\makebox(0,0)[r]{{\scriptsize LDC~F}}}
\put(1251,1265){\makebox(0,0)[r]{{\scriptsize LDC~A}}}
\put(808,1260){\makebox(0,0){(d)}}
\put(1468,51){\makebox(0,0){$\eta$}}
\put(220,805){%
\special{ps: gsave currentpoint currentpoint translate
270 rotate neg exch neg exch translate}%
\makebox(0,0)[b]{\shortstack{$1/N dn/d\eta$}}%
\special{ps: currentpoint grestore moveto}%
}
\put(2119,151){\makebox(0,0){4}}
\put(1685,151){\makebox(0,0){2}}
\put(1251,151){\makebox(0,0){0}}
\put(817,151){\makebox(0,0){-2}}
\put(540,1202){\makebox(0,0)[r]{0.3}}
\put(540,885){\makebox(0,0)[r]{0.2}}
\put(540,568){\makebox(0,0)[r]{0.1}}
\put(540,251){\makebox(0,0)[r]{0}}
\end{picture}
    }
  \caption[dummy]{{\it Comparison between different LDC strategies and
      data corresponding to \fig{fig:comp}e. In all cases the full
      line is the default LDC~A strategy. In (a) the dashed line is
      LDC~B using only DGLAP-like chains and dotted is LDC~C, i.e.\ 
      the same but allowing the virtuality of the link closest to the
      photon to be above \tq2. In (b) the long-dashed line is LDC~D
      with $k_{\perp 0}=1$ GeV, the short-dashed is LDC~A$_z$ with
      increased cutoff, $z_{\mbox{cut}}$, in the splitting functions
      and dotted is LDC~H, allowing some propagators below the cutoff.
      In (c) the dashed line is LDC~E without Sudakov form factors and
      disallowing all $q\rightarrow q$ and most $g\rightarrow q$
      splittings in the parton density fit and the dotted, LDC~A$_0$,
      is the same as LDC~A but using Sudakov only in the parton
      density fit. Finally in (d) the dashed line is LDC~F using a
      different form of the input densities at large $x$ and the
      dotted is LDC~G restricting the fit to \tf2\ data with
      $x<0.1$.}}
  \label{fig:comp3}
\end{figure}
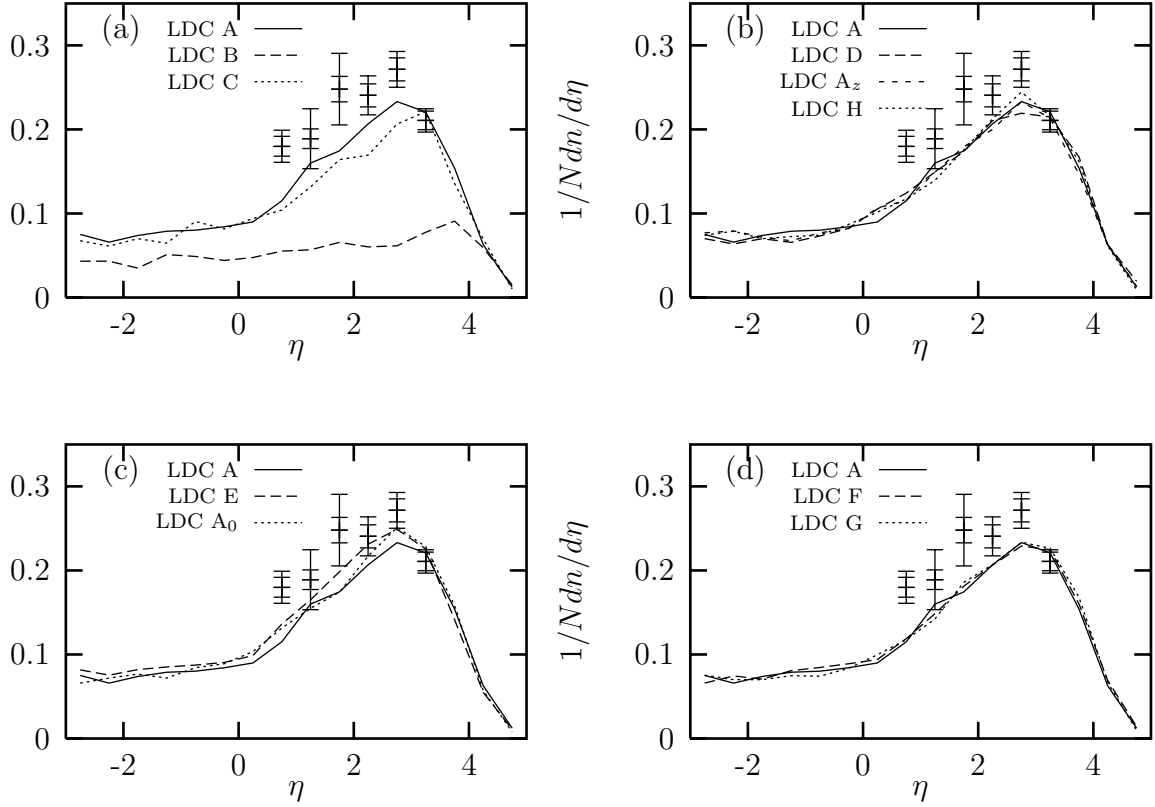

To study the hadronic final states we use the HZTOOL package
\cite{HZTOOL} developed jointly by H1, ZEUS and theoreticians for
comparison between event generators and published experimental data.
We have selected six different distributions which have been shown to
be sensitive to details in the models used in Monte Carlo event
generators. The distributions presented in \fig{fig:comp}, which are
all measured in the hadronic centre of mass system, are as follows.
\begin{itemize}
\item[(a)] The \tet-flow as a function of the pseudo rapidity for two
  bins in $x$ and \tq2, one with low $x$ and low \tq2, and one with
  moderate $x$ and \tq2\ in (b) \cite{ForwardJets}.  The large amount
  of \tet\ in the forward direction was previously claimed to be a
  good signal of \tkt\ non-ordering in the ISB, but it has been shown
  that this effect can also be obtained by the introduction of
  additional non-perturbative effects.
\item[(c)] The so-called seagull plot with the average \tk2t\ as a function
  of Feynman-$x$ \cite{HZ95221}, which at eg.\ EMC
  \cite{EMC} was shown to be difficult to reproduce with event
  generators.
\item[(d)] The \tkt-distribution of charged particles in a forward
  pseudo rapidity bin. This was recently proposed
  \cite{ForwardParticles,Kuhlen} as a new signal for perturbative
  activity in the forward region indicating \tkt\ non-ordering: a
  high-\tkt\ tail would be difficult to reproduce by non-perturbative
  models, where such tails would be exponentially suppressed.
\item[(e)] In \cite{ForwardParticles} was also shown that the
  pseudo rapidity distribution of charged particles with $\kt>1$ GeV
  also could be a good signal for \tkt\ non-ordering.
\item[(f)] Result for the two-jet ratio $R_2$ was recently reported
  \cite{R2H1} and showed large differences between the currently used
  event generators.
\end{itemize}

In \fig{fig:comp} we see the results from LDC with default settings
and using fit A, labeled LDC~A, compared with data and with the
results for \ariadne\ using the default soft radiation model, with
\lepto\ with and without the additional non-perturbative (soft colour
interactions (SCI), and perturbative-like treatment of remnants in the
case of sea quarks) assumptions presented in \cite{LeptoSCI}. For the
\ariadne\ and \lepto\ models we confirm previous results where
\ariadne, which until now was the only model implementing \tkt\ 
non-ordering, reproduces the data very well while the DGLAP-based
\lepto\ has difficulties, especially without the additional
non-perturbative models. We find that the result for LDC is quite
acceptable, although not reproducing data as well as \ariadne.

To compare different LDC strategies, we show in \fig{fig:comp3} only
the number of charged particles with transverse momentum larger than 1
GeV as a function of pseudo rapidity for small $x$ and \tq2\ 
(\fig{fig:comp3}). The effects on the other distributions in
\fig{fig:comp} are very similar.

In \fig{fig:comp3}a we see the results for LDC when restricting to
DGLAP-like chains. In this case, corresponding to the lines marked
LDC~B, the result is very poor as expected.  Allowing the virtuality
of the link closest to the photon to be above \tq2\ as for LDC~C,
makes things much better and only slightly worse than the default
LDC~A. Naïvely one may expect this to give the same result as \lepto\ 
which uses the exact \tordas\ Matrix Element for the emission closest
to the photon, also allowing the first link to have a virtuality
larger than \tq2, and which adds on parton-showers à la DGLAP on such
configurations. But in LDC, even though no ISB emissions are allowed
between the highest virtuality link and the photon, there is still a
resummation of diagrams which are then replaced by FSB emissions.
\lepto, however, uses the 'bare' matrix element and does not include
any resummation.

For the line marked LDC~D in \fig{fig:comp3}b, the cutoff in \tkt\ for
the ISB is set to 1 GeV. In the previous section we saw that the
parton density functions in this case became dramatically different.
For the final state, however, the reduction of ISB is compensated by
final state dipole emissions, which are allowed in the whole rapidity
range below $\kt=1$ GeV down to the cutoff fitted to LEP data,
$k_{\perp 0\mbox{\tiny LEP}}=0.6$.  The effect of increasing \tzcut\ 
is also shown in \fig{fig:comp3}b for the line LDC~A$_z$. The
dependence on this cutoff is small, which is expected as most of the
emissions with $z>0.5$ are counted as FSB as explained in
\fig{fig:largez} above. Also in \fig{fig:comp3}b is shown the effects
of allowing the virtuality of some links below the cutoff. Again the
differences are small.

In \fig{fig:comp3}c we see the effect of different regularizations of
the splitting functions. The line LDC~E uses the fit E in the previous
section, where all $q\rightarrow q$ and most of the $g\rightarrow q$
splittings are disallowed and all Sudakov form factors set to 1. For
the final state generation, all splittings are again included, and we
see a clear enhancement w.r.t. the default LDC~A strategy which
includes Sudakov form factors. The line LDC~A$_0$ uses the same parton
densities as the default strategy, but excludes the Sudakov form
factors when generating the final states. The main effect of the form
factor is to scale down all weights, and since in the total number of
events is fixed, the effects on the final state is small. The fact
that LDC~E is as different is then mostly due to the difference in
the input parton densities.

Finally in \fig{fig:comp3}d, we show the effects of using different
fitting procedures for the input parton densities and we see that the
differences are small.


\section{Conclusions}
\label{sec:sum}

We have here presented the first implementation of the Linked Dipole
Chain model in an event generator. Being based on the CCFM formalism
it represents one of the first attempts to correctly describe the
details of the hadronic final states in small-$x$ deep inelastic
scattering to leading-log accuracy.

The LDC model was originally formulated only for $g\rightarrow g$
splittings in a strict leading-log approximation. Going from this to a
full event generator is not trivial, and we have here described how we
implement massive quarks splittings, sub-leading corrections,
convolution with input parton densities, energy and momentum
conservation, Sudakov form factors, final-state radiation and
hadronization.

Our implementation still suffers from some uncertainties. One is the
input parton densities which are poorly constrained because only \tf2\ 
data can be fitted, where the gluon only enters indirectly. Another
issue is the uncertainty in how to deal with the final state of chains
where a link drops below the cutoff, surrounded by one perturbative
system on each side.  But from the results in the previous sections,
the main uncertainty is the regularization of the splitting functions
and the conservation of total momentum in the evolved parton density
functions. The correct way to treat this is with Sudakov form factors.
But in the current implementation these are only treated in an
approximate way and they need to be examined in more detail in the
future.

Despite these uncertainties, the result presented here allows for some
conclusions. Compared with the \lepto\ event generator, which is based
on DGLAP evolution, LDC is clearly better in describing the hadronic
activity in the forward region at HERA at small $x$. The description
is not perfect, however, and it seems that LDC is still
underestimating the perturbative activity in the forward region.

We have also shown the relative importance between DGLAP-like chains
with monotonically increasing virtualities and the unordered chains in
the full LDC model, and found that unordered chains indeed are very
important, but that at HERA, most of the activity can be attributed to
DGLAP-like chains where the link closest to the photon is allowed to
be above \tq2.

The main goals for the future developments of the LDC generator is to
investigate the exact form of the Sudakov form factors, and to include
a treatment of hadron--hadron collisions, to get a better constraint
on the input parton densities. Also further studies of the final state
of chains with two or more perturbative systems connected with
sub-cutoff links ought to be done. This is especially interesting in
connection with the large fraction of rapidity-gap events found at
HERA.

Despite the shortcomings of the current implementation, we feel that
the LDC event generator may become a very important tool for
understanding the small-$x$ hadronic final states at HERA.


\section*{Acknowledgements}

We would like to thank Bo Andersson and Gösta Gustafson for important
contributions and discussions. We would also like to thank Hannes Jung
for providing us with a preliminary HZTOOL implementation of the H1
results on di-jet production \cite{R2H1}.

\begin{appendix}
\section*{Appendix A: Colour connected matrix element corrections}
\label{clapp}
We will here describe the splitting functions that are used for the
local sub-collisions (see point~\ref{enum:splitfn} in
section~\ref{sec:MC}). These are derived from the corresponding {\em
  colour connected} $2\rightarrow 2$ QCD matrix elements presented in
\cite{HUB&co}.

\begin{figure}[t]
\begin{center}
\epsfig{figure=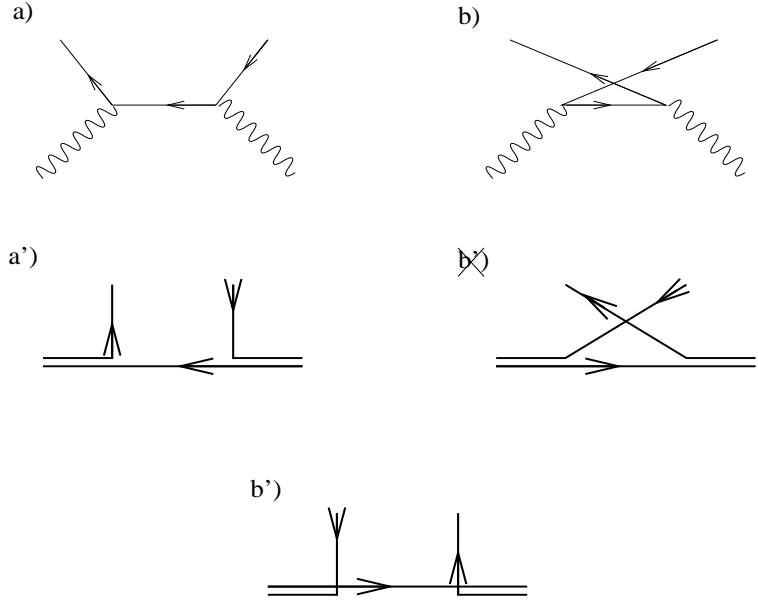,width=10cm, height=8cm}
\end{center}
\caption[dummy]{{\it Feynman vs.\ colour connection diagrams for the $g+g\rightarrow
q+q$ sub-process. If we do not average the colour states, the
diagrams a') and b') give separate contributions. The contribution
from diagram b') should be suppressed.}}
\label{colcon}
\end{figure}

In order to use matrix elements for deriving splitting functions, it
is necessary to take colour connections into account. We can
illustrate this statement with the matrix element for the
$g+g\rightarrow q+\bar{q}$ process. Suppose we are in a frame where
the incoming gluons are moving head on and parallel to the
longitudinal axis and let $\zeta_{+(-)}$ be the positive (negative)
light-cone momentum fraction for the quark and $\bar{\zeta}_{+(-)}$ be
the same for the anti-quark.  Because of symmetry between the quark
and the anti-quark it is clear that the matrix element must be
symmetric in these variables: $|{\cal
  M}|^2(\zeta_{+(-)},\bar{\zeta}_{+(-)})= |{\cal
  M}|^2(\bar{\zeta}_{+(-)},\zeta_{+(-)})$.  In the program, the
colours are connected randomly so that e.g. the quark has the same
probability to be connected with the gluon coming in from either the
proton or the photon side. This would mean that the two diagrams ${\rm
  a}')$ and ${\rm b}'\!\!\!\!\! \bigtimes \!\!)$ in \fig{colcon}
would be equally probable.

This is because the matrix element is calculated with an averaging of
the colour states of the gluons. The corresponding colour connected
matrix element, $|{\cal M}_{cc}|^2$ is on the other hand calculated
with the assumption that the colour states of the incoming gluons are
known. This means that we get two separate contributions, one from the
case that the quark is colour connected with the gluon coming in from
one side, say the proton side, and one contribution from the case that
it is connected with the gluon from the photon side. Each of these
contributions is non-symmetric with respect to the quark and anti-quark
variables, while the sum of them still is symmetric.

A similar procedure is used in the ${\cal O}(\alpha\as)$ matrix
element correction (see point~\ref{enum:wend} in
section~\ref{sec:MC}), to separate the $\gamma g$ matrix element into
two contributions corresponding to quark and anti-quark scattering
respectively as described in \refc{Mike}.

For a colour connected matrix element 
$|{\cal M}_{cc}|^2(\hat{s},\hat{t},\hat{u})$, the
corresponding splitting function is given by the formula
\begin{eqnarray}
P(z)\propto z(1-z)\cdot |{\cal M}_{cc}|^2(1,z,1-z) \ \ \ \ 
{\rm with}\ \ z=z_+\approx z_-.
\end{eqnarray}
The splitting functions $\Pslash_{ijk}(z)$ where the colour factor 
is divided out and with the successive flavours $ijk$ for
the propagators, become
\begin{eqnarray}
\Pslash_{qqq}(z)&=& (1-z)^2\left[z^2+(1-z)^2\right]\nonumber \\
\Pslash_{qqg}(z)=\Pslash_{gqq}(z)&=&0 \\
\Pslash_{qgq}(z)&=&\frac{1-z}{z}\left[1+(1-z)^2\right]\nonumber \\
\Pslash_{gqg}(z)&=& (1-z)^2\left[z^2+(1-z)^2\right]\nonumber \\
\Pslash_{qgg}(z)=\Pslash_{ggq}(z)&=&\frac{1}{2z}
\left[1+(1-z)^2\right]^2\nonumber \\
\Pslash_{ggg}(z)&=&\frac{1}{3}\nonumber 
\left\{
\frac{\left[1-z(1-z)\right]^2}{z(1-z)}\cdot\theta(0.5-z)+2\frac{1-z}{z}
\left[1-z+z^2\right]^2
\right\}.
\end{eqnarray}
The $\theta$-function in the first term of $\Pslash_{ggg}$ is a
cut-off to prevent divergences. This is needed since the Sudakov
form factors that are used regularize only regions with ordered
virtualities. The splitting functions $\Pslash_{qqg}$ and
$\Pslash_{gqq}$ are set to zero since they correspond to the
same ($g+q\rightarrow g+q$) scattering as the splitting functions
$\Pslash_{qgg}$ and $\Pslash_{ggq}$ which have been chosen to take
the whole contribution.

Here we have summed the contributions from different colour 
connections for each flavour combination. The choice of colour
connections has an effect on the multiplicity and transverse
momentum distributions at the hadronic level. Therefore, we will
in the future use this information also to choose the colour
connections for the local sub-collisions.
\end{appendix}



\begin{thebibliography}{10}
\itemsep -1mm

\bibitem{DGLAP} V.N.~Gribov, L.N.~Lipatov, \SJNP{15}{438}{1972} and 675;\\
  G.~Altarelli, G.~Parisi, \NPB{126}{298}{1977};\\
  Yu.L.~Dokshitzer, \JETP{46}{641}{1977}.

\bibitem{BFKL} E.A.~Kuraev, L.N.~Lipatov and V.S.~Fadin, \ZETF{72}{373}{1977},
  \JETP{45}{199}{1977}; \\
  Ya.Ya.~Balitsky and L.N.~Lipatov, \YF{28}{1597}{1978}, \SJNP{28}{822}{1978}.
  
\bibitem{HERAWS1} See eg.\ J.~Blümlein et al., Proceedings of the
  ``Future Physics at HERA'' workshop, Hamburg 1996, eds.\ 
  G.~Ingelman, A.~De~Roeck and R.~Klanner, vol.~1, p.~3,
  hep-ph/9609425 and references therein.
  
\bibitem{HERAWS2} See eg.\ M.~Erdmann et al., Proceedings of the
  ``Future Physics at HERA'' workshop, Hamburg 1996, eds.\ 
  G.~Ingelman, A.~De~Roeck and R.~Klanner, vol.~1, p.~500,
  hep-ph/9610327 and references therein.

\bibitem{CCFM} M.~Ciafaloni, \NPB{269}{49}{1988};\\
  S.~Catani, F.~Fiorani, G.~Machesini, \PLB{234}{339}{1990},
  \NPB{336}{18}{1990}.

\bibitem{HERWIG} G.~Marchesini et al., \CPC{67}{465}{1992}.
  
\bibitem{LEPTO} G.~Ingelman, A.~Edin and J.~Rathsman,
  \CPC{101}{108}{1997}, hep-ph/9605286.
  
\bibitem{ForwardJets} H1 Collaboration, S.~Aid et al.,
  \PLB{356}{118}{1995}, hep-ex/9506012.
  
\bibitem{ForwardParticles} H1 Collaboration, C.~Adloff et al.,
  \NPB{485}{3}{1997}, hep-ex/9610006.

\bibitem{ARIADNE} L.~Lönnblad, \CPC{71}{15}{1992}.

\bibitem{BGLP} B.~Andersson et al., \ZPC{43}{621}{1989}.

\bibitem{SMALLX} G.~Marchesini, B.~Webber, \NPB{386}{215}{1992}.

\bibitem{LDC96} B.~Andersson, G.~Gustafson, J.~Samuelsson, \NPB{463}{217}{1996}.

\bibitem{kharrDIPOLE} G.~Gustafson, \PLB{175}{453}{1986};\\
  G.~Gustafson, U.~Petterson, \NPB{306}{746}{1988};\\ B.~Andersson,
  G.~Gustafson, L.~Lönnblad, \NPB{339}{393}{1990}.

\bibitem{kharrLDC2} B.~Andersson et al., \ZPC{71}{613}{1996}.

\bibitem{HamidDIS97} H.~Kharraziha, ``The LDCMonte Carlo'', talk
  presented at the DIS~'97 conference, Chicago, April 1997, to be
  published in the proceedings.
  
\bibitem{HamidNext} B.~Andersson, G.~Gustafson and H.~Kharraziha,\\
  LU-TP~97-29, aps1997nov19\_001, hep-ph/9711403.

\bibitem{Kramer} R.D.~Peccei, R.~Rückl, \NPB{162}{125}{1980};\\
  Ch.~Rumpf, G.~Kramer, J.~Willrodt, \ZPC{7}{337}{1981}.

\bibitem{lambda} B.~Andersson, P.~Dahlqvist and G.~Gustafson,
  \PLB{214}{604}{1988}; \ZPC{44}{455}{1989}.

\bibitem{colrec} L.~Lönnblad, \ZPC{70}{107}{1996}.

\bibitem{JETSET} T.~Sjöstrand, \CPC{82}{74}{1994}
  
\bibitem{H1F2} H1 collaboration, S.~Aid et al., \NPB{470}{3}{1996},
  hep-ex/9603004.
  
\bibitem{ZEUSF2} ZEUS collaboration, M.~Derrick et al.,
  \ZPC{72}{399}{1996}, hep-ex/9607002.
  
\bibitem{NMCF2} NMC collaboration, M.~Arneodo et al.,
  \PL{B364}{107}{1995}, hep-ph/9509406.
  
\bibitem{E665F2} E665 collaboration, M.R.~Adams et al.,
  \PRD{54}{3006}{1996}
  
\bibitem{GRV} M.~Glück, E.~Reya and A.~Vogt, \ZPC{48}{471}{1990};
  \ZPC{53}{127}{1992}; \ZPC{67}{433}{1995}.

\bibitem{HZTOOL} J.~Bromley et al., Proceedings of the ``Future
  Physics at HERA'' workshop, Hamburg 1996, eds.\ G.~Ingelman,
  A.~De~Roeck and R.~Klanner, vol.~1, p.~611.
  
\bibitem{HZ95221} ZEUS Collaboration, M.~Derrick et al.,
  \ZPC{70}{1}{1996}, hep-ex/9511010.

\bibitem{EMC} EMC Collaboration, M.~Arneodo et al., \ZPC{36}{527}{1987}.
  
\bibitem{R2H1} H1 Collaboration, T.~Ahmed et al., Contributed paper,
  Abstract 247, HEP97, Jerusalem, Israel, August 1997 (unpublished).
  
\bibitem{Kuhlen} M.~Kuhlen, \PLB{382}{441}{1996}

\bibitem{LeptoSCI} A.~Edin, G.~Ingelman, J.~Rathsman,
  \PLB{366}{371}{1996}, hep-ph/9508386; \ZPC{75}{57}{1997},
  hep-ph/9605281.

\bibitem{HUB&co} H-U.~Bengtsson, \CPC{31}{323}{1984}.

\bibitem{Mike} M.H.~Seymour, \NPB{436}{443}{1995}

\end{thebibliography}
\end{document}